\documentclass[12pt,preprint]{aastex}

\usepackage[config, labelsep=none]{caption}
\usepackage{amsmath}
\usepackage{amssymb}
\usepackage{amsmath}
\usepackage{mathrsfs}
\usepackage{natbib}
\usepackage{subfig}
\usepackage{caption}

\shorttitle{Godunov Method for RMHD}
\shortauthors{Beckwith \& Stone}

\begin{document}

\title{A Second Order Godunov Method for Multidimensional Relativistic Magnetohydrodynamics}

\author{Kris Beckwith}
\affil{JILA\\
University of Colorado at Boulder\\
440 UCB\\
Boulder, CO 80301}

\email{kris.beckwith@jila.colorado.edu}

\and

\author{James M. Stone}
\affil{Department of Astrophysical Sciences\\
Princeton University\\
Princeton, NJ 08544}

\email{jstone@astro.princeton.edu}

\begin{abstract}

We describe a new Godunov algorithm for relativistic magnetohydrodynamics
(RMHD) that combines a simple, unsplit second order accurate
integrator with the constrained transport (CT) method for enforcing
the solenoidal constraint on the magnetic field.  A variety of 
approximate Riemann solvers are implemented to compute the
fluxes of the conserved variables.  The methods are tested with a
comprehensive suite of multidimensional problems.  These tests
have helped us develop a hierarchy of correction steps that are
applied when the integration algorithm predicts unphysical states
due to errors in the fluxes, or errors in the inversion between conserved and
primitive variables.  Although used exceedingly rarely, these
corrections dramatically improve the stability of the algorithm.
We present preliminary results from the application of these
algorithms to two problems in RMHD: the propagation of supersonic
magnetized jets, and the amplification of magnetic field by turbulence
driven by the relativistic Kelvin-Helmholtz instability (KHI).
Both of these applications reveal important differences between the
results computed with Riemann solvers that adopt different
approximations for the fluxes.  For example, we show that use of
Riemann solvers which include both contact and rotational
discontinuities can increase the strength of the magnetic field
within the cocoon by a factor of ten in simulations of RMHD jets,
and can increase the spectral resolution of three-dimensional RMHD
turbulence driven by the KHI by a factor of $2$.  This increase in
accuracy far outweighs the associated increase in computational cost.
Our RMHD scheme is publicly available as part of the Athena code.

\end{abstract}


\section{Introduction}\label{intro}

Study of the properties and behavior of magnetized fluids in the
relativistic limit is increasingly important for a wide variety of
astrophysical problems, such as accretion flows close to the event
horizon of a black hole \citep{Beckwith:2008a}; gamma ray bursts
\citep{Morsony:2007} and blazar jets \citep{Begelman:1998} to name
but three.  Often the inherent non-linearity of the underlying
equations and the need to account for multi-dimensional effects
means that only limited insight can be gained from purely analytic
studies.  As a result, there is a clear need for the development
of numerical algorithms to solve the equations of relativistic
magnetohydrodynamics (RMHD) in multi-dimensions.  Although algorithms
based on operator splitting have been very successful when applied
to RMHD \cite[e.g.][]{De-Villiers:2003}, in the past decade there
has been considerable effort devoted to the extension of Godunov
methods to RMHD, beginning with \cite{Komissarov:1999} and including,
for example,
\cite{Balsara:2001,Gammie:2003,Leismann:2005,Noble:2006,Del-Zanna:2007,Mignone:2006,Mignone:2009}.
Such methods have the advantage of not requiring an artificial
viscosity for shock capturing, and since they adopt the conservative
form, the coupling between total energy and momentum inherent in
relativistic flow is preserved directly.

In this paper, we describe a new Godunov scheme for RMHD.  There
are two critical ingredients to this algorithm which distinguish
it from previous work.  First and foremost is the method by which
the divergence-free constraint is enforced on the magnetic field.
Our algorithm combines the staggered, face-centered field version of
the constrained transport (CT) algorithm
with the method of \cite{Gardiner:2005,Gardiner:2008}
to compute the electric fields at cell edges.  This allows the
cell-centered, volume averaged discretization of the divergence to
be kept zero to machine precision.  Because of the tight coupling
between the conserved variables in RMHD, enforcing the divergence
free constaint through the integration algorithm using CT is likely
to offer advantages over {\em post facto} fixes to the field provided
by divergence cleaning methods \citep{Anninos:2005,Mignone:2010}.

The second crucial ingredient to our algorithm is the use of a
dimensionally unsplit integrator.  Algorithms for MHD based on
dimensional splitting require source terms that break the conservative
form \citep{Powell:1999}.  Thus, in this work we adopt the MUSCL-Hancock integrator described by \cite{Stone:2009}; hereafter we refer to the combination of this integrator and CT as the ``VL+CT" algorithm.  The VL+CT
integrator is particularly well suited to RMHD as it does not require
a characteristic decomposition of the equations of motion in the
primitive variables, which in RMHD is extremely complex
\cite[see][]{Anton:2010}, nor does it require the various source
terms necessary for integration of the MHD equations in multi-dimensions
that are required for the Corner Transport Upwind + Constrained
Transport (``CTU+CT") integrator described in
\cite{Gardiner:2005,Gardiner:2008}.  Previous experiments \cite[such
as supersonic MHD turbulence,][]{Lemaster:2009} have shown this
integrator to be more robust with only a small increase in diffusivity
compared to CTU+CT.

In addition to these two ingredients, other important aspects of
the algorithm described here are the choice of the Riemann solver
used to compute the fluxes of the conserved variables at cell edges,
and the method by which the pressure and velocity (the ``primitive"
variables) are recovered from the total energy and momentum (the
``conserved" variables).  We have implemented and tested a variety
of exact and approximate Riemann solvers for relativistic hydrodynamics
and RMHD, and we provide comparison of the accuracy and fidelity
of each on multidimensional applications in this paper.  To convert
the conserved into the primitive variables, we adopt the {\em $1D_W$}
approach of \cite{Noble:2006}, implemented as described by
\cite{Mignone:2007a} with some minor modifications.  Both the Riemann
solvers and inversion algorithm are described in more detail in the
following sections.

It is generally appreciated that numerical algorithms for RMHD are
more complex and less robust than similar methods for Newtonian
MHD, primarily because of the nonlinear couplings between the
conserved and primitive variables, and the possibility of unphysical
fluxes or superluminal velocities in approximate Riemann solvers.
We have developed a hierarchy of correction steps to control errors
introduced by these challenging aspects of the algorithm, ranging
from the use of less accurate but more robust Riemann solvers to
compute the fluxes, to the use of a first-order algorithm to integrate
individual problematic cells, to the use of approximate inversion
algorithms that break conservation when all else fails.  While these
corrections are required exceedingly rarely (in less than one in
$10^{9}$ cell updates in the most challenging cases), they are
crucial for improving the stability of the algorithm.  We document
all of our strategies in this paper with the expectation some of
them must be useful in other codes as well.

\newpage

This paper also presents a series of \emph{multi-dimensional} tests
for RMHD which have solutions that illustrate important properties
of the numerical method, such as its ability to hold symmetry, or to test
that the solenoidal constraint is preserved on the correct stencil.
Although one dimensional test suites (e.g.
\cite{Komissarov:1999,Balsara:2001,De-Villiers:2003}) are useful
for developing various elements of numerical algorithms for RMHD,
we have found multidimensional tests are far more exacting because
they require the tight couplings between components of four vectors
are handled with minimal errors, and they test whether the scheme
is free of pathologies related to (for example) violations of the
solenoidal constraint on the magnetic field.

Our algorithms for RMHD have been implemented within the Athena
code for astrophysical MHD \citep{Stone:2008}.  For this reason,
it can use features of the code that were originally developed for
Newtonian MHD, such as static mesh refinement (SMR).  The code is
freely available (with documentation) for download from the
web.\footnote{http://trac.princeton.edu/Athena}

The remainder of this paper is structured as follows. In \S\ref{theory},
we develop the equations of RMHD into a form suitable for numerical
integration. In \S\ref{numerics}, we describe the details of our
algorithm, including (1) the primitive variable inversion scheme,
(2) the various Riemann solvers we use for RMHD, (3) our method for
reconstructing the left- and right-states at cell interfaces, (4)
our methods for correcting unphysical states, (5) the
steps in the unsplit integration algorithm, and (6) the extension
of the SMR algorithm in Athena to RMHD. In \S\ref{tests}, we give
the details and results of the tests we have developed for
multi-dimensional RMHD, while in \S\ref{apps} we describe some
preliminary applications of the algorithm, using SMR, to two problems:
the propagation of supersonic jets and the development of turbulence
and magnetic field amplification in the Kelvin-Helmholtz instability.
Finally, in \S\ref{conclusion}, we summarize the work and point to
future directions of research.

\newpage
\section{Theoretical Background}\label{theory}

\subsection{Equations of RMHD}

The evolution of a relativistic, magnetized plasma is governed by
the conservation laws for particle number,
\begin{equation}
\label{mass}
\nabla_\mu [\rho U^\mu] = 0.
\end {equation}
and stress-energy,
\begin{equation}
\label{stress}
\nabla_\mu [T^{\mu}_{\nu}] = 0.
\end {equation}
The evolution of the electromagnetic field is described by
Maxwell's equations,
\begin{equation}
\label{maxwell}
\nabla_\mu [F^{\mu \nu}] = 4\pi J^\nu \;\; ; \;\;
\nabla_\mu [{\cal F}^{\mu \nu}] = 0
\end{equation}
supplemented by the equation of charge conservation
\begin{equation}
\label{charge}
\nabla_\mu [J^\mu] = 0
\end {equation}
In the above, $\rho$ is the mass density measured in the
comoving frame, $U^\mu$ is the four-velocity
(which is subject to the constraint $U^\mu U_\mu = -1$),
$T^\mu_\nu$ is the stress-energy tensor,  $J^\nu$ is the four-current density, $F^{\mu \nu}$ is the 
antisymmetic electromagnetic field tensor with
${\cal F}^{\mu \nu}$
its dual. The latter two are related to the electric, ${\cal E}^\mu$ and magnetic, ${\cal B}^\mu$ fields through:
\begin{equation}
\label{emfield}
F^{\mu \nu} = n^{\mu} {\cal E}^\nu - {\cal E}^\mu n^{\nu} + \epsilon^{\mu \nu \alpha \beta} {\cal B}_\alpha n_\beta
\;\; ; \;\;
{\cal F}^{\mu \nu} = n^{\mu} {\cal B}^\nu - {\cal B}^\mu n^{\nu} - \epsilon^{\mu \nu \alpha \beta} {\cal E}_\alpha n_\beta
\end{equation}
where $\epsilon^{\mu \nu \alpha \beta}$ is the contravariant form of the Levi-Civita tensor and $n^\nu$ is a future-pointing, time-like unit vector.
We adopt units with $c=1$ and adopt the usual convention that Greek indices run from $0$ to $3$ and are used in covariant expressions involving four-vectors and that latin indices run from $1$ to $3$ and are used to denote components of three-vectors. We adopt Lorentz-Heaviside notation for the electromagnetic fields so that factors of $\sqrt{4\pi}$ are removed. We work in a spatially flat, Cartesian coordinate system such that the line element, $ds^2$ and the metric tensor, $g_{\alpha \beta}$ are given by
\begin{equation}
ds^{2} = -(dt)^{2} + (dx^{i})^2 \;\; ; \;\;
g_{\alpha \beta} = \mathrm{diag} (-1,1,1,1)
\end{equation}
In this coordinate system, the divergence of a four-vector and tensor are
given by,
\begin{equation}
\label{divident}
\nabla_\mu ( x y^\mu) = \partial_t ( x y^t) +  \partial_i ( x y^i)\;\; ; \;\;
\nabla_\mu ( x y^\mu_\nu) = \partial_t ( x y^t_\nu) + \partial_i ( x y^i_\nu)
\end{equation}
respectively.

\newpage

We now use this set of conventions to transform the
conservations laws and Maxwell's equations \ref{mass}--\ref{maxwell}
into a form suitable for numerical integration.  We begin with
Maxwell's equations and work in the ideal MHD limit such that in the fluid rest frame, the Lorentz force on a charged particle is zero,
$F^{\mu \nu} U_\nu = 0$. Splitting the second of equation \ref{maxwell}
into its temporal and spatial components and exploiting the
antisymmetry of ${\cal F}^{\mu \nu}$ yields a constraint and an evolution equation,
\begin{equation}
\partial_i {\cal F}^{t i} = 0  \;\; ; \;\;
\partial_t {\cal F}^{t i} + \partial_j {\cal F}^{j i} = 0
\end{equation}
Substituting the definition of the electromagnetic field strength tensor and its dual (eqn. \ref{emfield}) into the second of Maxwell's equations and using the ideal MHD conditions yields:
\begin{equation}
\partial_i {\cal B}^{i} = 0 \;\; ; \;\;
\partial_t {\cal B}^{i} - \epsilon^{i j k} \partial_j {\cal E}_{k} = 0 \;\; ; \;\;
{\cal E}^{i} = - \epsilon^{i j k} {\cal B}_k V^{j}
\end{equation}
Here, $V^{i} = U^{i} / \Gamma$ is the velocity three-vector (``transport'' velocity) and $\Gamma = (1 - |V|^2)^{-1/2}$ is the Lorentz factor. Note that we can write the velocity four-vector as $U^\mu = \Gamma (1, \overrightarrow{V})$, the electric field four-vector as ${\cal E}^\mu = (0, \overrightarrow{{\cal E}})$ and the magnetic field four-vector as ${\cal B}^\mu = (0, \overrightarrow{{\cal B}})$. In terms of the three-vectors $\overrightarrow{V},\overrightarrow{{\cal E}},\overrightarrow{{\cal B}}$, the above equations take the form:
\begin{equation}
\nabla \cdot \overrightarrow{{\cal B}} = 0 \;\; ; \;\;
\partial_t {\overrightarrow{{\cal B}}} +
\nabla \times {\overrightarrow{{\cal E}}} = 0 \;\; ; \;\;
{\overrightarrow{{\cal E}}} =
- \overrightarrow{V} \times \overrightarrow{{\cal B}}
\end{equation}
These are the solenoidal
constraint, induction equation and ideal MHD condition familiar from Newtonian MHD.
To arrive at an equation describing the evolution of the momentum and total energy of the fluid, we begin by recalling that the stress-energy tensor can be decomposed into components describing the fluid :
\begin{equation}
T^{\mu}_{\nu} = \rho h u^\mu u_\nu + P_g \delta^\mu_\nu
\end{equation}
and the electromagnetic field \cite[see e.g.][]{Jackson:1975}:
\begin{equation}
\label{field-stress}
T^{\mu}_{\nu} = \frac{1}{4\pi} \left(
F^\mu_\alpha F^\alpha_\nu - \frac{1}{4} F^{\alpha \beta} F_{\alpha \beta} \delta^\mu_\nu \right)
\end{equation}
Here, $\delta^\mu_\nu$ is the Kronecker-delta symbol, $h$ is the relativistic enthalpy and $P_g$ is the gas pressure. Throughout the remainder of this work, we will assume an ideal gas equation of state, such that:
\begin{equation}
h = 1 + \frac{\gamma}{\gamma - 1} \frac{P_g}{\rho}
\end{equation}
where $\gamma$ is the adiabatic exponent (constant specific heat ratio).

\newpage

A simple expression for the stress-energy tensor of the electromagnetic field can be obtained by introducing the magnetic field four-vector:
\begin{equation}
\label{magfourvect}
b^\mu \equiv {\cal F}^{\mu \nu} U_\nu
\end{equation}
Expanding eqn. \ref{magfourvect}, substituting the definitions for the velocity three-vectors and using the ideal MHD condition yields:
\begin{equation}
b^\mu = \Gamma \left[  \overrightarrow{V} \cdot  \overrightarrow{{\cal B}}, \frac{{\cal B}^{i}}{\Gamma^2} + V^i \left( \overrightarrow{V} \cdot  \overrightarrow{{\cal B}} \right) \right]
 \;\; ; \;\;
|b|^{2} = b^\mu b_\mu = \frac{|{\cal B}|^2}{\Gamma^2} + (\overrightarrow{V} \cdot  \overrightarrow{{\cal B}})^2
\end{equation}
The stress-energy tensor for the electromagnetic field in ideal MHD then takes the form \cite[see e.g.][]{De-Villiers:2003}
\begin{equation}
\label{field-stress-energy}
T^{\mu}_{\nu} = |b|^2 u^\mu u_\nu + \frac{1}{2} |b|^2 \delta^\mu_\nu - b^\mu b_\nu
\end{equation}
In this notation, conservation of stress-energy is therefore expressed through:
\begin{equation}
\nabla_\nu [(\rho h  + |b|^2) u^\mu u_\nu + (P_g + \frac{1}{2} |b|^2) \delta^\mu_\nu - b^\mu b_\nu] = 0
\end{equation}

Finally, applying the identities for the divergence of a four-vector and tensor
given in eqn. \ref{divident} yields
\begin{equation}
\label{rmhdlaws}
\begin{split}
\partial_t (\rho U^t) + \partial_t (\rho U^i)= 0\\
\partial_t [(\rho h  + |b|^2) u^t u_j - b^t b_j] + \partial_i [(\rho h  + |b|^2) u^i u_j + (P_g + \frac{1}{2} |b|^2) \delta^i_j - b^i b_j] = 0\\
\partial_t [(\rho h  + |b|^2) u^t u_t + (P_g + \frac{1}{2} |b|^2) - b^t b_t] + \partial_i [(\rho h  + |b|^2) u^i u_t - b^i b_t]  = 0\\
\partial_t {\cal B}^i  - \epsilon^{i j k} \partial_j {\cal E}_{k}  = 0\\
\partial_i {\cal B}^i = 0
\end{split}
\end{equation}
This system of conservation laws can be cast in a standard form used for
numerical integration by defining vectors of conserved and primitive variables,
${\bf U}$ and ${\bf W}$ respectively,
\begin{equation}
\label{constate}
{\bf U} = \left(\begin{array}{c}D \\M_x \\M_y \\M_z \\E \\{\cal B}^x \\{\cal B}^y \\{\cal B}^z \end{array}\right) \;\; ; \;\;
{\bf W} = \left(\begin{array}{c}\rho \\V^x \\V^y \\V^z \\P_g \\{\cal B}^x \\{\cal B}^y \\{\cal B}^z \end{array}\right)
\end{equation}
We can express the conserved variables, ${\bf U}$ in terms of the
primitive variables, ${\bf W}$ by making use the definitions of
$\Gamma,V^i,{\cal B}^i,g_{\alpha \beta}$, yielding
\begin{equation}
\label{condef}
{\bf U}({\bf W}) = \left(\begin{array}{c}\rho \Gamma \\
(\rho h \Gamma^2 + |{\cal B}|^2) V^x - 
(\overrightarrow{V} \cdot  \overrightarrow{{\cal B}}) {\cal B}^x\\ 
(\rho h \Gamma^2 + |{\cal B}|^2) V^y - 
(\overrightarrow{V} \cdot  \overrightarrow{{\cal B}}) {\cal B}^y \\ 
(\rho h \Gamma^2 + |{\cal B}|^2) V^z - 
(\overrightarrow{V} \cdot  \overrightarrow{{\cal B}}) {\cal B}^z \\
\rho h \Gamma^2 - P_g + \frac{1}{2} |{\cal B}|^2
+ \frac{1}{2} |V|^2 |{\cal B}|^2 - 
\frac{1}{2} (\overrightarrow{V} \cdot  \overrightarrow{{\cal B}})^2 \\
{\cal B}^x \\{\cal B}^y \\{\cal B}^z \end{array}\right)
\end{equation}
Unfortunately, unlike the Newtonian case, analytic
expressions for ${\bf W}({\bf U})$ are not available and instead ${\bf W}$
must be obtained numerically. Our implementation of the required
procedure is described in \S\ref{invert}.

Defining a
vector of fluxes in the $x$ direction, ${\bf F}({\bf U})$ as
\begin{equation}
\label{flux}
{\bf F}({\bf U}) = \left(\begin{array}{c}
D V^x \\
M_x V^x - \Gamma^{-1} {\cal B}^x b_x  +P_g +\frac{1}{2}|b|^2\\
M_y V^x - \Gamma^{-1} {\cal B}^x b_y \\
M_z V^x - \Gamma^{-1} {\cal B}^x b_z \\
M_x \\
0 \\ {\cal B}^y V^x - {\cal B}^x V^y \\ {\cal B}^z V^x - {\cal B}^x V^z
\end{array}\right)
\end{equation}
where we have utilized the ideal MHD constraint, ${\cal E}^{i} = - \epsilon^{i j k} {\cal B}_k V^{j}$ in order to write (for example) ${\bf F}({\bf {\cal B}^y }) = {\cal B}^y V^x - {\cal B}^x V^y$, 
we can write the system of conservation laws eqn. \ref{rmhdlaws} in
the standard form
\begin{equation}
\label{balance}
\frac{\partial {\bf U}}{\partial t} + \frac{\partial {\bf F} ({\bf U})}{\partial x}
+ \frac{\partial {\bf G} ({\bf U})}{\partial y} + \frac{\partial {\bf H} ({\bf U})}{\partial z} = 0
\end{equation}
where expressions for ${\bf G}({\bf U})$ and ${\bf H}({\bf U})$ can
be found from ${\bf F}({\bf U})$ by cyclic permutation of indices.

\newpage

\subsection{Spatial and Temporal Discretization}

The system of conservation laws (eqn. \ref{balance}) are integrated
on a uniform Cartesian grid of dimensions $L_x,L_y,L_z$ which is
divided into $n_x,n_y,n_z$ cells such that $\delta_x = L_x / n_x$
and similarly for $\delta_y, \delta_z$. A cell centered at position
$(x_i,y_j,z_k)$ is denoted by indices $(i,j,k)$. Similarly, time
in the interval $t \in (t_0,t_f)$ is divided into $N$ \emph{uniform}
steps, determined by the requirement that a wave traveling at the
speed of light, $c=1$ crosses a fraction of a grid cell determined
by the Courant number, $C$, that is
\begin{equation}
\Delta t = C \mathrm{min} \left(\delta x^{-1}, \delta y^{-1}, \delta z^{-1} \right)
\end{equation}
Here, $C$ is determined from the stability requirements of the
algorithm, which for the second order accurate VL+CT
integrator adopted here is $C\le0.5$.

Conservation of mass (measured in the lab frame), $D$, momentum, $M_k$,
and energy, $E$ are
treated using a finite-volume discretization such that these
quantities are regarded as an average over the cell volume, $\delta
x \delta y \delta z$,
\begin{equation}
D^{n}_{i,j,k} = \frac{1}{\delta x \delta y \delta z}
\int^{z_{k+1/2}}_{z_{k-1/2}}
\int^{y_{j+1/2}}_{y_{i-1/2}}
\int^{x_{i+1/2}}_{x_{i-1/2}}
D(x,y,z,t^n) \; \mathrm{d}x \mathrm{d}y \mathrm{d}z
\end{equation}
whilst the associated fluxes are the time- and area- averaged flux through
the face of the cell,
\begin{equation}
\label{densityflux}
F(D)^{n+1/2}_{i-1/2,j,k} = \frac{1}{\delta y \delta z \delta t}
\int^{t^{n+1}}_{t^{n}}
\int^{z_{k+1/2}}_{z_{k-1/2}}
\int^{y_{j+1/2}}_{y_{i-1/2}}
\rho \Gamma V^x (x_{i-1/2},y,z,t) \; \mathrm{d}y \mathrm{d}z \mathrm{d}t
\end{equation}
These quantities are then updated according to (for example)
\begin{equation}
\begin{split}
D^{n+1}_{i,j,k} = D^{n}_{i,j,k}
- \frac{\delta t}{\delta x} \left[ F(D)^{n+1/2}_{i+1/2,j,k} 
- F(D)^{n+1/2}_{i-1/2,j,k} \right] \\
- \frac{\delta t}{\delta y} \left[ G(D)^{n+1/2}_{i,j+1/2,k} 
- G(D)^{n+1/2}_{i,j-1/2,k} \right] \\
- \frac{\delta t}{\delta z} \left[ H(D)^{n+1/2}_{i,j,k+1/2} 
- H(D)^{n+1/2}_{i,j,k-1/2} \right]
\end{split}
\end{equation}

The induction equation is integrated using a finite-area discretization
such that the magnetic field three-vector, ${\cal B}^i$ is regarded
as an area-average over the surface of the cell,
\begin{equation}
({\cal B}^x)^{n}_{i-1/2,j,k} = \frac{1}{\delta y \delta z}
\int^{z_{k+1/2}}_{z_{k-1/2}}
\int^{y_{j+1/2}}_{y_{i-1/2}}
{\cal B}^x(x_{i-1/2},y,z,t^n) \; \mathrm{d}y \mathrm{d}z
\end{equation}
whilst the associated emfs are averaged along the appropriate line element are,
\begin{equation}
({\cal E}^z)^{n+1/2}_{i-1/2,j+1/2,k} = \frac{1}{\delta z \delta t}
\int^{t^{n+1}}_{t^{n}}
\int^{z_{k+1/2}}_{z_{k-1/2}}
{\cal E}^z(x_{i-1/2},y_{j+1/2},z,t) \; \mathrm{d}z \mathrm{d}t
\end{equation}
The magnetic field is then updated according to (for example)
\begin{equation}
\begin{split}
\left({\cal B}^x \right)^{n+1}_{i-1/2,j,k} = \left({\cal B}^x \right)^{n}_{i-1/2,j,k}
- \frac{\delta t}{\delta y} \left[ ({\cal E}^z)^{n+1/2}_{i-1/2,j+1/2,k}
- ({\cal E}^z)^{n+1/2}_{i-1/2,j-1/2,k} \right] \\
+ \frac{\delta t}{\delta z} \left[ ({\cal E}^y)^{n+1/2}_{i-1/2,j,k+1/2} 
- ({\cal E}^y)^{n+1/2}_{i-1/2,j,k-1/2}  \right]
\end{split}
\end{equation}
There are therefore two sets of magnetic field three-vectors utilized
in this scheme, the face-centered, surface area averaged fields
$({\cal B}^x)^{n}_{i-1/2,j,k}, ({\cal B}^y)^{n}_{i,j-1/2,k}, ({\cal
B}^z)^{n}_{i,j,k-1/2}$ which are updated using CT, and a set of
cell-centered, volume-averaged fields, $({\cal B}^x)^{n}_{i,j,k},
({\cal B}^y)^{n}_{i,j,k}, ({\cal B}^z)^{n}_{i,j,k}$ which are
computed using second-order accurate averages (for example)
\begin{equation}
({\cal B}^x)_{i,j,k} = \frac{1}{2} \left[
({\cal B}^x)_{i+1/2,j,k} + ({\cal B}^x)_{i-1/2,j,k} \right]
\end{equation}
The face centered fields are always regarded as the primary representation
of the magnetic field.

\section{Numerical Method}\label{numerics}

\subsection{Primitive variable inversion}\label{invert}

Many of the elements developed for numerical schemes for Newtonian
MHD carry across directly to the relativistic case. The major
exception to this is the method by which the vector of primitive
variables, ${\bf W}$ is recovered from the vector of conservative
variables, ${\bf U}$. In Newtonian physics, there are simple algebraic
relationships between these two sets of quantities so that one can
express ${\bf W}({\bf U})$ analytically. Unfortunately, this is not
the case in relativistic MHD and as a result, the method by which
the primitive variables are recovered from conserved quantities
lies at the heart of any numerical scheme. Detailed examination of
a variety of methods to accomplish this procedure are presented in
\cite{Noble:2006}. Our chosen method corresponds to the $1D_W$
scheme described by these authors, implemented as described by
\cite{Mignone:2007a} with the modification that we use the total
energy, $E$ as one of our conserved quantities rather than the
difference of the total energy and the rest mass, $E-D$. We take this
approach for the sake of simplicity and for compatibility with
the SMR algorithm detailed in \S\ref{smr}. The algorithm implemented
within Athena is compatible with the equation of state for an ideal gas;
extension to more general equations of state can be accomplished
by modification of this algorithm as described by (e.g.) \cite{Mignone:2007a}
for the Synge gas. We give
a brief overview of the details of our method below; the interested
reader is referred to the above references for
further details.

In our version of the scheme described by \cite{Mignone:2007a}, the
primitive variables, ${\bf W}$ are found by finding the root of a
single non-linear equation in the variable $Q= \rho h \Gamma^2$
\begin{equation}
\label{primsolve}
f(Q) = Q - P_g + \frac{(1+|V|^2)|{\cal B}|^2}{2} - \frac{S^2}{2Q^2} - E
\end{equation}
which arises directly from the definition of the total energy, $E$
(see eqn. \ref{condef}). Here, $S = M_i {\cal B}^{i}$ and the
remaining unknowns, $P_g,\Gamma$ can be written in terms of $Q$
via
\begin{equation}
\label{unknowns}
|V|^2 = \frac{|M|^2 + \frac{|S|^2}{|Q|^2} \left( 2Q + |{\cal B}|^2 \right)}
{\left( Q + |{\cal B}|^2 \right)^2} \;\; ; \;\;
\Gamma = \sqrt{\frac{1}{1-|V|^2}} \;\; ; \;\;
P_g = \frac{\gamma - 1}{\gamma}(Q - D\Gamma) (1-|V|^2)
\end{equation}
Finding the root of eqn. \ref{primsolve} is accomplished via a
Newton-Raphson (NR) iteration scheme \cite[see e.g.][]{Press:1992} for which
it is necessary to supply derivatives of$f(Q)$ with respect to $Q$
\begin{equation}
\label{primderiv}
\frac{df(Q)}{dQ} = 1 - \frac{dP_g}{dQ} + \frac{|{\cal B}|^2}{2}\frac{d|V|^2}{dQ}
+ \frac{S^2}{Q^3}
\end{equation}
where
\begin{equation}
\begin{split}
\frac{d|V|^2}{dQ}  = -\frac{2}{Q^3 (Q + |{\cal B}|^2)^3}
\left[S^2 \left( 3 Q (Q + |{\cal B}|^2) + |{\cal B}|^4 \right) + |M|^2 Q^3 \right]\\
\frac{dP_g}{dQ} = \frac{\gamma - 1}{\gamma} \left[
1 - |V|^2 -\frac{\Gamma( D + 2 (Q - D\Gamma) (1-|V|^2) \Gamma}{2}
\frac{d|V|^2}{dQ} \right]
\end{split}
\end{equation}
The NR root finder requires an initial guess for the
independent variable, ${\bf W}$. This is obtained in a similar
fashion to that described in \cite{Mignone:2007a} by finding the
positive root of the quadratic equation
\begin{equation}
f(Q) = |M|^2 - Q^2 + (2Q + |{\cal B}|^2) (2Q + |{\cal B}|^2 - 2E)
\end{equation}
which guarantees a positive initial guess for the pressure, $P_g$.
The ability to accurately predict an initial guess for the NR
iterations is a major advantage to this scheme because it means
values for the primitive variables from the previous time step do
not need to be saved for use as the first guess..  This results in a
significantly simplified code structure, particularly with regard
to implementation of algorithms for SMR (see \S\ref{smr}). Once
$f(Q) = 0$ has been determined within some desired tolerance (typical
$\sim10^{-10}$), the velocity three-vector, $V^i$ is determined
via
\begin{equation}
V^{i} = \frac{1}{Q + |{\cal B}|^2} \left( M_{i} + \frac{S {\cal B}^i}{Q} \right)
\end{equation}

\subsection{Computing the Interface States}\label{reconstruct}

The conserved variables to the left, ${\bf U}^L_{i-1/2}$ and right,
${\bf U}^R_{i-1/2}$ of the cell interface at $i-1/2$ are reconstructed
from cell-centered values using second-order accurate piecewise
linear interpolation as described in \S$4.2$ of \cite{Stone:2009}
with two important differences. Firstly, we perform limiting solely
on the primitive variables, ${\bf W}$, rather than the characteristic
variables. Secondly, we replace the velocity three-vector contained
in the primitive state, ${\bf W}$ at the cell-centers with the
four-velocity, $U^\mu$ and then recalculate the three-velocity based
on the reconstructed components of this four-vector at the cell
interfaces. This procedure helps to ensure that reconstruction does
not result in an unphysical primitive state, characterized in this
case by $|V|^2 > 1$ and is particularly important for strongly
relativistic shocks \cite[for example, the $\Gamma=30$ colliding
shock described in][]{Mignone:2006}. In the case where reconstruction
\emph{does} result in an unphysical primitive state, we revert to
first order spatial reconstruction. We note that the scheme can
easily be extended to third order spatial accuracy by implementation
of (for example) the Piecewise Parabolic Method of \cite{Colella:1984}.
We note though that improving the order of convergence of the
reconstruction algorithm is not always the best approach to improve the overall accuracy of the solution as demonstrated in \cite{Stone:2008}.

\subsection{Riemann Solvers} \label{solvers}

Computation of the time- and area-averaged fluxes (e.g. eqn.
\ref{densityflux}) is accomplished via a Riemann solver, which provides
the solution (either exact or approximate) to the initial value problem
\begin{equation}
\label{riemann}
{\bf U}(x,0) = \left\{\begin{array}{cc} 
{\bf U}^L_{i-1/2} & \mathrm{if} \;  x<x_{i-1/2}\\
{\bf U}^R_{i-1/2} & \mathrm{if} \; x>x_{i-1/2}
\end{array}\right.
\end{equation}
Here, ${\bf U}^{L,R}_{i-1/2}$ are the left- and right-states
at the zone interface located at $i-1/2$ computed in the
reconstruction step described in \S\ref{reconstruct}. A variety of Riemann
solvers of varying complexity can be used.  To date, we have
implemented three such solvers, all of which belong to the
Harten-Lax-van Leer (HLL) family of non-linear solvers.  Approximate
HLL-type solvers require knowledge of the outermost wavespeeds of
the Riemann fan, $\lambda^{L,R}$, which correspond to the fast
magnetosonic waves. Accurate calculations of speed of these waves
involve finding the roots of the quartic polynomial \citep{Anile:1989}
\begin{equation}
\rho h \left( 1 - c^2_s \right) \Gamma^4 \left( \lambda^4 - V^x  \right)
- \left(1 - \lambda^2 \right) \left[ \left( |b|^2 + \rho h c^2_s \right) \Gamma^2
\left( \lambda - V^x \right)^2 -c^2_s \left( b^x - \lambda b^t \right)^2 \right]
= 0
\end{equation}
where $c^2_s = \gamma P_g / \rho h$ is the sound speed. This quartic
is solved by standard numerical techniques \cite[see e.g.][]{Mignone:2006},
which we have found to provide an accurate solution for $\lambda$
provided a physical state is input. The roots of the quartic are
sorted to find the smallest, $\lambda^-$ and largest, $\lambda^+$
roots for both ${\bf U}^L$ and ${\bf U}^R$. Finally,  $\lambda^{L,R}$
are then found from \cite{Davis:1988}
\begin{equation}
\lambda^L =
\mathrm{min}\left[ \lambda^- ({\bf U}^L),  \lambda^- ({\bf U}^R) \right]
\;\; ; \;\;
\lambda^R =
\mathrm{max}\left[ \lambda^+ ({\bf U}^L) , \lambda^+ ({\bf U}^R) \right]
\end{equation}
We have found that the robustness of the code is greatly improved
by using accurate calculations of the wavespeed, rather than estimates
based on quadratic approximations \cite[as described in, for
example][]{Gammie:2003}.

\subsubsection{HLLE Solver}\label{HLLE-solver}

The simplest Riemann solver that we have implemented in Athena for
RMHD is the HLLE solver \cite{Harten:1983}, which computes the
solution to eqn. \ref{riemann} as
\begin{equation}
{\bf \mathsf{U}}(0,t) = \left\{\begin{array}{cc} 
{\bf U}^L & \mathrm{if} \;  \lambda^L \ge 0\\
{\bf U}^{hll} & \mathrm{if} \;  \lambda^L \le 0 \le \lambda^R\\
{\bf U}^R & \mathrm{if} \; \lambda^R \le 0 
\end{array}\right.
\end{equation}
where $\lambda^{L,R}$ are the slowest, fastest wave speeds and ${\bf
U}^{hll}$ is the state integral average of the solution of the
Riemann problem \cite{Toro:1999}
\begin{equation}
{\bf U}^{hll} =
\frac{\lambda^R {\bf U}^L - \lambda^L {\bf U}^R
+ {\bf F}({\bf U}^R) - {\bf F}({\bf U}^L)} {\lambda^R - \lambda^L}
\end{equation}
The interface flux associated with this solution is
\begin{equation}
{\bf \mathsf{F}}({\bf U}^L,{\bf U}^R) = \left\{\begin{array}{cc} 
{\bf F}({\bf U}^L) & \mathrm{if} \;  \lambda^L \ge 0\\
{\bf F}^{hll} & \mathrm{if} \;  \lambda^L \le 0 \le \lambda_R\\
{\bf F}({\bf U}^R) & \mathrm{if} \; \lambda^R \le 0 
\end{array}\right.
\end{equation}
where ${\bf F}^{hll}$ is the flux integral average of the solution of 
the Riemann problem
\begin{equation}
{\bf F}^{hll} =
\frac{\lambda^R {\bf F}({\bf U}^L) - \lambda^L {\bf F}({\bf U}^R)
+ \lambda^L \lambda^R ({\bf U}^R - {\bf U}^L)} {\lambda^R - \lambda^L}
\end{equation}
Note that setting $|\lambda^L| = |\lambda^R| = c = 1$ reverts the
interface fluxes defined above to the Lax-Friedrichs prescription,
thereby applying maximal dissipation (which for RMHD is set by the
speed of light) to the solution of the Riemann problem.

\subsubsection{HLLC Solver}\label{HLLC-solver}

Whilst the non-linear HLLE solver described above provides a robust,
simple and computationally efficient method to calculate the upwind
interface fluxes required for the solution of eqn. \ref{balance},
it has a major drawback in that contact and rotational discontinuities
are diffused even when the fluid is at rest.  \cite{Mignone:2006}
describe the extension of the non-linear HLLC (HLL ``contact'',
denoting the restoration of the contact discontinuity to the Riemann
fan) solver \citep{Toro:1999} to relativistic MHD and we have
implemented this solver within Athena.

This is accomplished by solution of the RankineÐHugoniot jump conditions
across the left- and right-going waves, as well as across a contact
wave intermediate between the two.  This requires solving a single
quadratic equation for the speed of the contact discontinuity and
then using this quantity to solve the jump conditions across the
left- and right-going waves for the intermediate states \cite[further
details can be found in][]{Mignone:2006}.  We have implemented this
solver following the version in the publicly available Pluto code
\cite{Mignone:2007} and we encourage the interested reader to refer
to both this code and Athena for algorithmic details not found in
\cite{Mignone:2006}. We have found that the HLLC solver involves
little increase in cost or complexity compared to HLLE, whilst
greatly increasing the accuracy of the resulting interface fluxes.
However, the tests in \S\ref{tests} show that this solver does
possess pathologies relating to separate treatments of the case
where ${\cal B}^x = 0$ and ${\cal B}^x \ne 0$, in addition to
exhibiting singular behavior in the case where ${\cal B}^x \to 0$
with ${\cal B}^z \ne 0$ or ${V}^z \ne 0$ (i.e.  for truly
three-dimensional MHD flows); see the discussion in \S3.3 of
\cite{Mignone:2006}.

\subsubsection{HLLD Solver}\label{HLLD-solver}

The shortcomings of the HLLC solver for truly three-dimensional MHD
flows led \cite{Mignone:2009} to extend the non-linear HLLD solver
\citep{Miyoshi:2005} to relativistic MHD.  The name is chosen to
indicate that both the contact and the rotational discontinuities
are restored to the Riemann fan.

For this solver, the intermediate states and wavespeeds are determined
by solution of the RankineÐHugoniot jump conditions across the left-
and right-going fast waves and left- and right-going rotational
discontinuities (Alfven waves), and then matching solutions are
applied across the contact discontinuity.  Unlike the case of
Newtonian MHD, the solution to this problem admits discontinuities
in the normal component of the velocity in the intermediate states
due to the effects of relativistic aberration \cite[see
e.g.][]{Anton:2010}. Despite these complexities, the solution of
the problem is determined matching the normal velocity associated
with the states to the left and right of the contact discontinuity
across this discontinuity, which amounts to solving a one-dimensional,
non-linear equation in the total pressure, accomplished by standard
numerical techniques to a typical accuracy of $\sim10^{-7}$
\cite[further details can be found in][]{Mignone:2009} We have
implemented this solver, again following the version in the publicly
available Pluto code \citep{Mignone:2007}, and we again encourage
the interested reader to refer to both this code and Athena for
algorithmic details not found in \cite{Mignone:2009}. The formulation
of the HLLD solver removes the flux singularity suffered by the
HLLC solver in the truly three-dimensional case and as such, we
find the HLLD solver is better suited than HLLC for truly
multi-dimensional MHD problems \citep{Mignone:2009}.

\subsection{When Everything Goes Wrong}\label{fixups}

The most significant challenge presented in extending a Newtonian
integration algorithm to relativistic MHD is developing a strategy
to resolve the case where the integration algorithm produces a
conserved state, ${\bf U}$ that does not correspond to a physical
primitive state, ${\bf W}$.  Such a failure can take place in one
of five different ways; firstly, the Godunov fluxes derived from
the Riemann solver can be non-real valued; secondly, the algorithm
outlined in \S\ref{invert} can fail to converge; thirdly, the density
can become negative, $\rho < 0$; fourthly, the gas pressure can
become negative, $P_g < 0$ and finally, the velocity can become
superluminal, $|V|^2 > 1$. To resolve the first of these failure
modes, one can simply verify that the Godunov fluxes obtained from
the Riemann solver are real valued and replace those that are not
with a more diffusive estimate. For the remaining failure modes,
several approaches are possible; for example, one can revert to a
first order update which applies enhanced \emph{numerical} dissipation
(derived from the Riemann solver) to the affected cells whilst
retaining the conservative properties of the algorithm \cite[see
e.g.][]{Lemaster:2009}. Alternatively, one can break the conservative
properties of the algorithm and derive the gas pressure from the
entropy, which guarantees the gas pressure to be positive definite
\cite[see e.g.][]{Noble:2008}. A third approach is to set the density
and gas pressure to floor values and derive an estimate for $|V|^2$
satisfying $|V|^2<1$ \cite[see e.g.][]{Mignone:2007}. We have
implemented all of these methods within Athena and use them
sequentially, as outlined below.

To give an indication of the frequency with which these
fixes are required for real applications, of the problems
described in \S\ref{apps}, the high resolution computation of
relativistic magnetized jet using the HLLD solver (see \S\ref{srjet})
required the greatest use of the fallback methods described here.
In this case, the first order flux correction was required approximately
once in $10^9$ updates, the entropy correction was required
approximately once in $10^{10}$ updates and the final correction
to ensure that $|V|^2<1$ was required approximately once in $10^{11}$
updates. 

\subsubsection{Correcting Non-Real Valued Fluxes}\label{nonrealflux}

In some circumstances, the HLLC and HLLD approximate Riemann solvers
described in \S\ref{solvers} can produce non-real valued fluxes.
To handle this eventuality, we test inside the Riemann solver itself
for real valued fluxes; in the circumstance that they are not, we
replace these fluxes with those derived from the HLLE solver. We
have found the HLLE solver always returns fluxes that are real
valued, provided that the input left- and right-states
correspond to a physical primitive state.

\subsubsection{First Order Flux Correction}\label{fofc}

Our primary strategy for fixing unphysical primitive variable states
within the algorithm is that of the ``first-order flux-correction".
This approach reverts to a first order update in the affected cells,
a strategy that \emph{preserves} the conservation properties of the
algorithm.  This adds a small amount of numerical dissipation,
derived directly from the Riemann solver, to the affected cells.
It has been successfully used in simulations of supersonic (Newtonian)
MHD turbulence by \cite{Lemaster:2009}. We find that this method
fixes most incidences of unphysical states resulting from the
primitive variable routine outlined in \S\ref{invert}. Only in the
rare cases where this first-order flux correction fails do we resort
to the inversion methods described in S\ref{invertent} or
\S\ref{invertvel} and break strict conservation within the code.

Adopting the notation of \cite{Stone:2009}, let us denote the first
order cell-interface fluxes used in the predict step of the VL+CT
integrator as ${\bf F}^*_{i-1/2,j,k},{\bf G}^*_{i,j-1/2,k}$ and
${\bf H}^*_{i,j,k-1/2}$ in the $x$- $y$- and $z$-directions
respectively and similarly the second-order cell-interface fluxes
used in the correct step as ${\bf F}^{n+1/2}_{i-1/2,j,k, }{\bf
G}^{n+1/2}_{i,j-1/2,k}$ and ${\bf H}^{n+1/2}_{i,j,k-1/2}$.  The
first order flux correction is applied to an affected cell denoted
by $(i_b,j_b,k_b)$ by first computing flux differences in all
three-dimensions, e.g.
\begin{equation}
\begin{split}
\delta F_{i_b-1/2,j_b,k_b} =
{\bf F}^{n+1/2}_{i_b-1/2,j_b,k_b} - {\bf F}^*_{i_b-1/2,j_b,k_b}\\
\delta G_{i_b,j_b-1/2,k_b} =
{\bf G}^{n+1/2}_{i_b,j_b-1/2,k_b} - {\bf G}^*_{i_b,j_b-1/2,k_b}\\
\delta H_{i_b,j_b,k_b-1/2} =
{\bf H}^{n+1/2}_{i_b,j_b,k_b-1/2} - {\bf H}^*_{i_b,j_b,k_b-1/2}
\end{split}
\end{equation}
These corrections are then applied to cell-centered hydrodynamic quantities 
after a full timestep update via, for example
\begin{equation}
\begin{split}
D^{n+1}_{i_b,j_b,k_b} = D^{n+1}_{i_b,j_b,k_b}
- \frac{\delta t}{\delta x} \left[ \delta F(D)_{i_b+1/2,j_b,k_b} 
- \delta F(D)_{i_b-1/2,j_b,k_b} \right] \\
- \frac{\delta t}{\delta y} \left[ \delta G(D)_{i_b,j_b+1/2,k_b} 
- \delta G(D)_{i_b,j_b-1/2,k_b} \right] \\
- \frac{\delta t}{\delta z} \left[ \delta H(D)_{i_b,j_b,k_b+1/2} 
- \delta H(D)_{i_b,j_b,k_b-1/2} \right]
\end{split}
\end{equation}
Conservation also requires corrections to cells adjacent to the unphysical
cell located at $(i_b,j_b,k_b)$, e.g.
\begin{equation*}
\begin{split}
D^{n+1}_{i_b-1,j_b,k_b} = D^{n+1}_{i_b-1,j_b,k_b}
+ \frac{\delta t}{\delta x} \delta F(D)_{i_b-1/2,j_b,k_b} \;\; ; \;\;
D^{n+1}_{i_b+1,j_b,k_b} = D^{n+1}_{i_b+1,j_b,k_b}
- \frac{\delta t}{\delta x} \delta F(D)_{i_b+1/2,j_b,k_b} \\
D^{n+1}_{i_b,j_b-1,k_b} = D^{n+1}_{i_b,j_b-1,k_b}
+ \frac{\delta t}{\delta y} \delta G(D)_{i_b,j_b-1/2,k_b} \;\; ; \;\;
D^{n+1}_{i_b,j_b+1,k_b} = D^{n+1}_{i_b,j_b+1,k_b}
- \frac{\delta t}{\delta y} \delta G(D)_{i_b,j_b+1/2,k_b} \\
D^{n+1}_{i_b,j_b,k_b-1} = D^{n+1}_{i_b,j_b,k_b-1}
+ \frac{\delta t}{\delta z} \delta H(D)_{i_b,j_b,k_b-1/2} \;\; ; \;\;
D^{n+1}_{i_b,j_b,k_b+1} = D^{n+1}_{i_b,j_b,k_b+1}
- \frac{\delta t}{\delta z} \delta H(D)_{i_b,j_b,k_b+1/2}
\end{split}
\end{equation*}

The first-order flux corrections to the cell-interface magnetic fields are
applied by first computing flux differences to the emfs via (for example)
\begin{equation}
\delta {\cal E}^z_{i_b-1/2,j_b-1/2,k_b} =
({\cal E}^z)^{n+1/2}_{i_b-1/2,j_b-1/2,k_b}
- ({\cal E}^z)^{*}_{i_b-1/2,j_b-1/2,k_b}
\end{equation}
The cell-interface fields surrounding the affected cell are then corrected
according to (for example)
\begin{equation}
\begin{split}
\left({\cal B}^x \right)^{n+1}_{i_b-1/2,j_b,k_b} =
\left({\cal B}^x \right)^{n+1}_{i_b-1/2,j_b,k_b}
+ \frac{\delta t}{\delta y} \left[ \delta {\cal E}^z_{i_b-1/2,j_b+1/2,k_b}
- \delta {\cal E}^z_{i_b-1/2,j_b-1/2,k_b} \right] \\
- \frac{\delta t}{\delta z} \left[ \delta {\cal E}^y_{i_b-1/2,j_b,k_b+1/2} 
- \delta {\cal E}^y_{i_b-1/2,j_b,k_b-1/2}  \right] \\
\left({\cal B}^x \right)^{n+1}_{i_b+1/2,j_b,k_b} =
\left({\cal B}^x \right)^{n+1}_{i_b+1/2,j_b,k_b}
+ \frac{\delta t}{\delta y} \left[ \delta {\cal E}^z_{i_b+1/2,j_b+1/2,k_b}
- \delta {\cal E}^z_{i_b+1/2,j_b-1/2,k_b} \right] \\
- \frac{\delta t}{\delta z} \left[ \delta {\cal E}^y_{i_b+1/2,j_b,k_b+1/2} 
- \delta {\cal E}^y_{i_b+1/2,j_b,k_b-1/2}  \right] \\
\end{split}
\end{equation}
Finally, conservation of magnetic flux requires corrections to the
cell-interface fields around the affected cell, for example
\begin{equation}
\begin{split}
\left({\cal B}^x \right)^{n+1}_{i_b-1/2,j_b-1,k_b} =
\left({\cal B}^x \right)^{n+1}_{i_b-1/2,j_b-1,k_b} -
\frac{\delta t}{\delta y} \delta {\cal E}^z_{i_b-1/2,j_b-1/2,k_b} \\
\left({\cal B}^x \right)^{n+1}_{i_b+1/2,j_b-1,k_b} =
\left({\cal B}^x \right)^{n+1}_{i_b+1/2,j_b-1,k_b} -
\frac{\delta t}{\delta y} \delta {\cal E}^z_{i_b+1/2,j_b-1/2,k_b} \\
\left({\cal B}^x \right)^{n+1}_{i_b-1/2,j_b+1,k_b} =
\left({\cal B}^x \right)^{n+1}_{i_b-1/2,j_b+1,k_b} +
\frac{\delta t}{\delta y} \delta {\cal E}^z_{i_b-1/2,j_b+1/2,k_b} \\
\left({\cal B}^x \right)^{n+1}_{i_b+1/2,j_b+1,k_b} =
\left({\cal B}^x \right)^{n+1}_{i_b+1/2,j_b+1,k_b} +
\frac{\delta t}{\delta y} \delta {\cal E}^z_{i_b+1/2,j_b+1/2,k_b} \\
\end{split}
\end{equation}
and
\begin{equation}
\begin{split}
\left({\cal B}^x \right)^{n+1}_{i_b-1/2,j_b,k_b-1} =
\left({\cal B}^x \right)^{n+1}_{i_b-1/2,j_b,k_b-1} +
\frac{\delta t}{\delta z} \delta {\cal E}^y_{i_b-1/2,j_b,k_b-1/2} \\
\left({\cal B}^x \right)^{n+1}_{i_b+1/2,j_b,k_b-1} =
\left({\cal B}^x \right)^{n+1}_{i_b+1/2,j_b,k_b-1} +
\frac{\delta t}{\delta z} \delta {\cal E}^y_{i_b+1/2,j_b,k_b-1/2} \\
\left({\cal B}^x \right)^{n+1}_{i_b-1/2,j_b,k_b+1} =
\left({\cal B}^x \right)^{n+1}_{i_b-1/2,j_b,k_b+1} -
\frac{\delta t}{\delta z} \delta {\cal E}^y_{i_b-1/2,j_b,k_b+1/2} \\
\left({\cal B}^x \right)^{n+1}_{i_b+1/2,j_b,k_b+1} =
\left({\cal B}^x \right)^{n+1}_{i_b+1/2,j_b,k_b+1} -
\frac{\delta t}{\delta z} \delta {\cal E}^y_{i_b+1/2,j_b,k_b+1/2} \\
\end{split}
\end{equation}
Similar corrections are applied to ${\cal B}^y,{\cal B}^z$. Once
completed, cell-centered values of the fields for \emph{all} of the corrected
cells are recomputed using second-order accurate averages.

\subsubsection{Inversion Scheme Utilizing Entropy}
\label{invertent}

In some (rare) circumstances, the method for computing the
primitive from the conserved variables outlined in \S\ref{invert} fails to
converge to a physical state. One possible solution to this problem
is to regard the plasma as a locally ideal fluid (i.e. no shocks
or dissipation) such that the total energy conservation law is
equivalent to the equation of entropy conservation
\begin{equation}
\nabla_\mu [ \rho s u^\mu ] = 0
\end{equation}
where, for the ideal gas equation of state considered here $s = P_g
/ \rho^\gamma$ \cite[see e.g.][]{Del-Zanna:2007}. This equation can
be integrated numerically in a similar fashion to the mass continuity
equation, where the conserved quantity is ${\cal S} =  \rho s \Gamma$
and the flux in the $x$-direction is $\rho s \Gamma V^x$. A modified
primitive variable inversion algorithm that utilizes the entropy
is found by replacing eqn. \ref{primsolve} and \ref{primderiv} with
\begin{equation}
\begin{split}
f(Q) = \frac{D P_g}{\rho^\gamma} - {\cal S}\\
\frac{df(Q)}{dQ} = \frac{D}{\rho^\gamma}\frac{dP_g}{dQ} - \gamma P_g \rho^{\gamma+1}\frac{d\rho}{dQ}
\end{split}
\end{equation}
Here, $dP_g/dQ$ is calculated as above and $d\rho/dQ$ is given by
\begin{equation}
\frac{d\rho}{dQ} = -\frac{D\Gamma}{2}\frac{d|V|^2}{dQ}
\end{equation}
Once $f(Q)$ has been found to some desired accuracy, the velocity
three-vector is determined as previously.

\subsubsection{Inversion Scheme to Resolve Superluminal Velocities}
\label{invertvel}

The method for determining the primitive variables utilizing the
entropy, ${\cal S}$, guarantees that $P_g$ is positive definite.
However, it is still possible to obtain a primitive state for which
$|V|^2>1$, or for the scheme to fail to converge. To resolve this
(even rarer) eventuality, we set the density, $\rho$ and pressure, $P_g$
to floor values, set $|V|^2 = 1 - \eta$ (where $\eta$ is some small
number, typically $\eta = 10^{-8}$) and find the root of \cite[see
e.g.][]{Mignone:2007}
\begin{equation}
f(|V|^2) = |V|^2 \left( Q + |{\cal B}|^2 \right)^2 - |M|^2 - \frac{|S|^2}{|Q|^2} \left( 2Q + |{\cal B}|^2 \right)
\end{equation}
where we calculate $Q = \rho h \Gamma^2$ from the density, pressure 
floors and the current value of $|V|^2$.
As mentioned above, this final resort is only required in one in every
$10^{11}$ cell updates for the most challenging applications we have tried
to date, which is exceedingly rarely.

\subsection{Integration Algorithm}\label{integrate}

At the heart of the Godunov method for RMHD we have developed in this
work is an extension to the directionally unsplit VL+CT integrator described by
\cite{Stone:2009}.  Below we outline all the steps in the RMHD versions
of this integrator.

\begin{enumerate}

\item Compute ${\bf W}({\bf U})$ at cell centers using the algorithm
described in \S\ref{invert} and recompute ${\bf U}({\bf W})$.
Form the conserved entropy variable, ${\cal S}
= P_g \rho^{1-\gamma} \Gamma$, from cell centered primitive variables.

\item Using a Riemann solver (such as those described in \S\ref{solvers}),
construct first order upwind fluxes using ${\bf U}({\bf W})$
calculated in step 1
\begin{equation}
{\bf F}^*_{i-1/2,j,k} = {\bf \mathsf{F}} ({\bf U}_{i-1,j,k},{\bf U}_{i,j,k})
\end{equation}
and similarly for ${\bf G}^*_{i,j-1/2,k}$ and ${\bf H}^*_{i,j,k-1/2}$.
Form first order fluxes for the entropy, ${\cal S}$ using an HLLE-type average.

\item Apply the algorithm of \S$4.3$ of \cite{Stone:2009} to calculate
the CT electric field at cell-corners, $({\cal
E}^x)^{*}_{i,j-1/2,k-1/2};({\cal E}^y)^{*}_{i-1/2,j,k-1/2};({\cal
E}^z)^{*}_{i-1/2,j-1/2,k}$ from the face-centered flux returned by
the Riemann solver in step 2 and a cell center reference field
calculated using the initial data at time level $n$, i.e. ${\cal E}^{i}
= - \epsilon^{i j k} {\cal B}_k V^{j}$.

\item Verify (in addition to the step of \S\ref{nonrealflux}) that
the the first order fluxes and CT electric fields are real valued
and store for use in steps 10 and 13. If the first order fluxes are
\emph{not} real valued, abort the calculation.

\item Update the cell-centered hydrodynamical variables (including
${\cal S}$) for one-half time step, $\delta t / 2$ using flux
differences in all three-dimensions. Update the face-centered
component of the magnetic field for one-half time step using CT as
described in \cite{Stone:2009}.

\item Compute the cell-centered magnetic field at the half-time
step from the average of the face centered field computed in step
5.

\item Compute ${\bf W}^{n+1/2}({\bf U}^{n+1/2})$ using the algorithm
described in \S\ref{invert} and verify that the state is physical;
that is the primitive variable inversion routine converged and
returned a state ${\bf W}$ with the properties $\rho > 0$, $P_g >
0$, $|V|^2<1$. For those cells with unphysical ${\bf W}^{n+1/2}({\bf
U}^{n+1/2})$, compute a new primitive state using the entropy ${\cal
S}$ in place of the total energy, $E$ using the algorithm from
\S\ref{invertent}. Verify that the primitive variable inversion
routine converged and returned a state ${\bf W}$ with the properties
$\rho > 0$, $P_g > 0$, $|V|^2<1$.  For cells with where ${\bf
W}^{n+1/2}({\bf U}^{n+1/2})$ remains unphysical, replace ${\bf
U}^{n+1/2}$ with ${\bf U}^{n}$. This renders the update first order
for these cells.

\item Using the second-order (piecewise linear) reconstruction
algorithm described in \S\ref{reconstruct}, compute left- and right-
state quantities at the half-time step at cell interfaces in the
$x$-direction, $[({\bf W}^L)^{n+1/2}_{i-1/2,j,k},({\bf
W}^R)^{n+1/2}_{i-1/2,j,k}]$ and verify that the reconstructed
three-velocity vector satisfies $|V|^2<1$ for both. In cells where
this constraint is violated, replace the second-order $L,R$ states
with spatially first-order states, $[({\bf W}^L)^{n+1/2}_{i-1,j,k},({\bf
W}^R)^{n+1/2}_{i,j,k}]$. Finally, recompute $[{\bf U}({\bf
W^L})^{n+1/2}_{i-1/2,j,k},{\bf U}({\bf W^R})^{n+1/2}_{i-1/2,j,k}]$
and $[{\cal S}({\bf W^L})^{n+1/2}_{i-1/2,j,k},{\cal S}({\bf
W^R})^{n+1/2}_{i-1/2,j,k}]$. Repeat for the $y$-direction and
$z$-direction.

\item Using a Riemann solver, construct 1D fluxes at cell interfaces
in all three dimensions:
\begin{equation}
{\bf F}^{n+1/2}_{i-1/2,j,k} =
{\bf \mathsf{F}} ({\bf U}^{n+1/2}_{i-1/2,j,k},{\bf U}^{n+1/2}_{i-1/2,j,k})
\end{equation}
and similarly for ${\bf G}^{n+1/2}_{i,j-1/2,k}$ and ${\bf
H}^{n+1/2}_{i,j,k-1/2}$. In all cases, the logitudinal component
of the magnetic field in the vector of left and right states is set
equal to the face-centered value at the interface. Form second order
fluxes for the entropy, ${\cal S}$ using an HLLE-type average.

\item Verify (in addition to the step of \S\ref{nonrealflux}) that
the second order fluxes computed in step 9 are real valued, replacing
with the first order fluxes saved in step 4 if not.

\item Compute a cell-centered reference electric field, ${\cal E}^{i} = - \epsilon^{i j k} {\cal B}_k V^{j}$ at $t^{n+1/2}$
using the cell-centered velocities and magnetic field computed in
Steps $3$--$5$. Then apply the algorithm of \S$4.3$ of \cite{Stone:2009}
to calculate the CT electric fields at cell-corners, $({\cal
E}^x)^{n+1/2}_{i,j-1/2,k-1/2};({\cal E}^y)^{n+1/2}_{i-1/2,j,k-1/2};({\cal
E}^z)^{n+1/2}_{i-1/2,j-1/2,k}$.

\item Update the cell-centered hydrodynamical variables (including
${\cal S}$) for a full timestep using flux differences in all
three-dimensions and the fluxes calculated in steps 9 and 10. Update
the face-centered components of the magnetic field using CT and the
emfs from step 11.

\item 
\begin{enumerate}
\item Compute ${\bf W}^{n+1}({\bf U}^{n+1})$ using the algorithm
described in \S\ref{invert} and verify that the state is physical;
that is the primitive variable inversion routine converged and
returned a state ${\bf W}$ with the properties $\rho > 0$, $P_g >
0$, $|V|^2<1$.
\item For cells with unphysical primitive states revert to a first-order
update using the algorithm described in \S\ref{fofc}.
\end{enumerate}

\item
\begin{enumerate}
\item Recompute ${\bf W}^{n+1}({\bf U}^{n+1})$ using the algorithm
described in \S\ref{invert} and verify that the state is physical;
that is the primitive variable inversion routine converged and
returned a state ${\bf W}$ with the properties $\rho > 0$, $P_g >
0$ $|V|^2<1$.
\item For those cells with unphysical primitive states, compute a
new primitive state using the entropy ${\cal S}$ in place of the
total energy, $E$ using the algorithm from \S\ref{invertent}. Verify
that the primitive variable inversion routine converged and returned
a state ${\bf W}$ with the properties $\rho > 0$, $P_g > 0$,
$|V|^2<1$. If true, recalculate the conserved state ${\bf U}$ using
this new primitive state and overwrite the hydrodynamical variables.
\item If the primitive state arising from the entropy remains
unphysical, calculate a new primitive state using the algorithm
described in \S\ref{invertvel}, recalculate ${\bf U}$ based on this
new primitive state and overwrite the hydrodynamical variables.
\end{enumerate}

\item Repeat step $1$--$14$ until the stopping criterion is reached,
i.e. $t^{n+1} \ge t_f$.

\end{enumerate}

\subsection{Static Mesh Refinement}\label{smr}

In \S\ref{srjet}, we give details of an example application of the
integration algorithm outlined above, namely that of a $\Gamma =
7$,  Mach number ${\cal M} = v_{jet} / c_{s_{jet}} = 4$ magnetized
jet computed using SMR. The SMR algorithms used here are based on
those described by \cite{Berger:1989}.  For MHD, they utilize the
second order divergence- and curl-preserving prolongation and
restriction formulas of \cite{Toth:2002}. The implementation and
testing of the SMR algorithms in Athena will be described in detail
in a future publication.  We note here that we have found the
circularly polarized Alfven wave and field loop advection test
described in \cite{Gardiner:2005,Gardiner:2008} to be essential in
verifying that the prolongation operator does not introduce magnetic
field divergence into the grid at fine/coarse boundaries.  The
prolongation algorithms for Newtonian MHD can be used in RMHD without
alteration, with the caveat that they are applied
solely to the conserved variables, $\bf{U}$.  We also have found
it necessary to verify that the conserved variables, $\bf{U}$
resulting from second order prolongation correspond to a physical
primitive state, ${\bf W}$. In the circumstance that the resulting
${\bf W}$ is unphysical, we use a first order prolongation instead.

\section{Test Problems}\label{tests}

In this section, we present a series of primarily multi-dimensional
tests of the algorithm presented in \S\ref{numerics}, along with
quantitative diagnostics that will hopefully make such tests useful
for future workers.  As we have argued in the introduction,
multi-dimensional tests are essential for MHD.  All of the tests
presented in this section were performed with the HLLD solver using
a Courant No. of $0.4$ and adiabatic index $\gamma = 5/3$ unless otherwise stated, and none require \emph{ad-hoc} numerical fixes to be run successfully.

\subsection{Large Amplitude Circularly Polarized Alfven Wave}

\begin{figure}
\begin{center}
\includegraphics[width=0.3\textwidth]{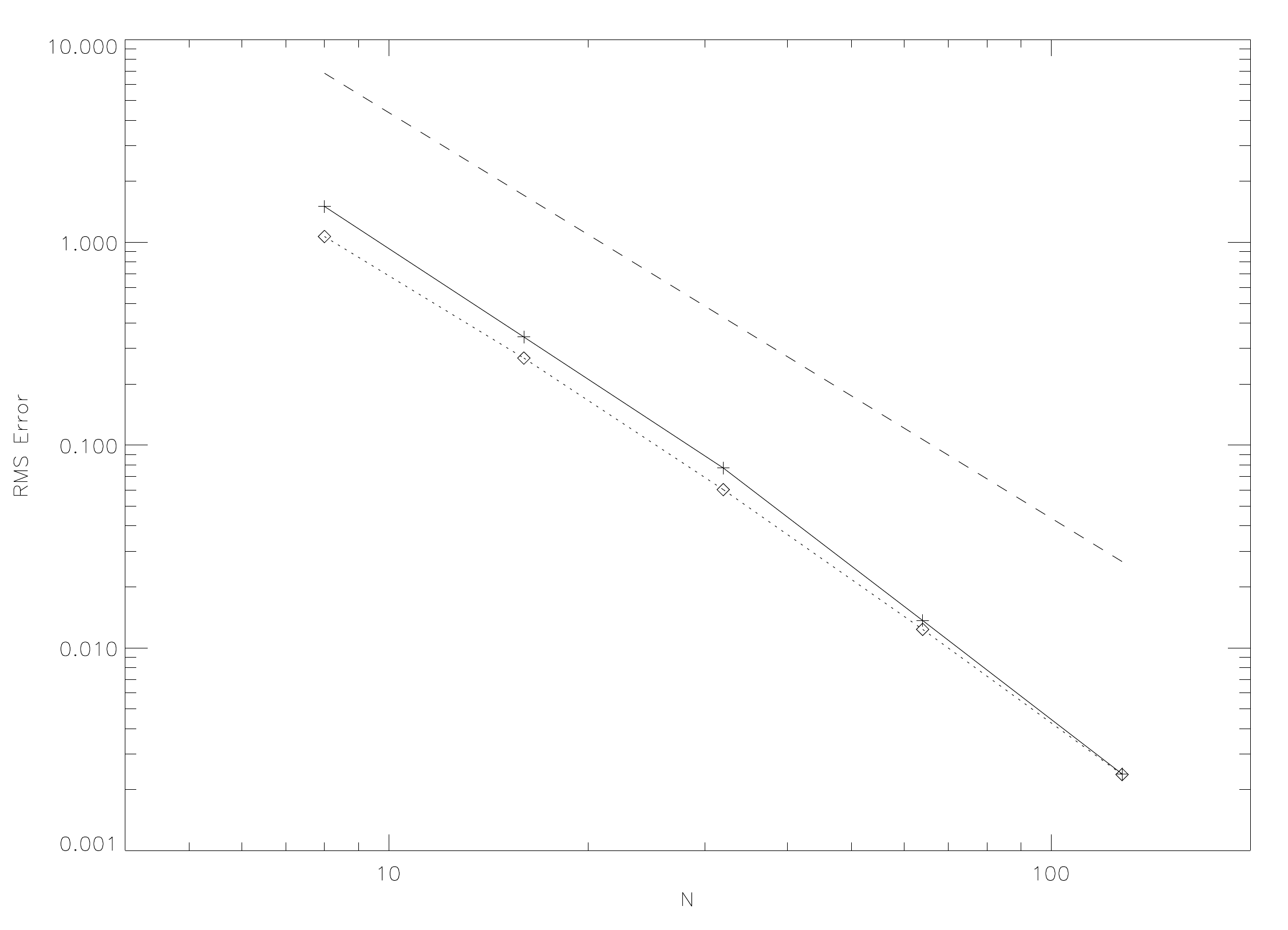}
\includegraphics[width=0.3\textwidth]{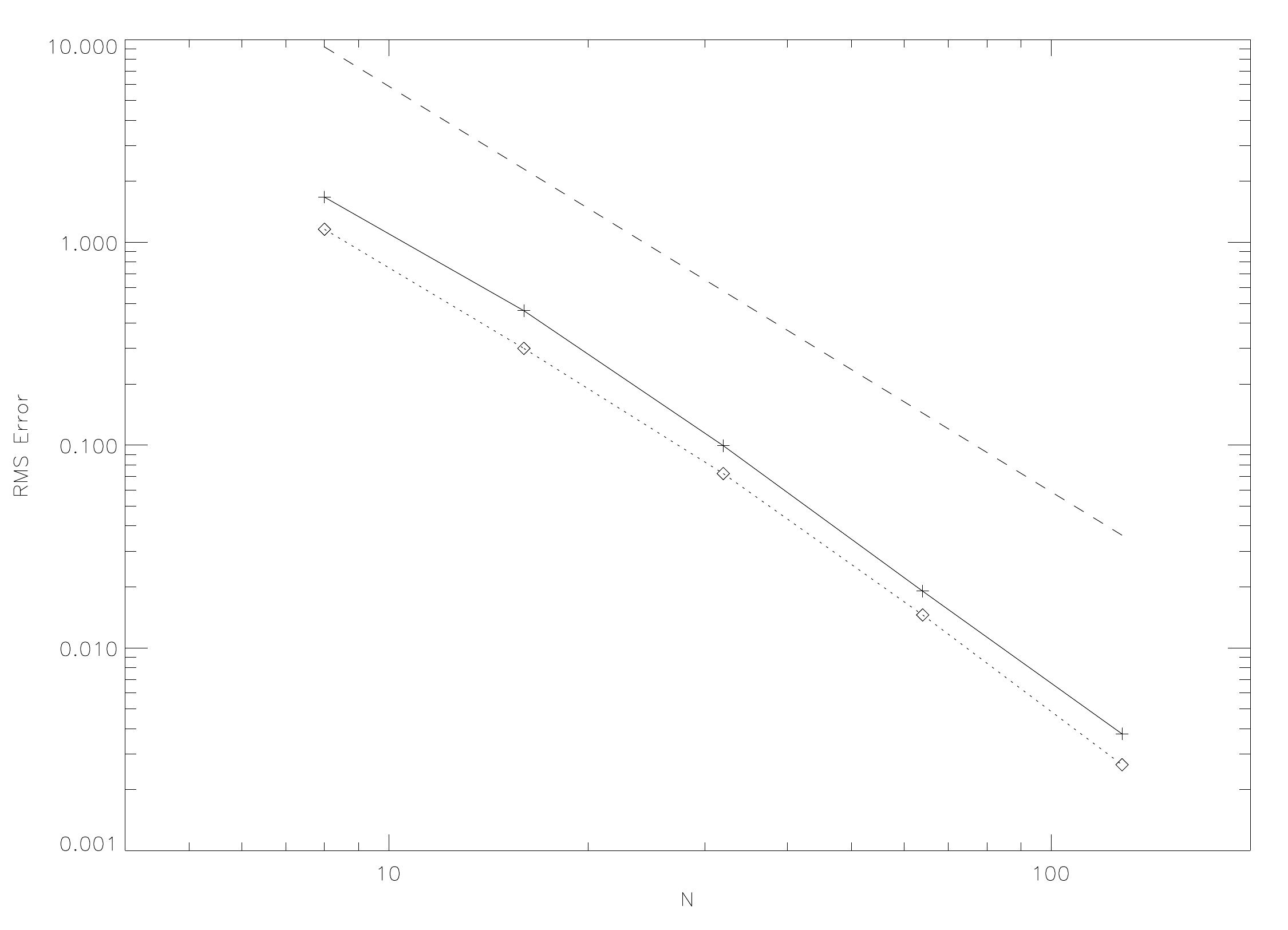}
\includegraphics[width=0.3\textwidth]{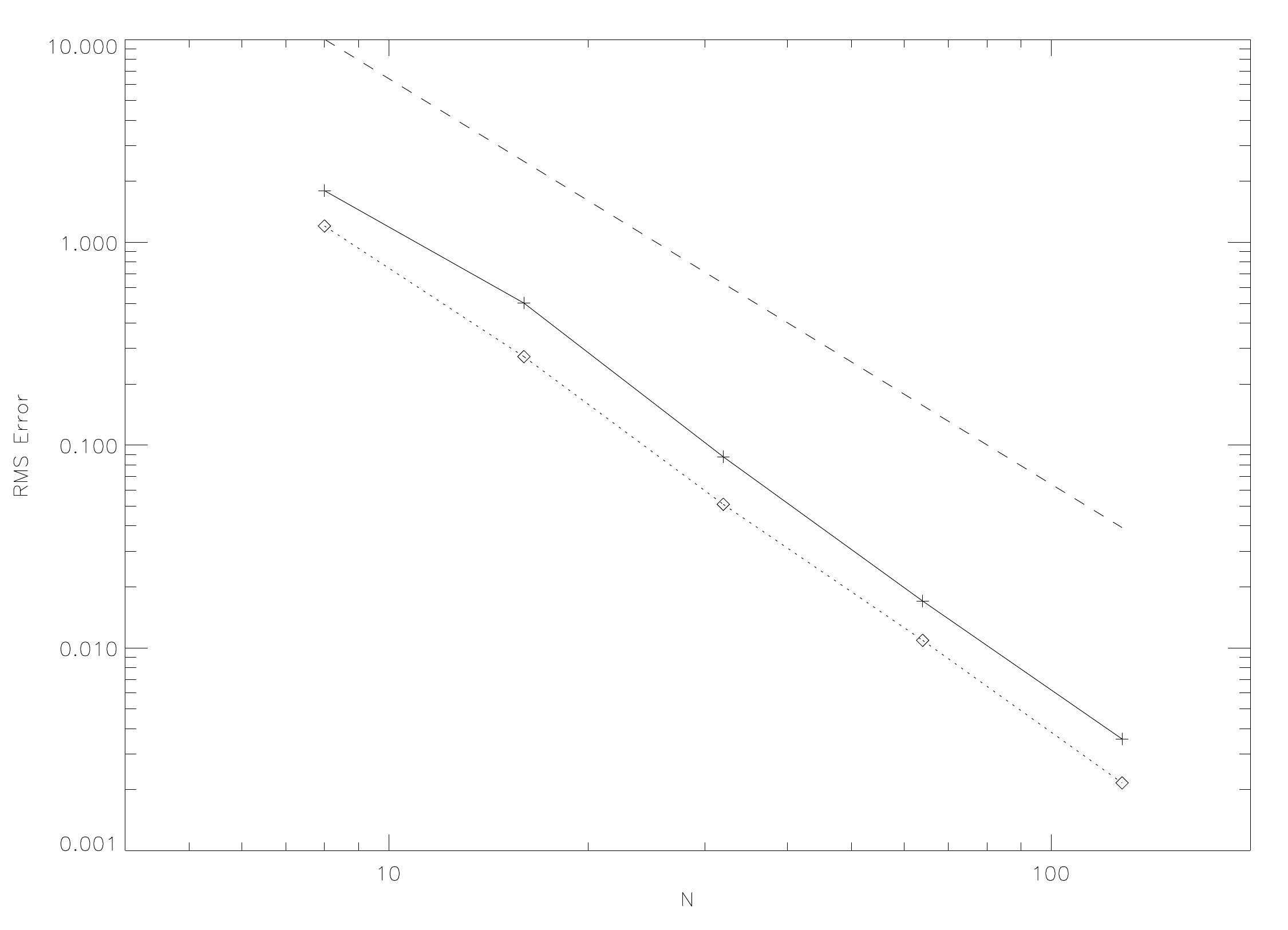}
\end{center}
\caption[]{Convergence of the large amplitude circularly polarized
Alfven wave test due to \cite{Del-Zanna:2007}. From left to right,
the panels show the overall RMS-error (see text) for one-, two- and three-dimensional versions of this test. In each panel, solid lines show results for the HLLE solver, dotted lines results for the HLLD solver and dashed
lines the expected dependence for overall 2nd order convergence.}
\label{cpaw} 
\end{figure}

A test that was found extremely useful in the development of the
Newtonian algorithm \citep{Stone:2009} is the propagation of
circularly polarized Alfven waves, which are an exact solution of
the Newtonian MHD equations. This remains the case for the relativistic
MHD equations, but here the speed at which the wave propagates is
modified by finite contributions to the fluid inertia by kinetic
and electromagnetic energies and the presence of electric fields
within the momentum equation \citep{Del-Zanna:2007}. The one-dimensional
form of this test is initialized in a similar fashion to
\cite{Del-Zanna:2007} with
\begin{equation}
\begin{split}
{\cal B}^y = A_0 {\cal B}^x \cos (kx) \;\; ; \;\;
{\cal B}^z = A_0 {\cal B}^x \sin (kx) \\
V^y = - v_A A_0 \cos (kx) \;\; ; \;\;
V^z = - v_A A_0 \sin (kx)
\end{split}
\end{equation}
where $A_0$ is the wave amplitude, $k = 2\pi / L_x$ is the wavevector
and $v_A$ is the Alfven speed
\begin{equation}
v^2_A = \frac{\left({\cal B}^x\right)^2}
{\rho h + \left({\cal B}^x\right)^2 \left(1 + A^2_0 \right)}
\left\{ \frac{1}{2}\left[1+ \sqrt{1-\left(\frac{2 A_0 \left({\cal B}^x\right)^2}
{\rho h + \left({\cal B}^x\right)^2 \left(1 + A^2_0 \right)} \right)^2}
\right] \right\}^{-1}
\end{equation}
To make this a truly multi-dimensional, the wave is placed at an
oblique angle to the grid, as in \cite{Gardiner:2005,Gardiner:2008}.
The wave is initialized with parameters $\rho = 1, P_g = 1, \eta =
1, {\cal B}^x = 1$ on a unit cell, $L_x=1.0$. The solution is evolved
for one grid crossing time in one-, two- and three-dimensions and
the modulus of the mean L1-norm errors between the evolved solution
and initial condition is measured via (for example)
\begin{equation}
\delta {\bf U}^n = \frac{1}{N^3}\sum_{i,j,k} | {\bf U}^n_{i,j,k} - {\bf U}^0_{i,j,k}|
\end{equation}
Results for the HLLE and HLLD solvers in one-, two- and three-dimensions
are shown in Figure \ref{cpaw}. We find overall second order
convergence for both these solvers in each test. The HLLD solver
exhibits increased accuracy over the HLLE solver at a given resolution;
at the highest resolution in three-dimensions, the HLLD solver is
a factor of $1.65$ more accurate than the HLLE solver, whilst the
overall code performance is reduced by a factor of $1.32$ for the
HLLD solver.  We conclude that for this test the HLLD solver yields
the best compromise between accuracy and computational cost.

The HLLC solver fails this test in multi-dimensions for the choice
of parameters used here.  As discussed in \S\ref{HLLC-solver}, this
is due to the pathologies associated with the lack of rotational
discontinuities in the assumed solution.  The HLLC solver can be
made to pass this test if ${\cal B}^x \le 0.1$, or if the Alfven
wave is aligned (rather than oblique) to the computational grid.
When the Alfven wave is obliquely with respect to the grid, there are
cells where the conditions for the flux singularity exhibited by
this solver are fulfilled, i.e. that the field normal to the cell
interface tends to zero, whilst the transverse fields and velocities
remain non-zero. Evolutions with the Alven wave aligned to the grid
remove the conditions for the flux singularity, whilst evolutions
with weaker fields reduce the severity of this singularity. This
test therefore emphasizes the importance of performing \emph{truly}
multi-dimensional tests for MHD algorithms; simply executing this
test in one-dimension, or with grid-aligned Alfven waves in
multi-dimensions would not reveal this particular failure mode of
the HLLC solver.

\subsection{Field Loop Advection}\label{field_loop}

A particularly discriminating multi-dimensional test for Newtonian
MHD is the advection of a magnetic field loop,
\cite{Gardiner:2005}. In the Newtonian limit, this test problem
probes whether $\nabla \cdot {\bf B} = 0$ on the appropriate numerical
stencil \citep{Gardiner:2008} by monitoring the evolution of $B^z$,
which in our notation is given by (assuming ${\cal B}^z = 0$ and
$V^z = \mathrm{cons.} \ne 0$)
\begin{equation}
\partial_t {\cal B}^z =
V^z \left( \partial_x  {\cal B}^x
+ \partial_y {\cal B}^y \right)
\end{equation}
Therefore, if $V^z = \mathrm{cons.} \ne 0$ and $\partial_x  {\cal
B}^x + \partial_y {\cal B}^y = 0$ (as required from the solenoidal
constraint), then if ${\cal B}^z = 0$ initially, it must remain so
for all time.

In relativistic MHD, this test also probes the ability of the primitive variable inversion scheme to maintain uniform $V^z\ne0$. This test is non-trivial since the $z$-component of the fluid velocity is recovered from the momentum via (assuming ${\cal B}^z = 0$)
\begin{equation}
V^{z} = \frac{M_z}{Q + |{\cal B}|^2}
\end{equation}
where $Q = \rho h \Gamma^2$ is determined by the (numerical) solution of a non-linear equation, as described in \S\ref{invert}. $Q$ is therefore known only within some tolerance, $\delta = 10^{-10}$, which will result in errors in $\delta V^z / V^z \sim \delta / (Q + |{\cal B}|^2)$ (where $\delta$ has the dimensions of inertia). As a result, it is impossible to maintain uniform $V^z \ne 0$ to machine accuracy in RMHD\footnote{Note that this is not to say that we would be unable to recover the case $V^z=0$ when $M^z, {\cal B}^z = 0$; this is guaranteed by the equation defining $V^i$ in terms of these quantities.}, which in turn can drive the evolution of ${\cal B}^z$ \emph{even} if $\nabla \cdot {\bf B} = 0$ on the appropriate numerical
stencil. This serves to highlight the importance of first developing numerical schemes for Newtonian MHD \emph{before} the relativistic case.

One useful strategy
to assess the ability of the algorithm to evolve a field
loop in RMHD is to compare the evolution of identical
field loops on the $x-y$ plane with either $V^z = 0$ or $V^z =
\mathrm{cons.} \ne 0$.  We have run such tests on a grid with a $2:1$
aspect ratio using $256\times128$ zones in $(x,y)$ with $\delta x
= \delta y = 7.8125\times10^{-3}$. The initial condition for the
hydrodynamic variables consists of a uniform density, $\rho=1$ and
gas pressure, $P_g=3$ medium with either $V^i = (0.2,0.1,0)/\sqrt{6}$
or $V^i = (0.2,0.1,0.1)/\sqrt{6}$. The magnetic field was initialized
as in \cite{Gardiner:2005}
\begin{equation}
A_z = \left\{ \begin{array}{cc}A_0 (R - r) & r \le R \\0 & r > R\end{array}\right.
\end{equation}
where $A_0 = 10^{-3}$, $R=0.3$ and $r = \sqrt{x^2 + y^2}$. With these parameters, $1000$ complete cycles of the primitive variable inversion scheme over the entire grid (where the result of each cycle is fed back into the conserved variables, but no fluid evolution takes place) produces a maximum fractional error $\delta V^z / V^z  = 5\times10^{-14}$.

\begin{figure}
\begin{center}
\includegraphics[width=0.3\textwidth]{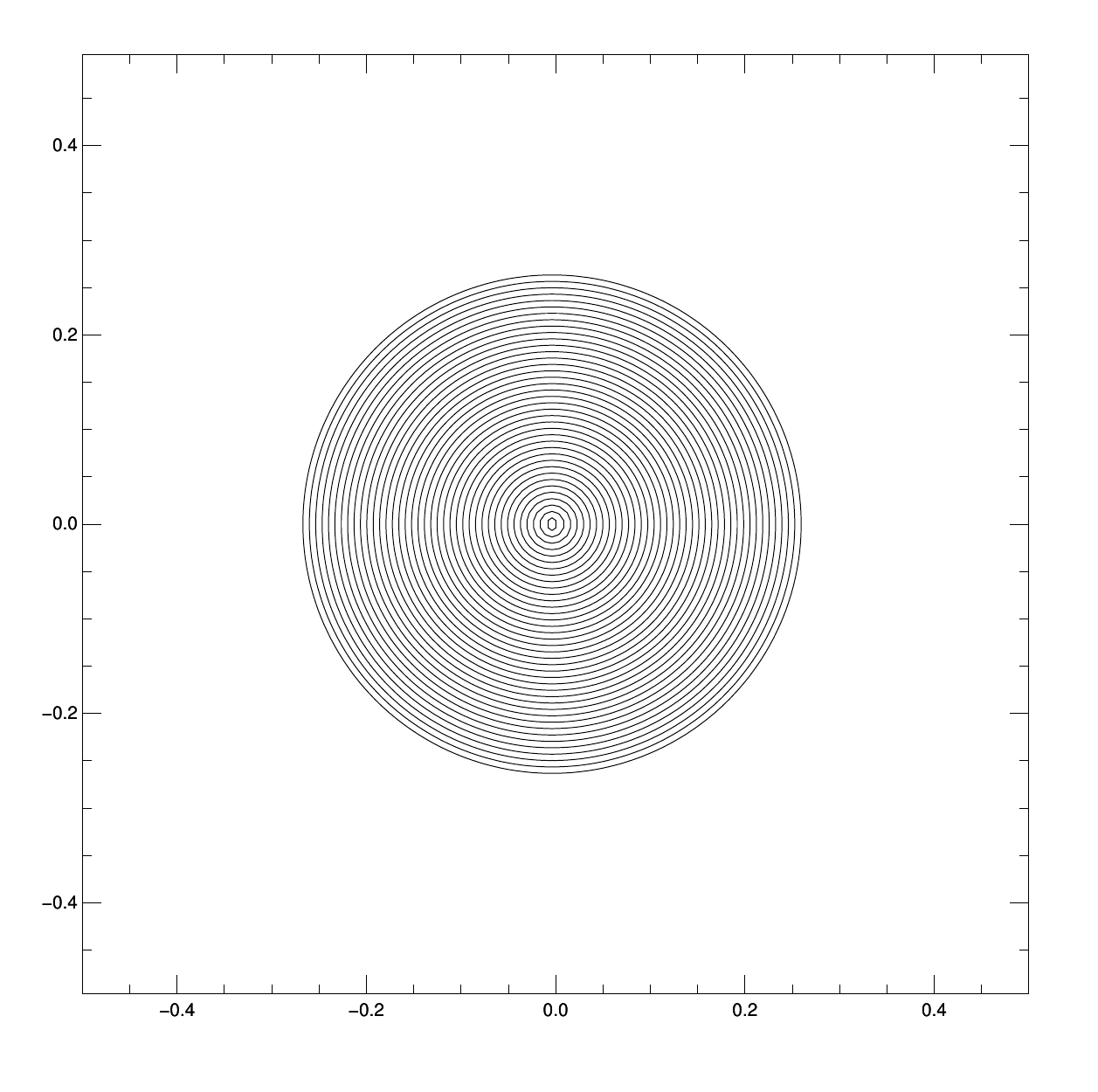}
\includegraphics[width=0.3\textwidth]{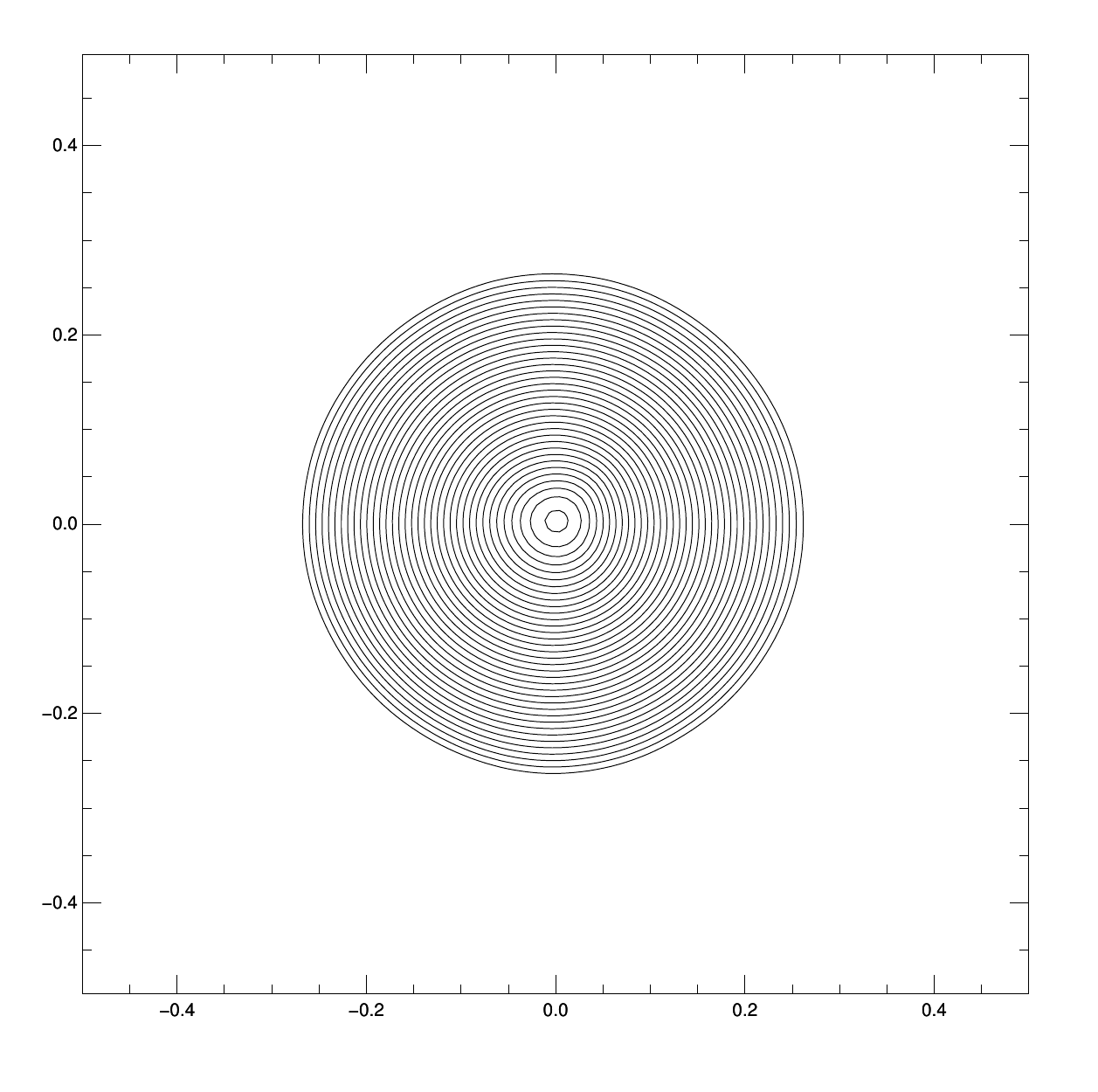}
\includegraphics[width=0.3\textwidth]{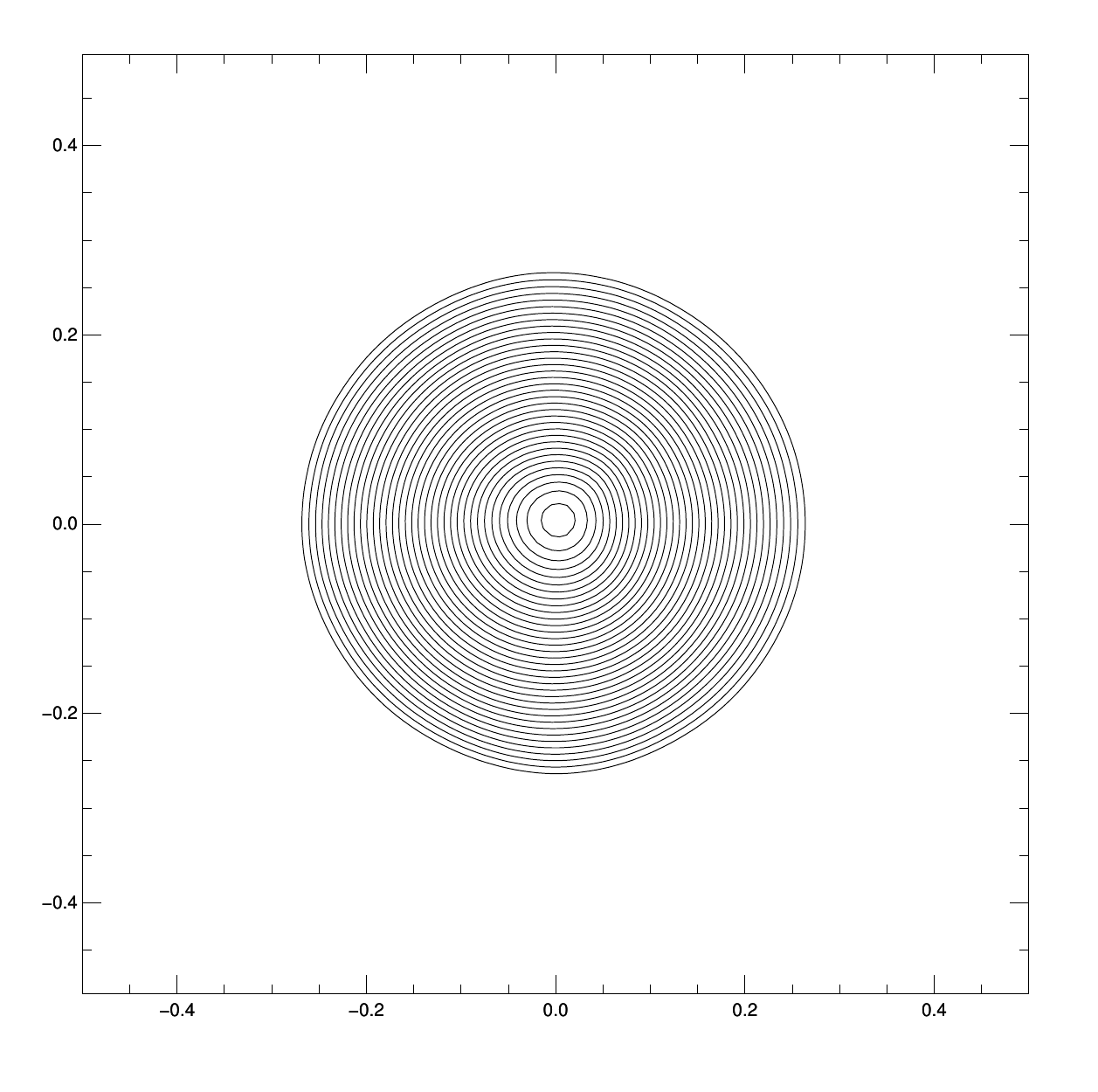}
\includegraphics[width=0.3\textwidth]{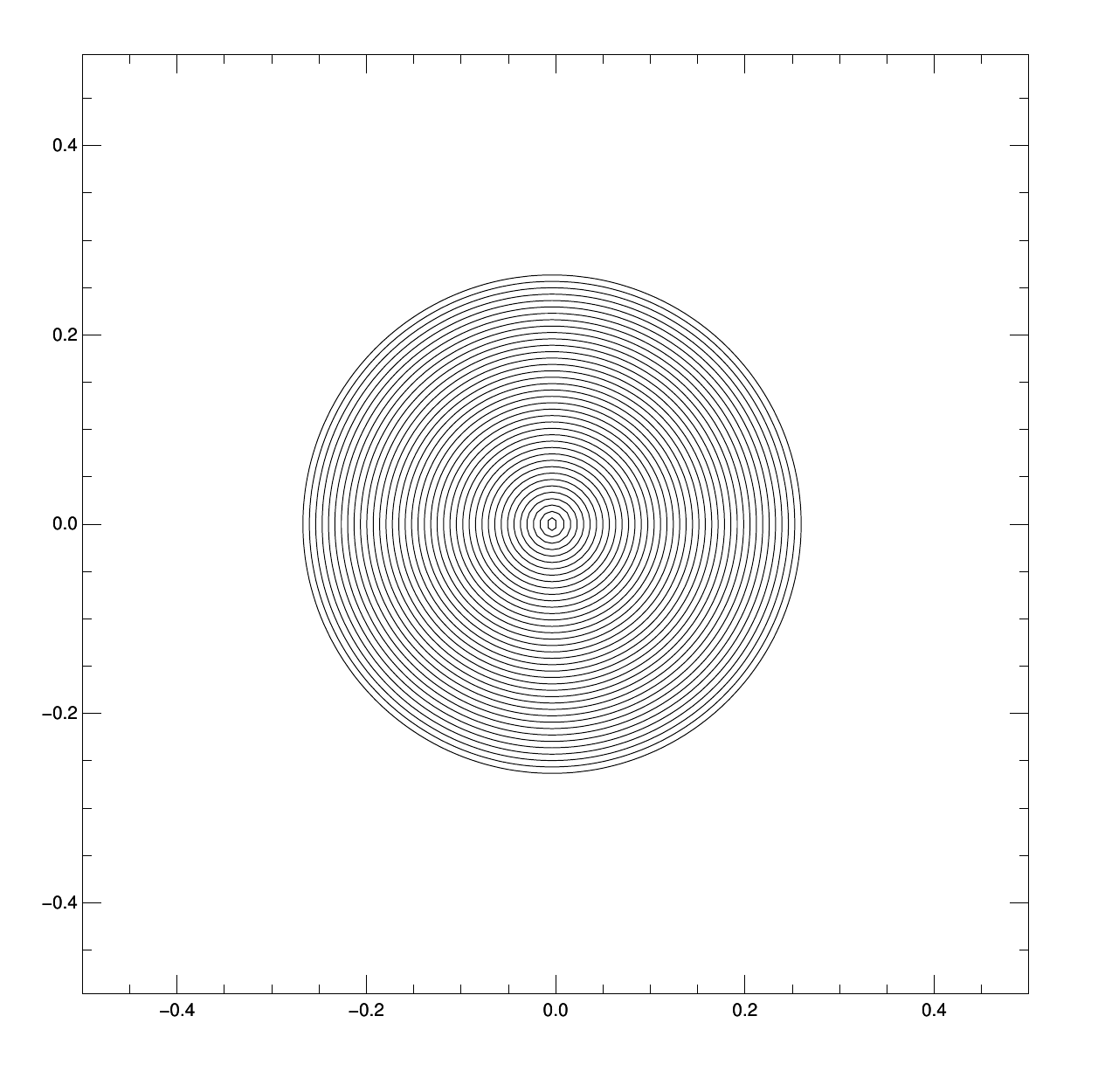}
\includegraphics[width=0.3\textwidth]{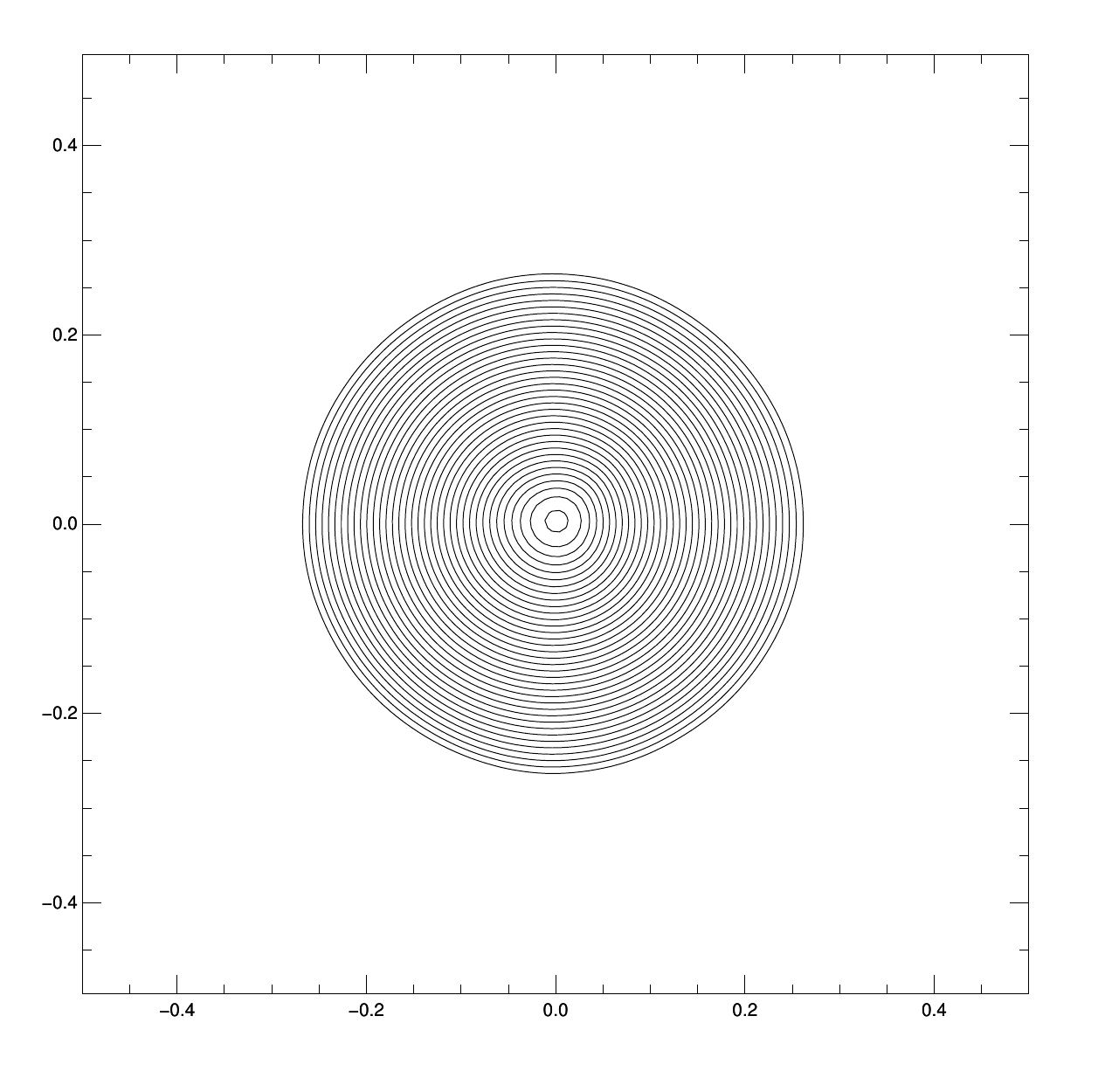}
\includegraphics[width=0.3\textwidth]{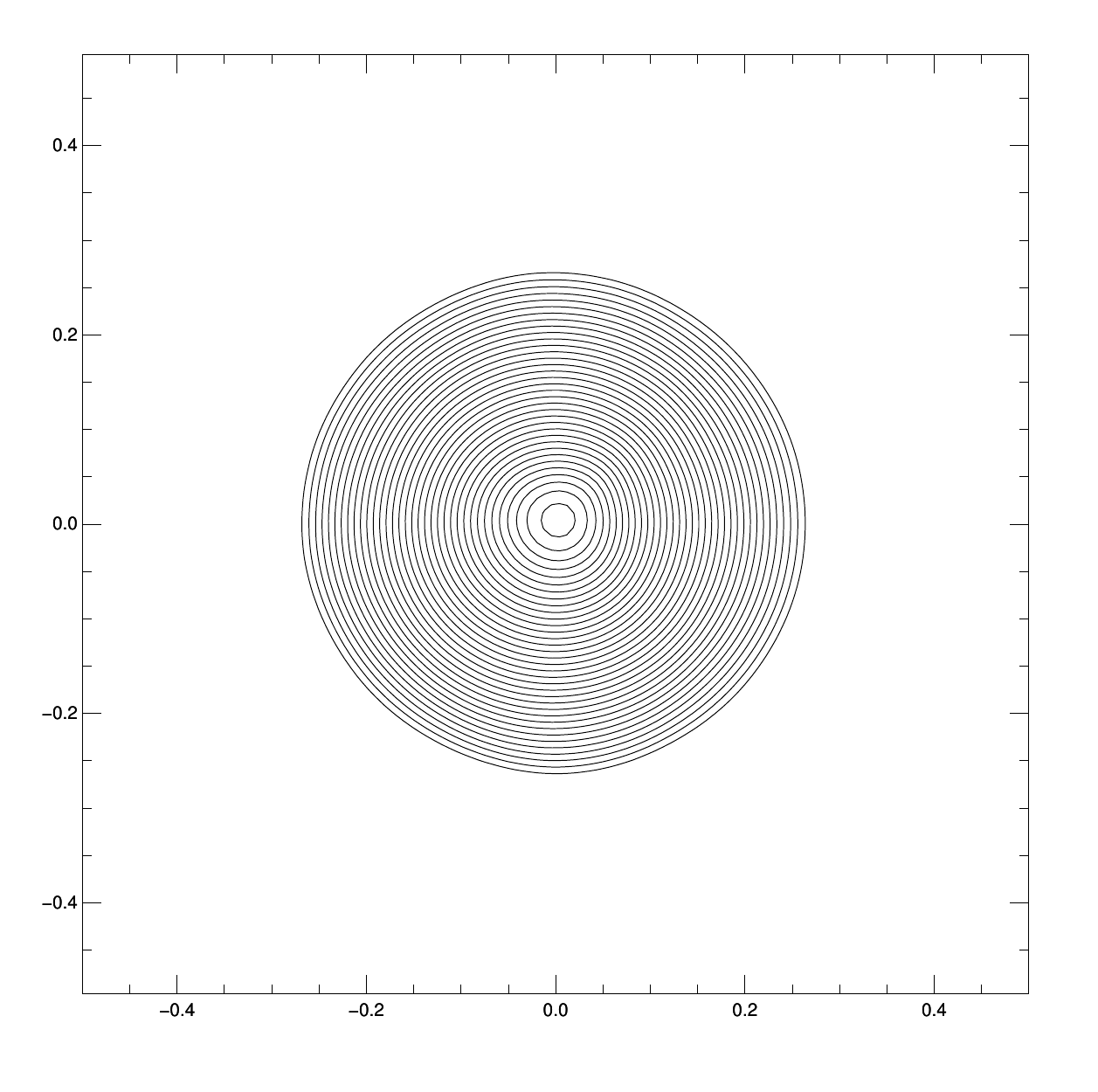}
\end{center}
\caption[]{Two-dimensional field loop advection test from \cite{Gardiner:2005}. The panels show contours of $A_z$ (magnetic field lines) after zero (left panels), one (center panels) and two (right panels) grid crossings. The top tow shows the case with $V^z=0$, the bottom row shows the case with $V^z\ne0$. In each panel, $A_z$ is plotted using $40$ levels arranged linearly between $3.0\times10^{-5}$ and $3.0\times10^{-4}$.}
\label{field_loop_aphi} 
\end{figure}

\begin{figure}
\begin{center}
\includegraphics[width=0.3\textwidth]{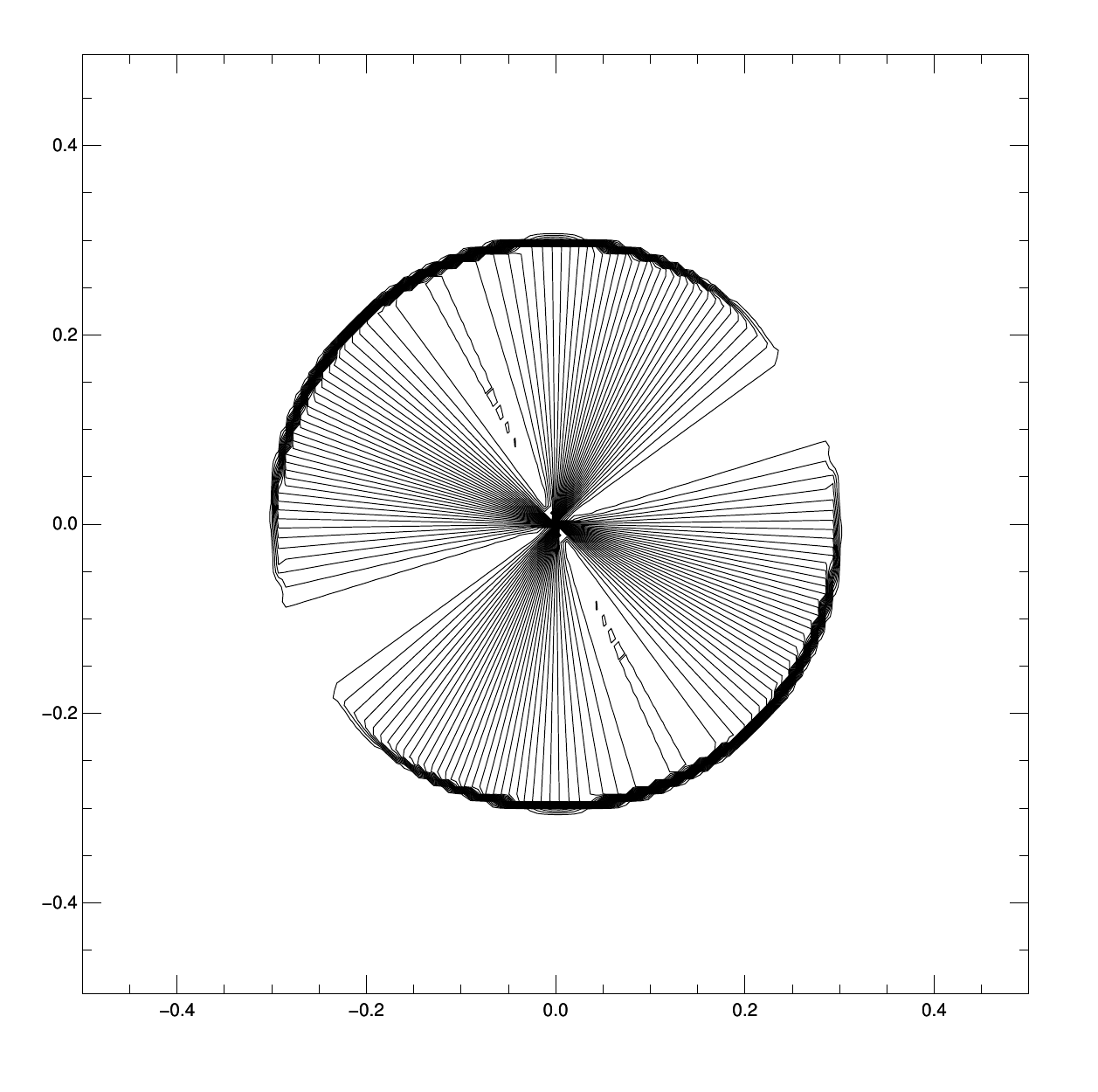}
\includegraphics[width=0.3\textwidth]{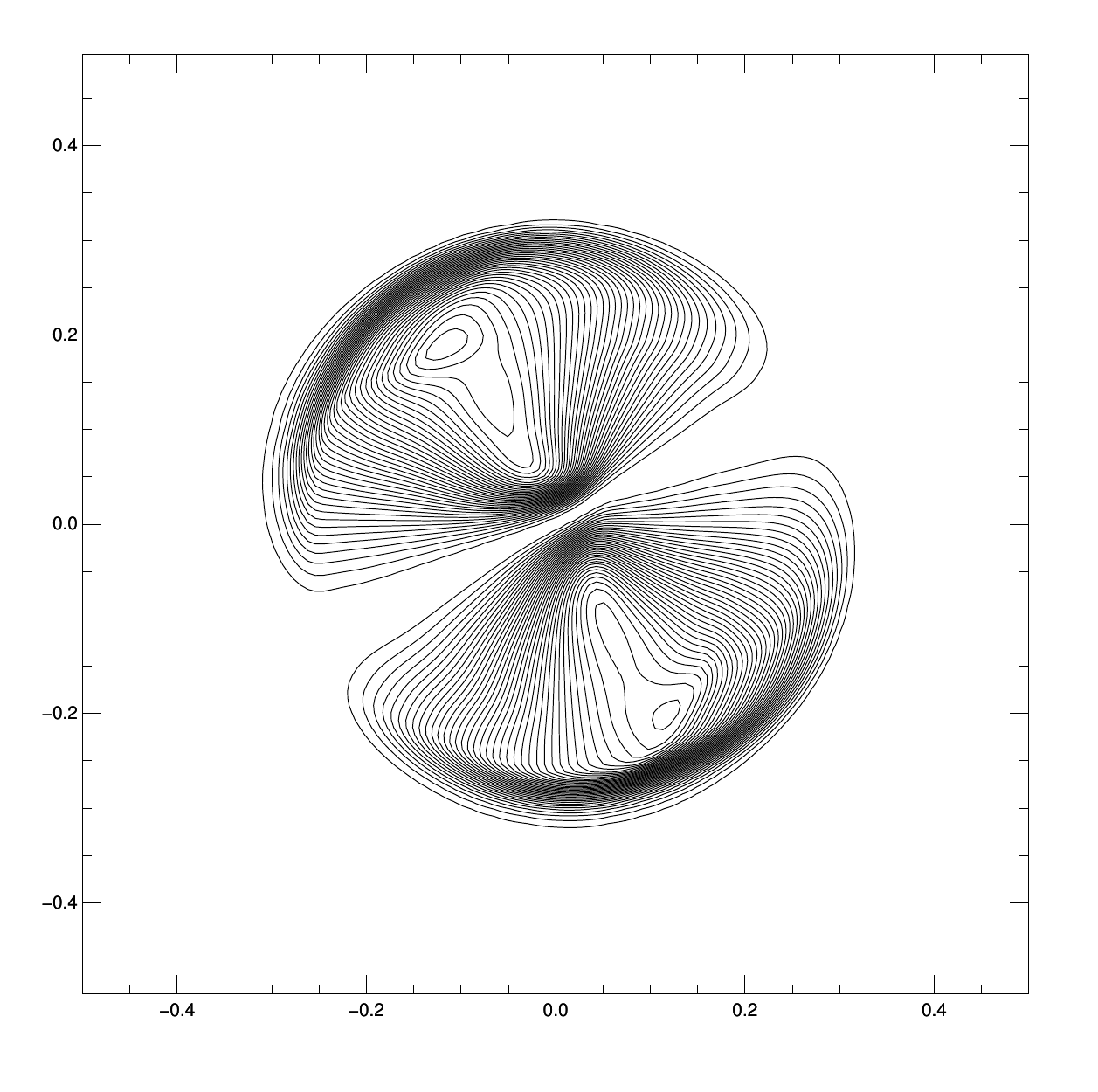}
\includegraphics[width=0.3\textwidth]{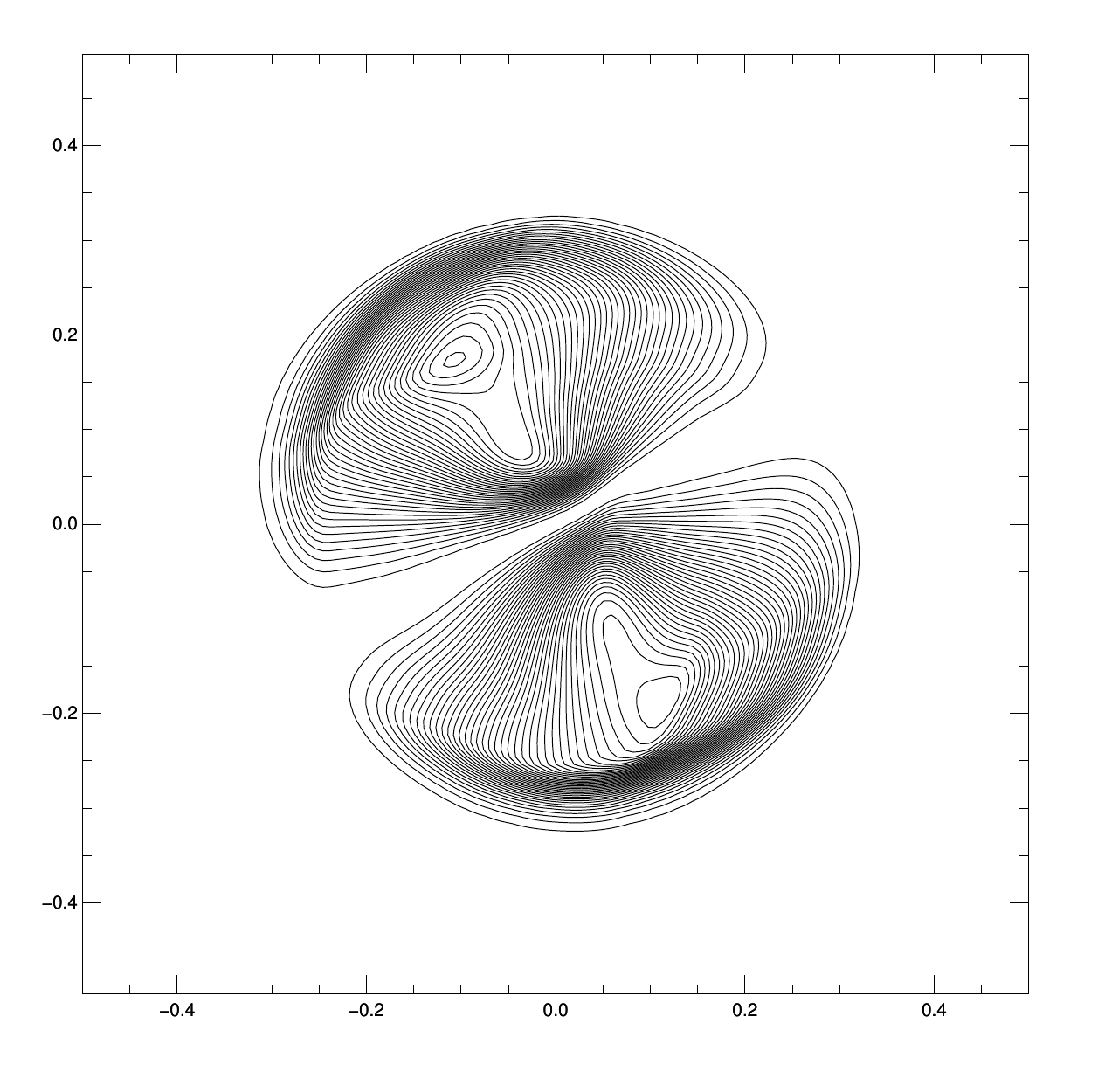}
\includegraphics[width=0.3\textwidth]{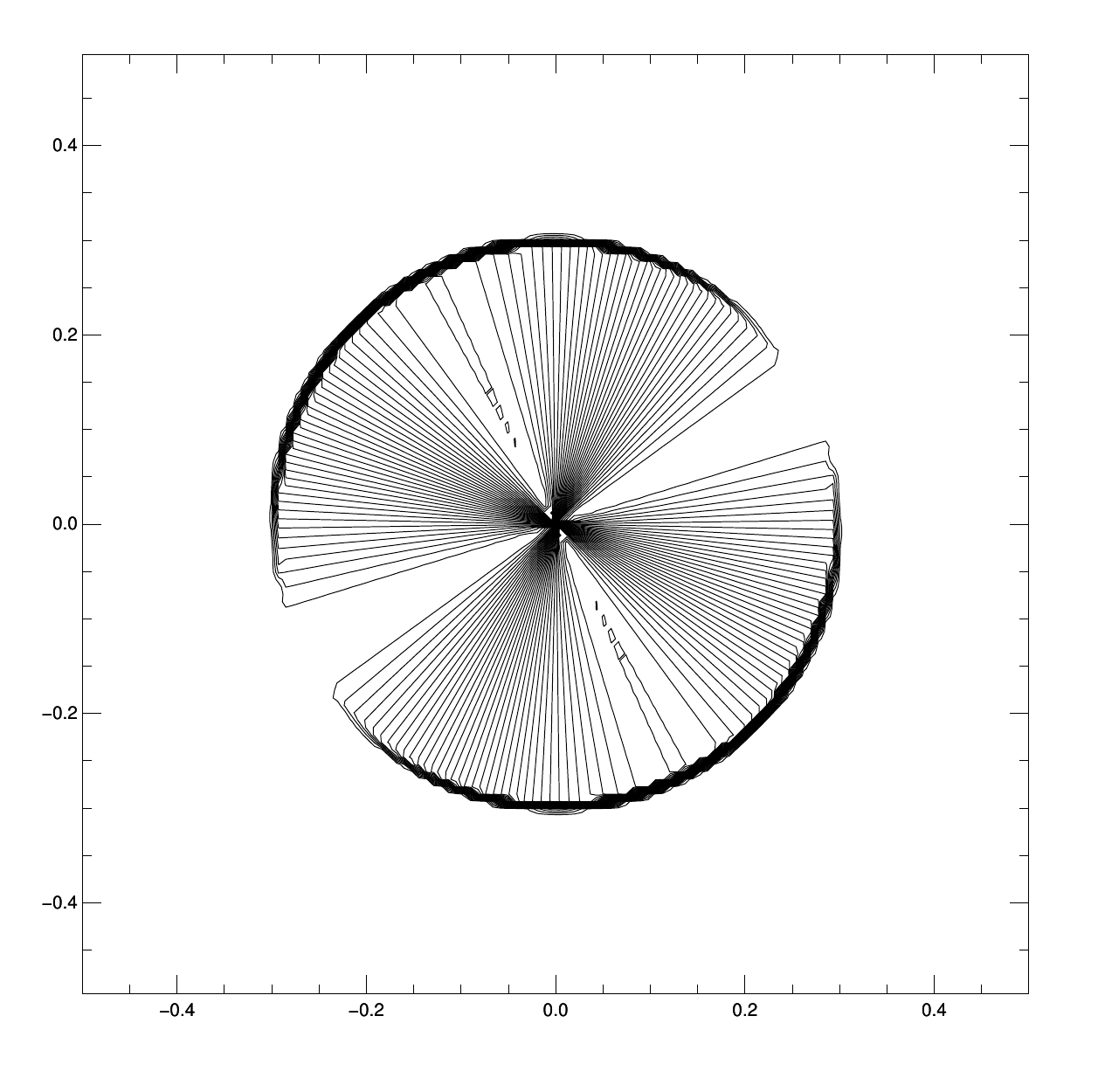}
\includegraphics[width=0.3\textwidth]{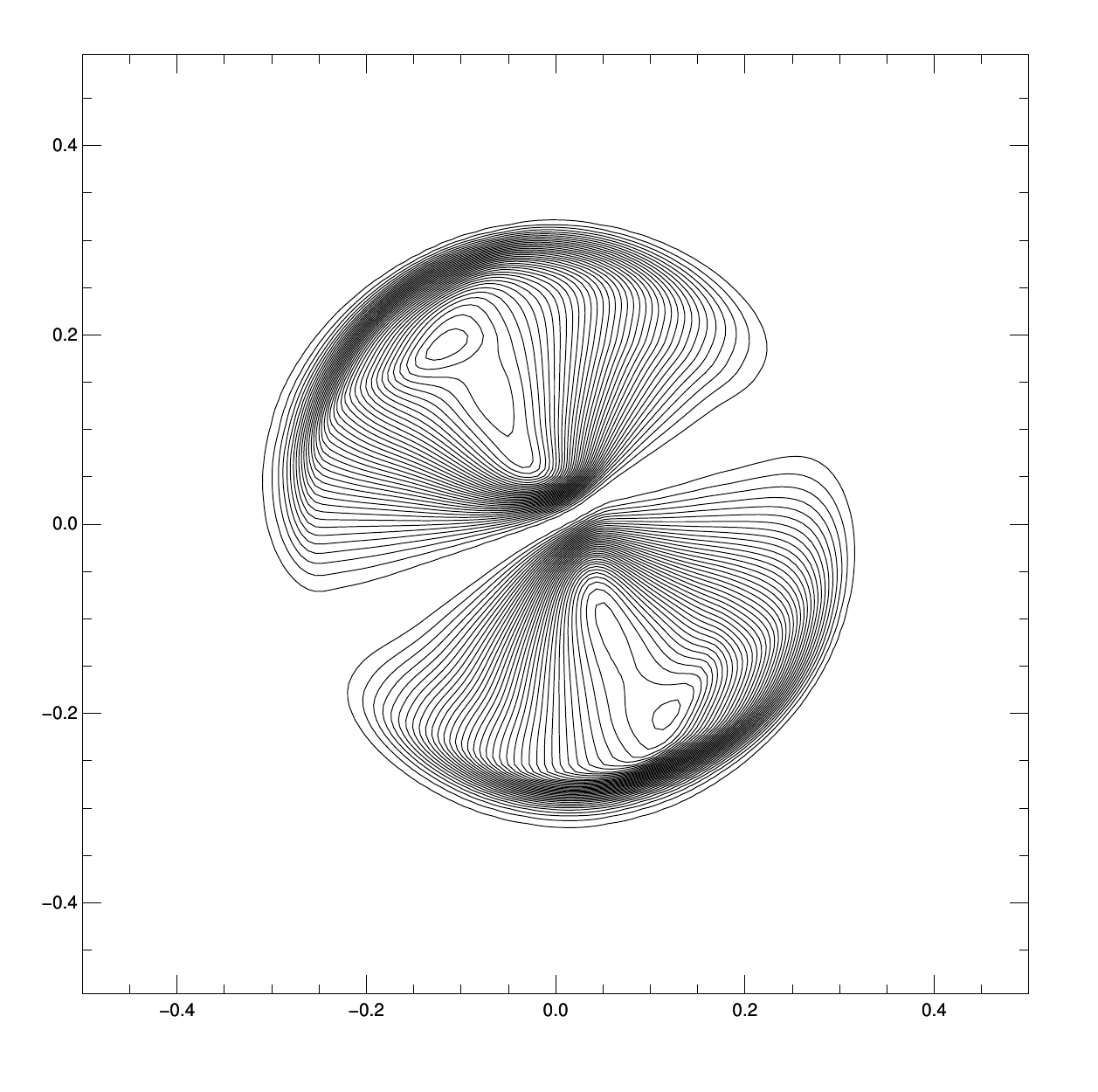}
\includegraphics[width=0.3\textwidth]{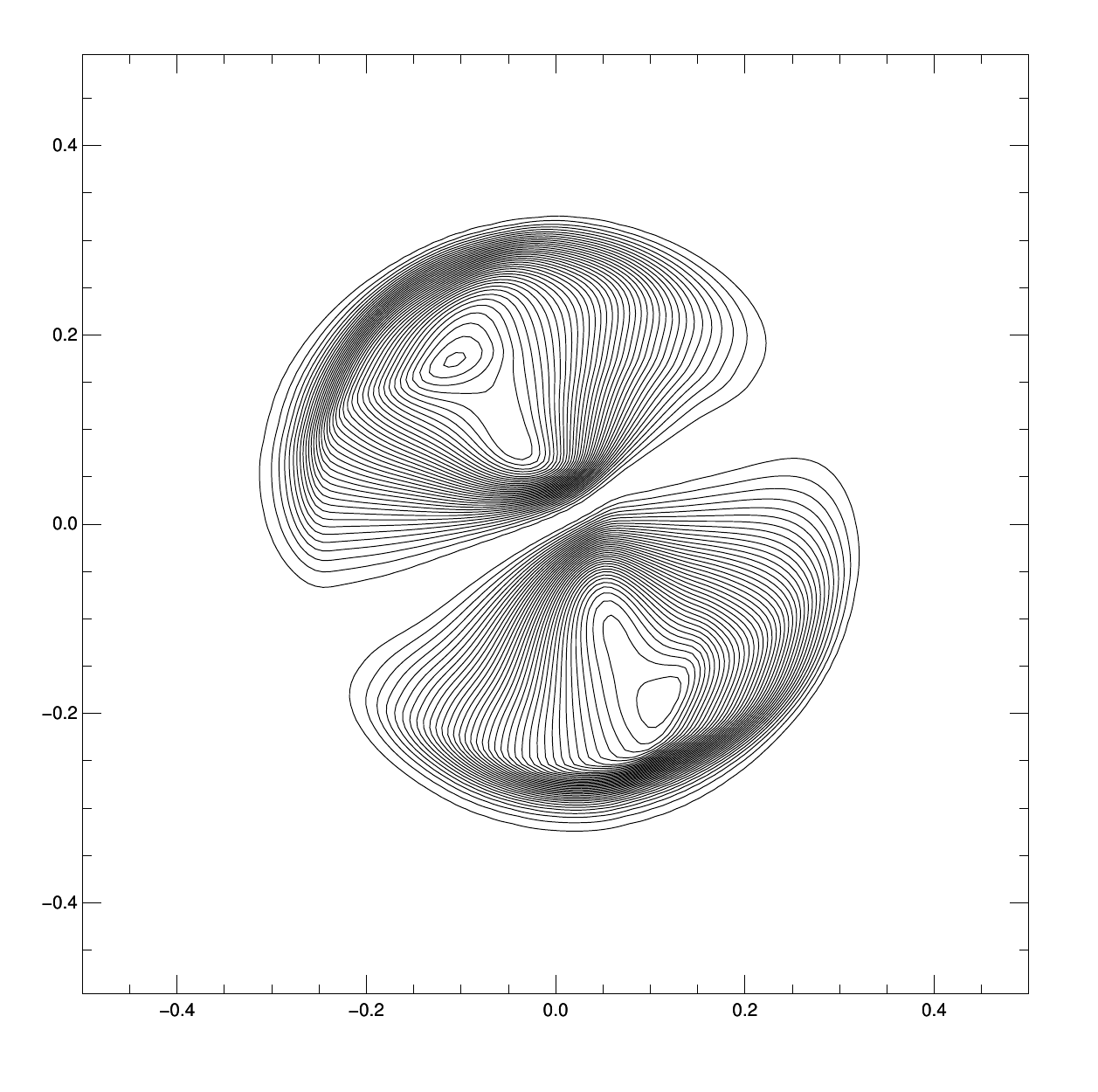}
\end{center}
\caption[]{Two-dimensional field loop advection test from \cite{Gardiner:2005}. The panels show contours of $P_m$ after zero (left panels), one (center panels) and two (right panels) grid crossings. The top tow shows the case with $V^z=0$, the bottom row shows the case with $V^z\ne0$. In each panel, $P_m$ is plotted using $40$ levels arranged linearly between zero and $6.0\times10^{-9}$.}
\label{field_loop_pm} 
\end{figure}

Figures \ref{field_loop_aphi} and \ref{field_loop_pm}  compare the
distribution of $A_z$ and $P_m$  respectively for the two different
three-velocity vectors after $0$, $1$, and $2$ grid crossing times. Inspection
of these figures reveals identical distributions in $A_z$ and $P_m$
for these two calculations, implying that for the particular set
of parameters chosen here, the primitive inversion scheme is able to maintain uniform $V^z$ to high accuracy; we find the maximum fractional error $\delta V^z / V^z = 10^{-7}$. In the case where $V^z \ne 0$, the fractional errors in $V^z$ are confined within the field loop and distributed such that on the leading edge of the loop, the fractional error in $V^z$ is positive and on the trailing edge, the fractional error is negative. As a result the volume integrated kinetic energy in the $z$-direction is conserved over the course of the evolution. In the case of a weak magnetic field loop, $\delta V^z / V^z \propto Q^{-1}$, suggesting that higher values of $Q = \rho h \Gamma^2$ will lead to smaller fractional errors in $V^z$; we have recomputed this test either using significantly higher densities ($\rho = 10^3$), pressures ($P=3\times10^3$) or Lorentz factors ($\Gamma^2 = 10^3$) and have found that increasing any of these parameters independently decreases the fractional error in $V^z$ to $\delta V^z / V^z \sim 10^{-10}$.

In the case where $V^z = 0$ initially, we have verified that $|b^z|$ remains exactly zero for the entire evolution. In the case
where $V^i \ne 0$ using the HLLD solver, the energy density in the $z$ component of the magnetic field, $|b^z|/|b|_0 = 3.54\times10^{-3}$ initially (recall that $b^i = {\cal B}^i / \Gamma + \Gamma V^i [\overrightarrow{V} \cdot  \overrightarrow{{\cal B}}]$), decreasing to $|b^z|/|b|_0 = 3.29\times10^{-3}$ after the field
loop as been advected twice around the grid, a fractional decrease
of $\sim7\%$. For comparison, the magnetic of the magnetic field
four-vector, $|b|$ decreases from $7.45\times10^{-3}$ to
$7.01\times10^{-3}$ over this same period, a decrease of $\sim6\%$.
We have further verified that in both cases $\partial_i {\cal B}^i
= 0$ to machine accuracy. 

Executing the $V^i \ne 0$ test using the HLLC solver, we find that
the energy density in the $z$ component of the magnetic field,
$|b^z|/|b|_0 = 3.54\times10^{-3}$ initially, decreasing to $|b^z|/|b|_0
= 2.97\times10^{-3}$ at the end of the evolution, a fractional
decrease of $\sim16\%$ whilst the magnetic of the magnetic field
four-vector, $|b|$ decreases from $7.45\times10^{-3}$ to
$6.32\times10^{-3}$ over this same period, a decrease of $\sim15\%$.
Overall, the HLLD solver gives a factor of $3.75$ increase in
accuracy for this test compared to the HLLC, whilst the overall
code performance is reduced by a factor of $1.65$.

\begin{figure}
\begin{center}
\includegraphics[width=0.1\textwidth, viewport=0 500 100 700,clip]{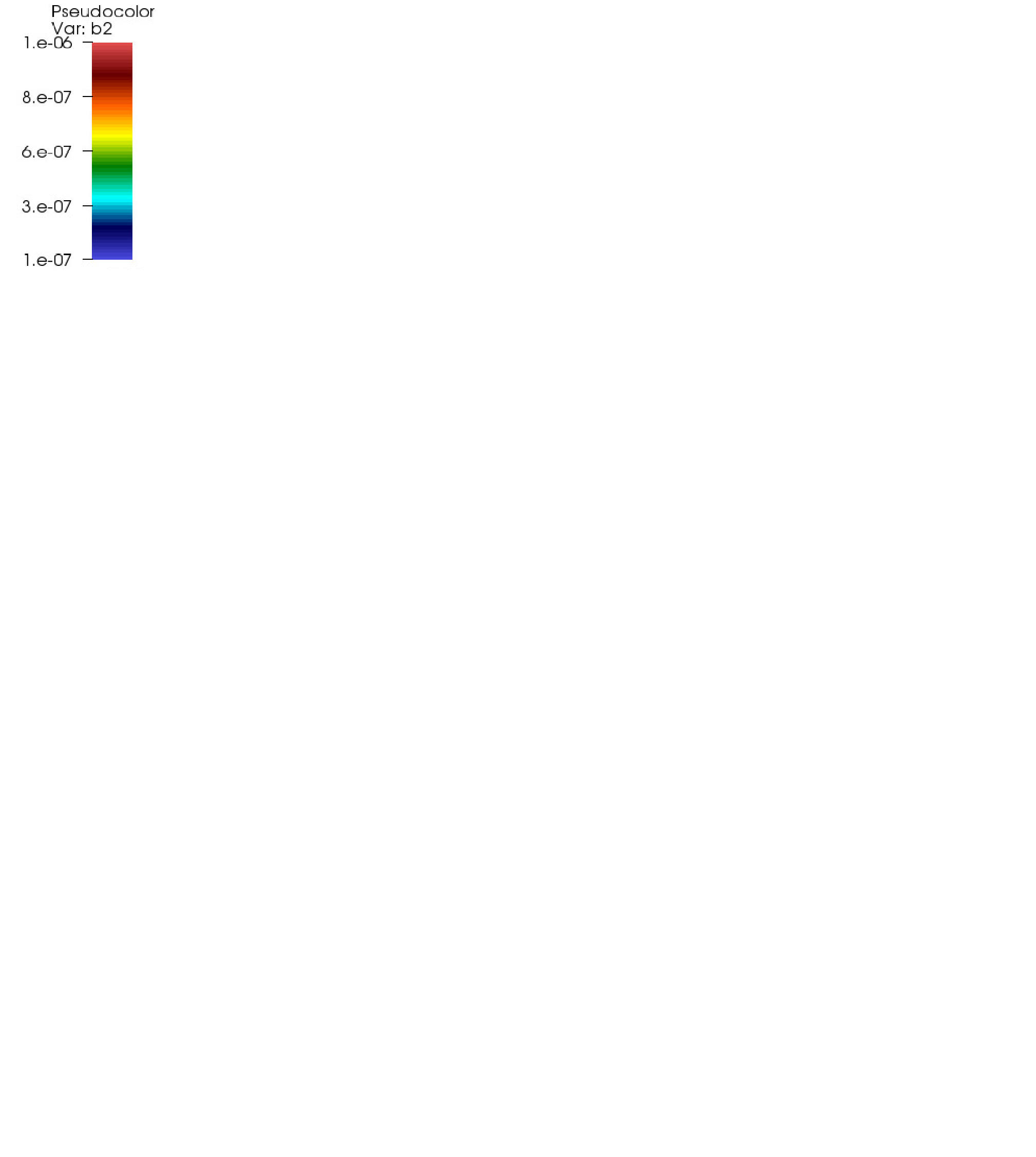}
\includegraphics[width=0.29\textwidth, viewport=40 40 560 550,clip]
{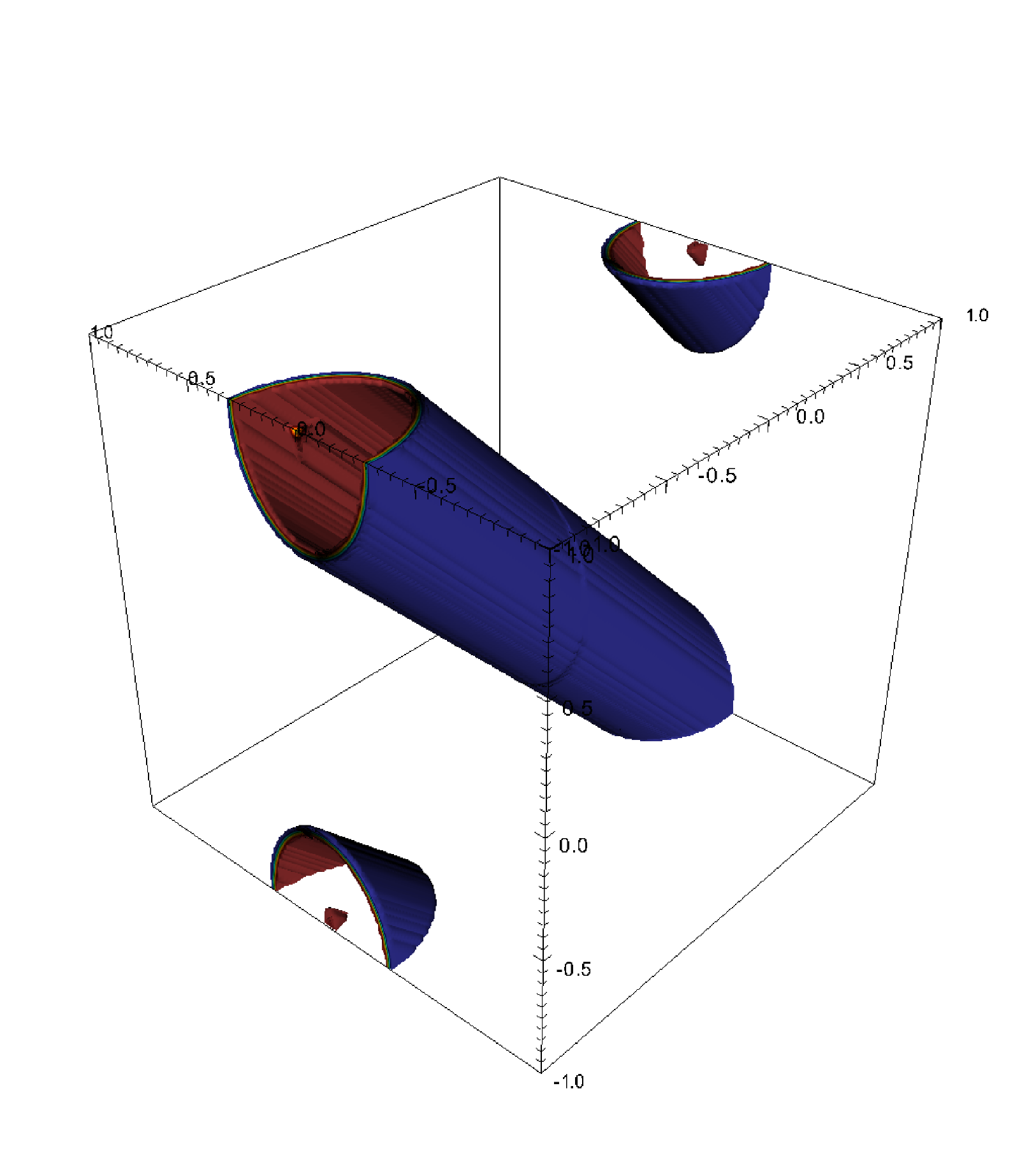}
\includegraphics[width=0.29\textwidth, viewport=40 40 560 550,clip]
{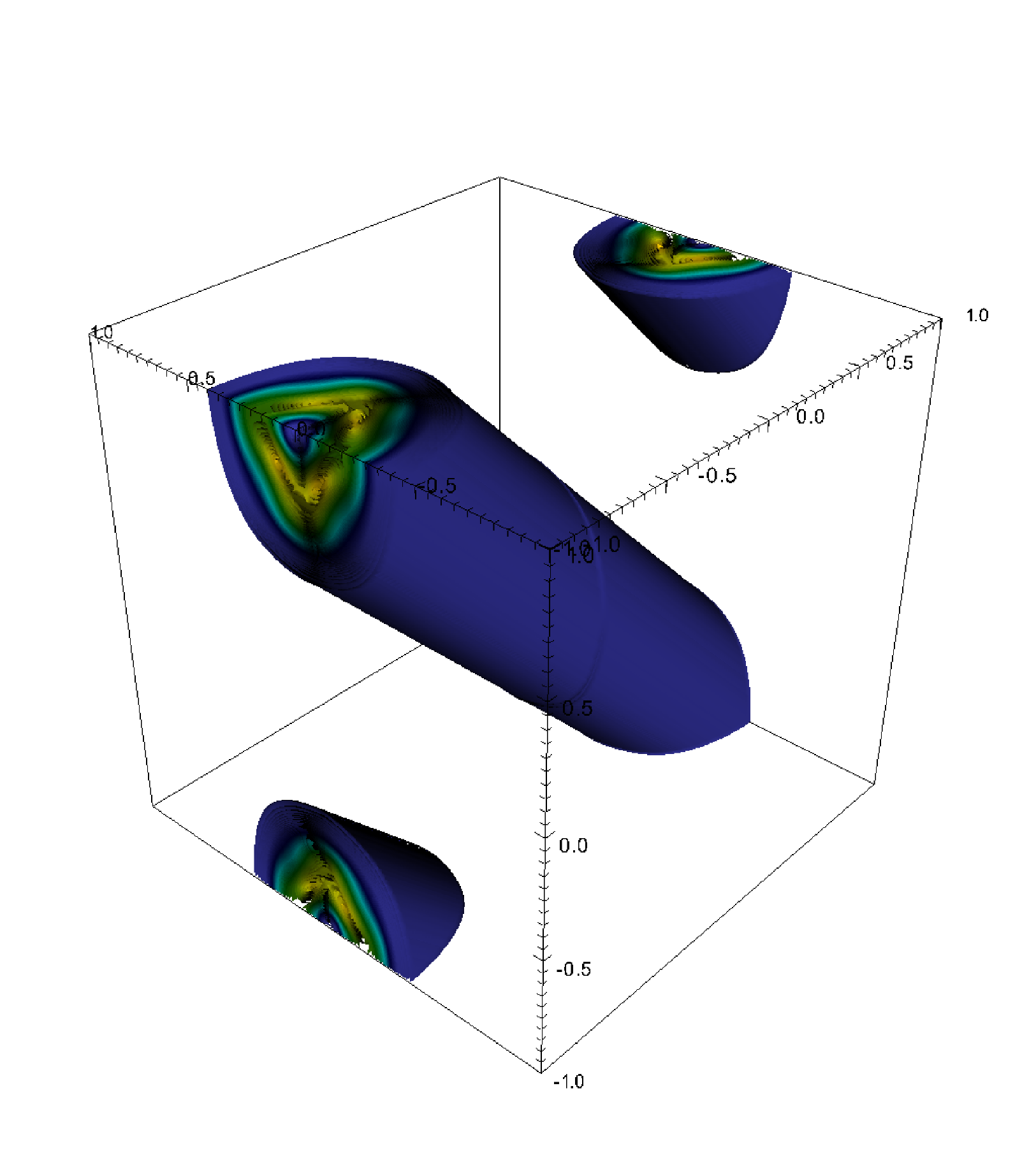}
\includegraphics[width=0.29\textwidth, viewport=40 40 560 550,clip]
{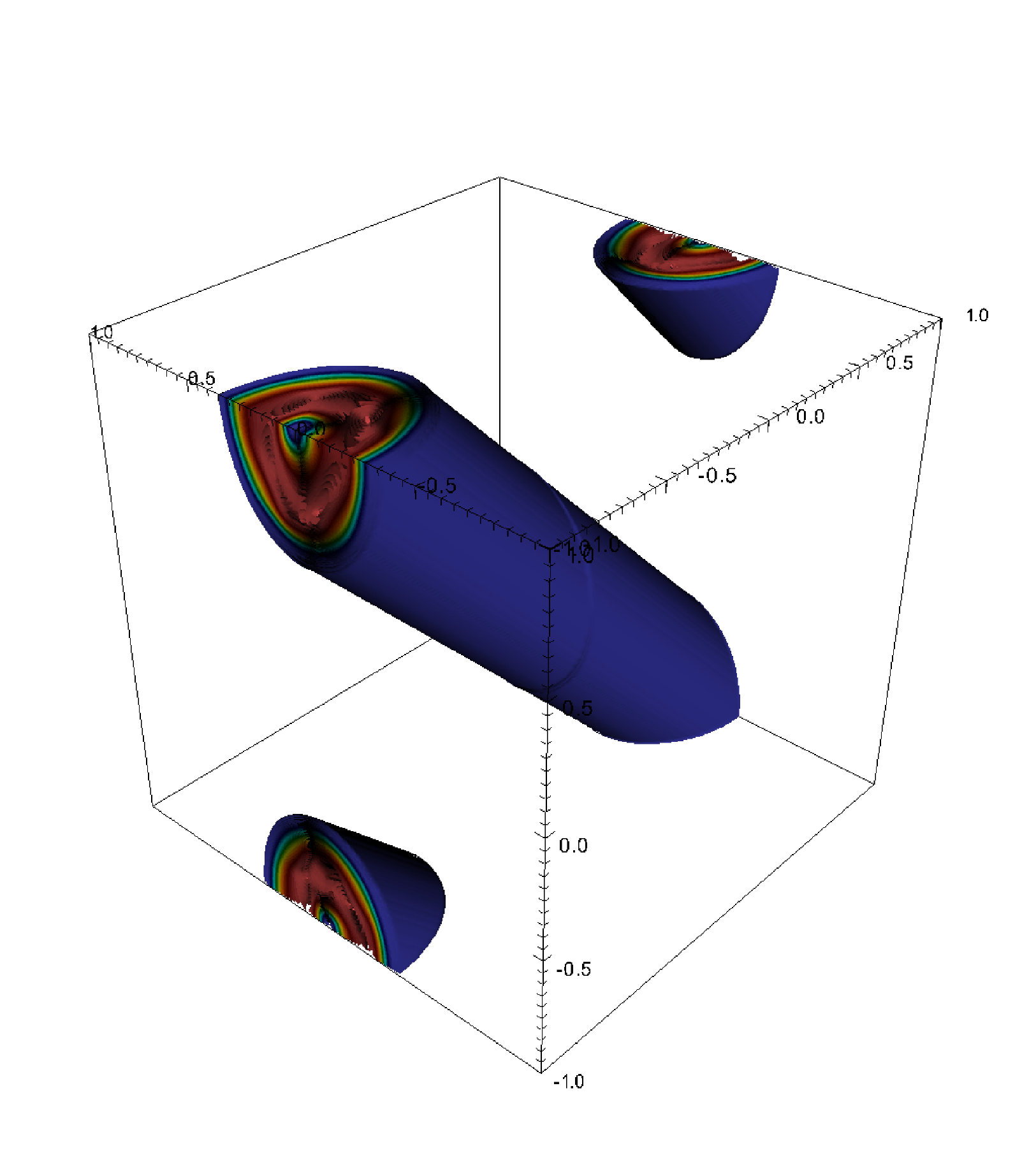}
\end{center}
\caption[]{Magnetic field strength distribution, $|b|^2$ in the three-dimensional field loop advection problem for the initial state (left panel) and after one grid crossing time for the HLLC solver (center panel) and the HLLD solver (right panel).}
\label{loop3d} 
\end{figure}

A fully three-dimensional version of this test is obtained by placing
a column of field loops at an oblique angle to the grid, as is
described in \cite{Gardiner:2008}. In Newtonian MHD, the component
of the magnetic field parallel to the axis of the cylinder should
remain zero for all time; however, in RMHD this is not
the case as outlined above. Nevertheless, the three-dimensional
field-loop advection test is useful to measure the ability of the algorithm
to handle truly multi-dimensional MHD problems as well as providing
a method to estimate the diffusivity of a given Riemann solver. The
test is initialized as described above and the solution rotated so
that the field loop column lies oblique to the grid as shown in
Figure \ref{loop3d}. The grid covers $0\le (x,y,z) \le 1$ using
$128^3$ and is tri-periodic. Figure \ref{loop3d} also shows the
structure of the field loops after one complete advection around
the grid for both the HLLC and HLLD solver. The solution computed with
the HLLC solver is significantly diffused compared to both the
initial state and the HLLD solution. More quantitatively, we find
that after one grid crossing time, the magnitude of the magnetic
field four-vector has decreased by $14\%$ compared to its initial
value for the HLLC solver, whilst for the HLLD solver, this same
quantity has decreased by $9\%$.  The overall code performance
is again reduced by a factor of $1.65$ by using the HLLD solver.
As in the Alfven wave test, these results suggest that
at a given resolution the HLLD solver provides more accurate
results than the HLLC solver.

\subsection{Current Sheets}\label{current_sheet}

The next multidimensional test that we consider is the evolution
of a current sheet. Whilst this test has no analytic solution, it has
been found to be a good test of the robustness of multidimensional
algorithms for MHD in strongly magnetized media  \citep{Hawley:1995a}. The
test is run on a grid of $200\times200$ zones covering a domain
$0.0<x<2.0$, $0.0<y<2.0$. The initial condition is uniform in
density, $\rho=1.0$ and pressure, $P_g = \beta/2$ where $10^{-1}
\ge \beta \ge 10^{-3}$. The fluid three-velocity is initialized
according to $V^i = \left[ A \mathrm{cos} \left(\pi y \right), 0.0,
0.0 \right]$ where $A=0.2$. Finally the magnetic field three-vector
is given by ${\cal B}^i = \left[ 0.0, -1.0, 0.0 \right]$ for $0.5
\le x \le 1.5$ and ${\cal B}^i = \left[ 0.0, 1.0, 0.0 \right]$
otherwise.

\begin{figure}
\begin{center}
\includegraphics[width=0.3\textwidth]{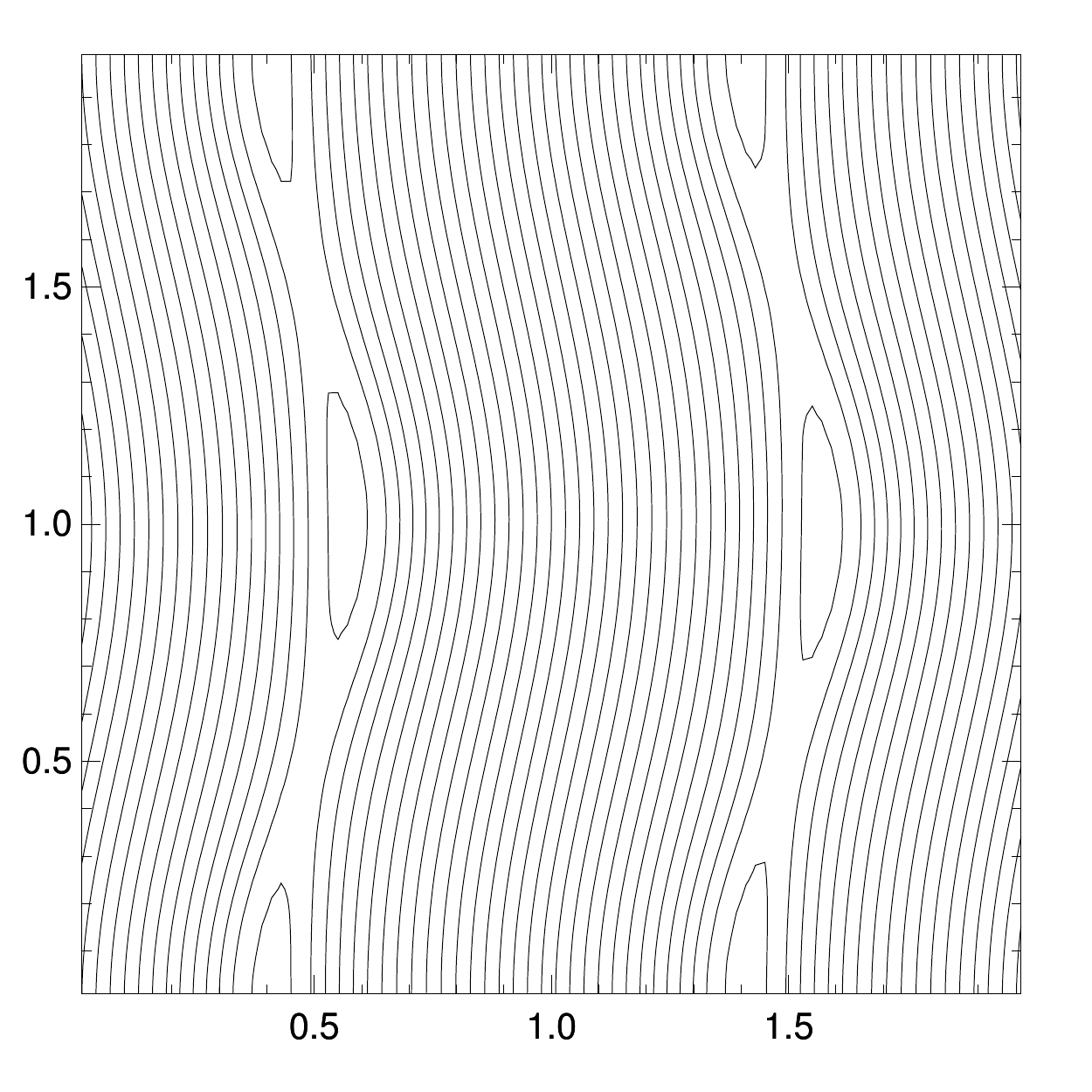}
\includegraphics[width=0.3\textwidth]{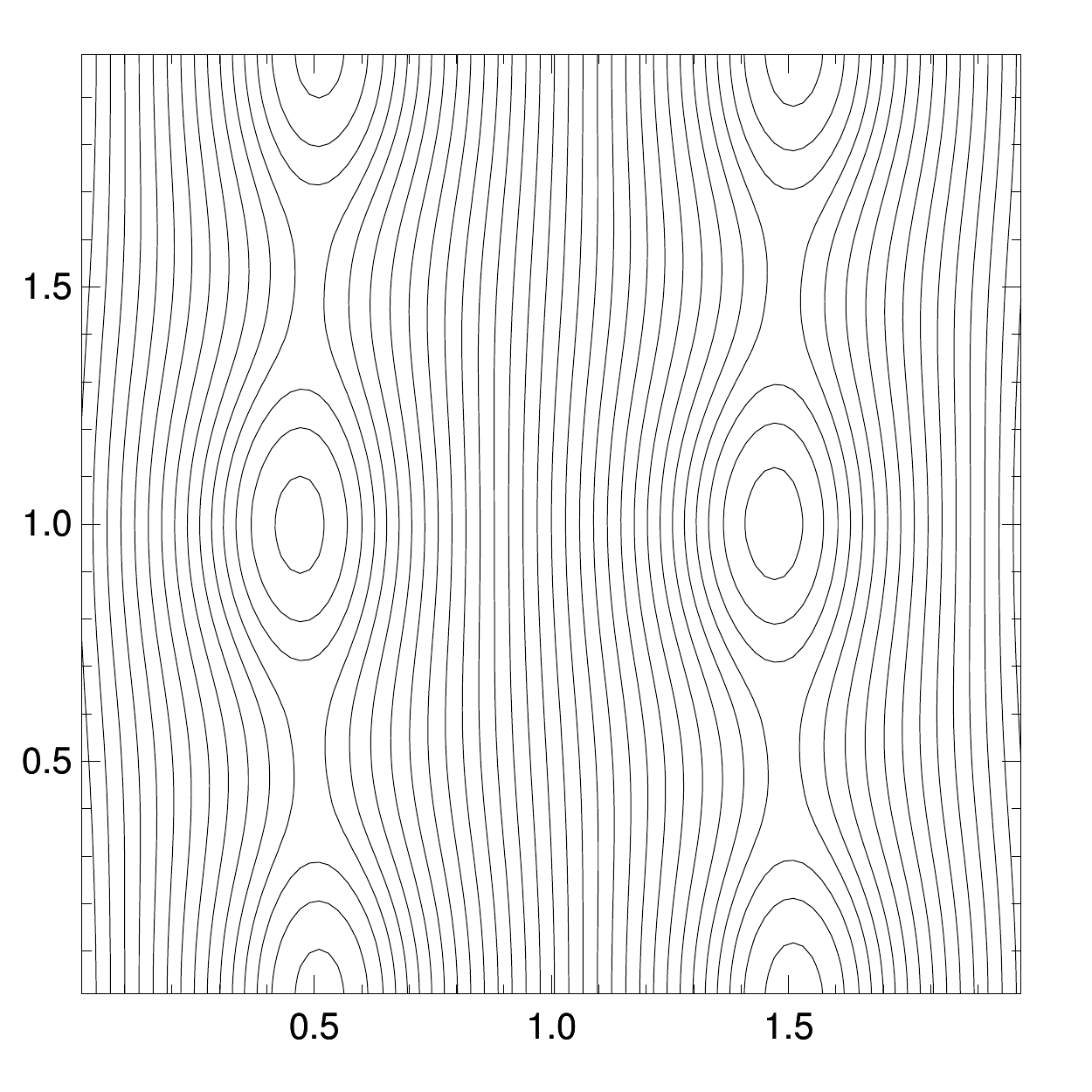}
\includegraphics[width=0.3\textwidth]{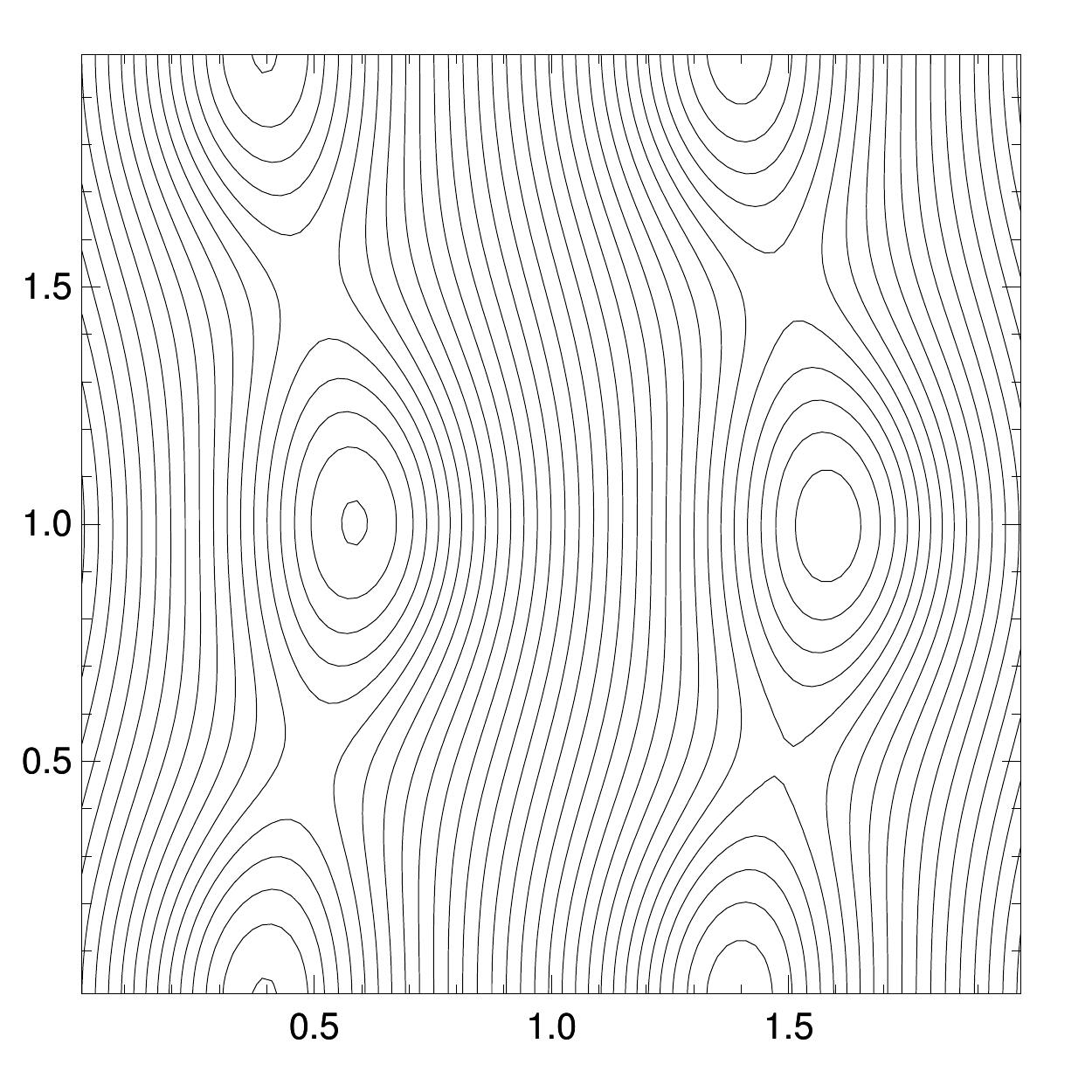}
\includegraphics[width=0.3\textwidth]{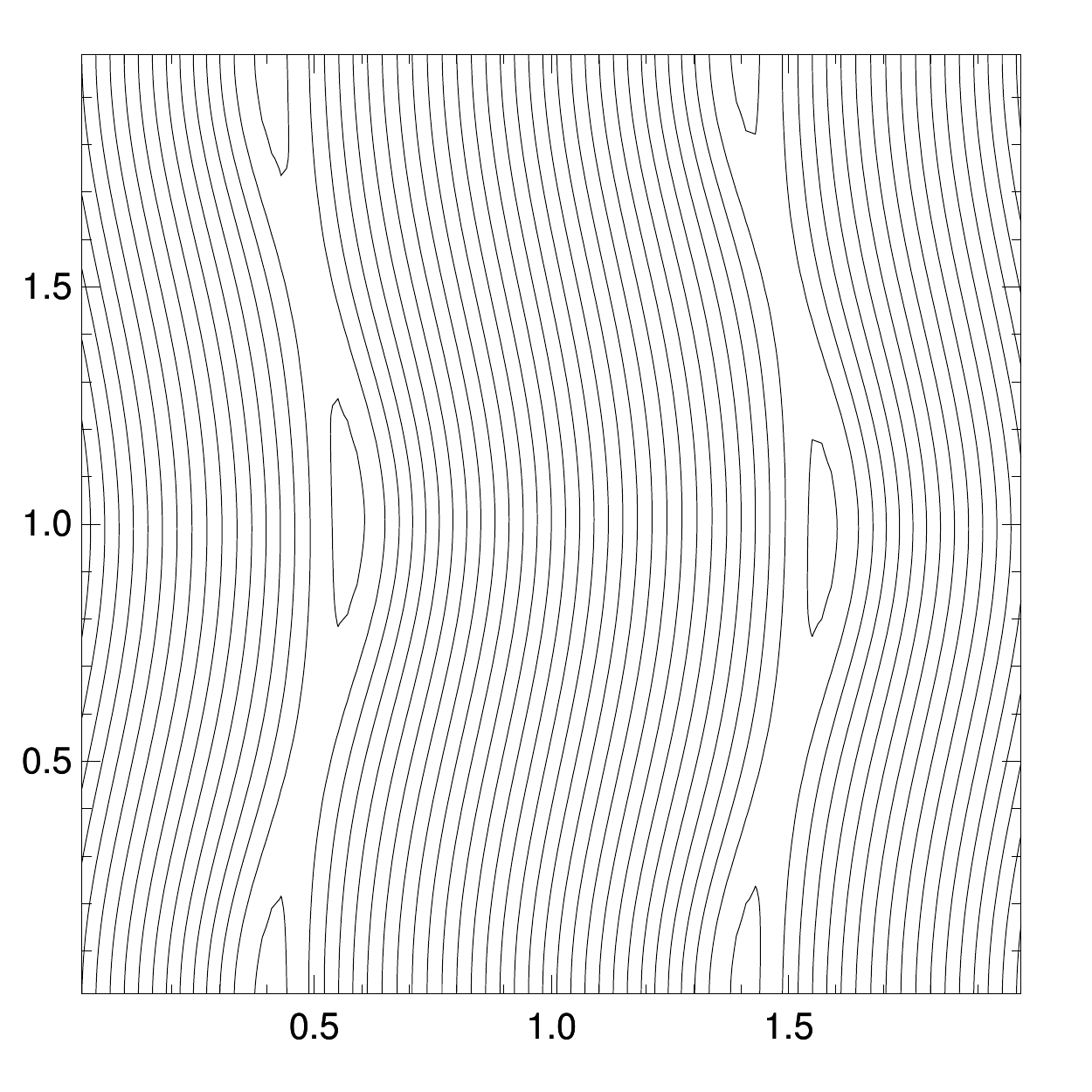}
\includegraphics[width=0.3\textwidth]{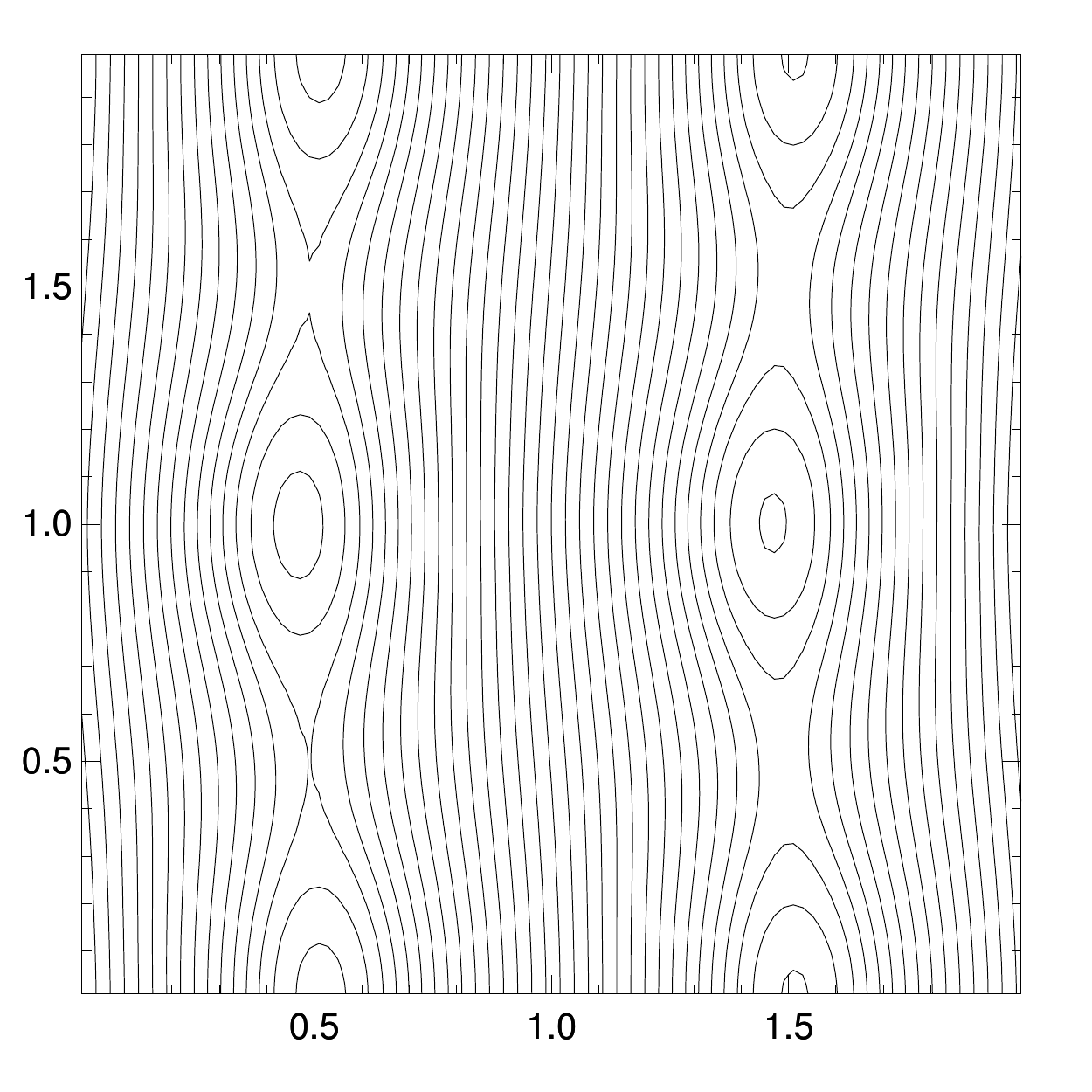}
\includegraphics[width=0.3\textwidth]{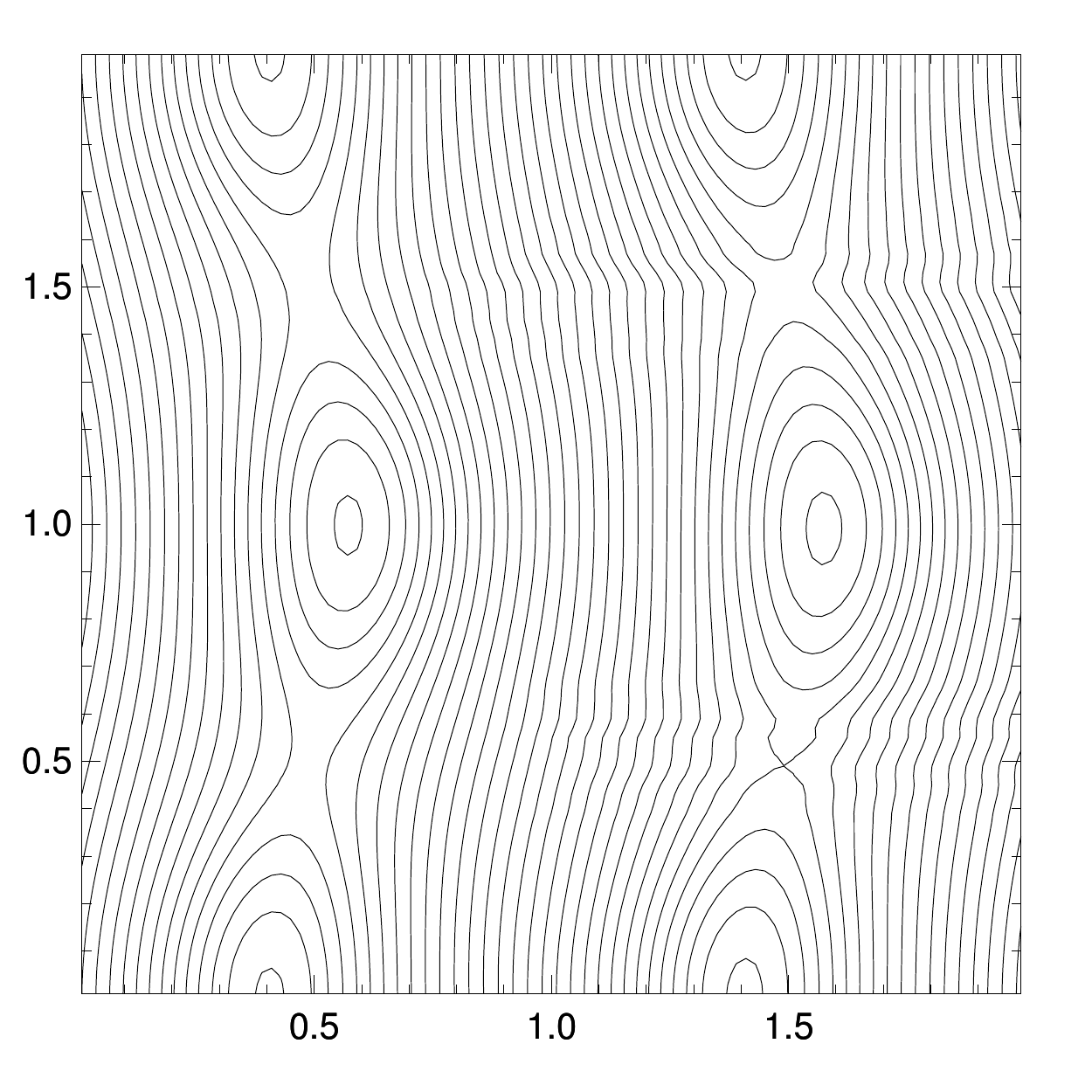}
\end{center}
\caption[]{Evolution of $\beta=0.1$ current sheet (top row) and $\beta=0.01$ current
sheet (bottom row) on the $x-y$ plane at time $t=2.0$ (left panels), $t=5.0$ (center
panels) and $t=10.0$ (right panels). Panels show the evolution of the $z$-component of
the magnetic vector potential (magnetic field lines).}
\label{CurrentSheet} 
\end{figure}

The evolution of this system for the two different values of $\beta$
is shown in Fig. \ref{CurrentSheet}. Reconnection occurs via a
tearing-mode-like instability mediated by grid-scale reconnection in the
two current sheets initially located at $x=0.5$ and $1.5$, forming a
series of magnetic islands. These islands assemble into progressively
larger structures as the simulation proceeds, until a stable
configuration is reached. The object of this test is to probe the
region of parameter space which the code can robustly evolve to
late times $t\ge10$. The outcome of this test depends on the
dissipation properties of the Riemann solver; we have found that,
with $A=0.2$ the HLLE solver is able to probe $\beta\sim10^{-3}$,
the HLLC solver is able to probe $\beta\sim10^{-2}$ and the HLLD
solver $\beta\sim10^{-1}$. Increasing the amplitude of the perturbation
to $A\ge0.5$ breaks the algorithm for $\beta<1.0$ for any of the Riemann solvers.

\subsection{Orszag Tang vortex}

\begin{figure}
\begin{center}
\includegraphics[width=0.3\textwidth]{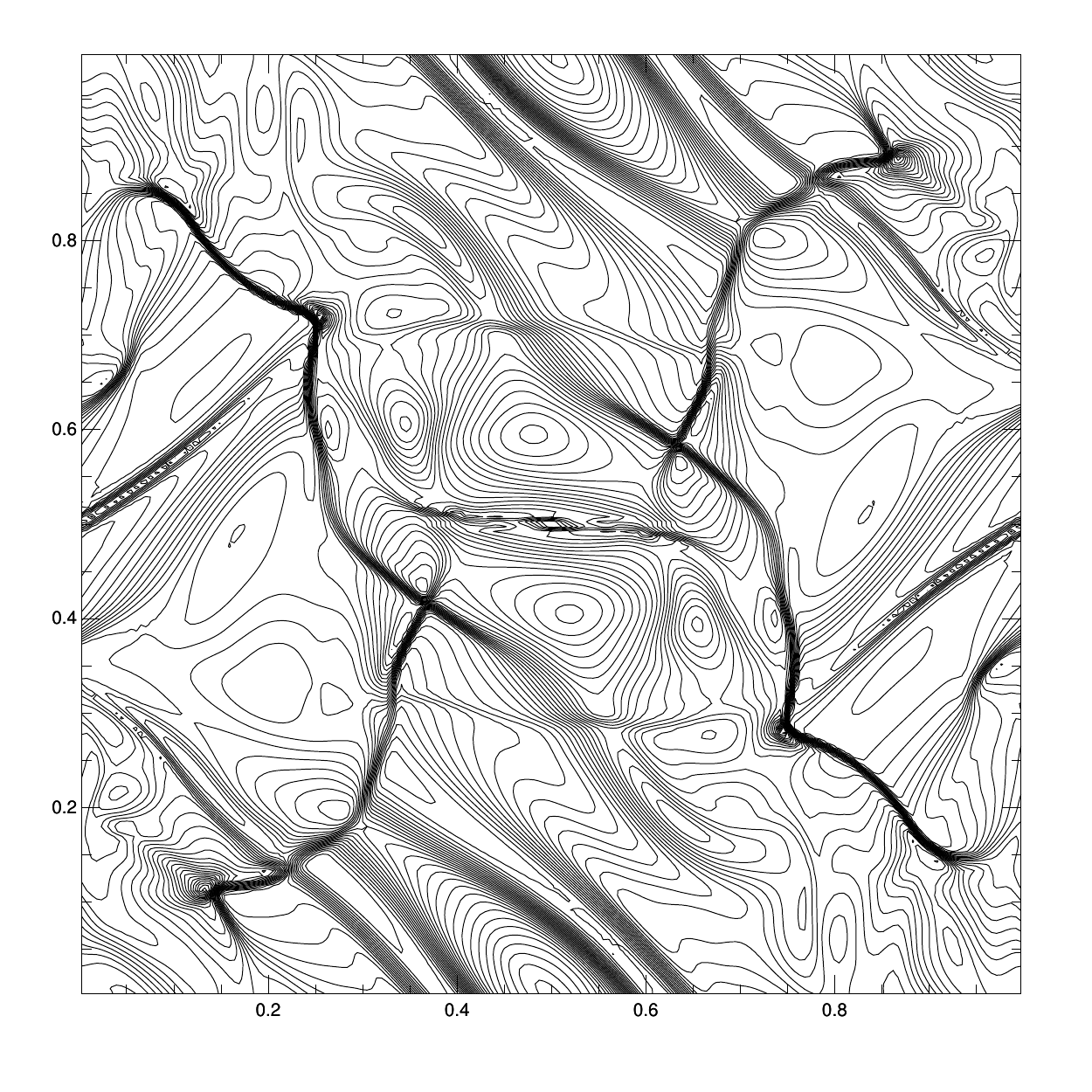}
\includegraphics[width=0.3\textwidth]{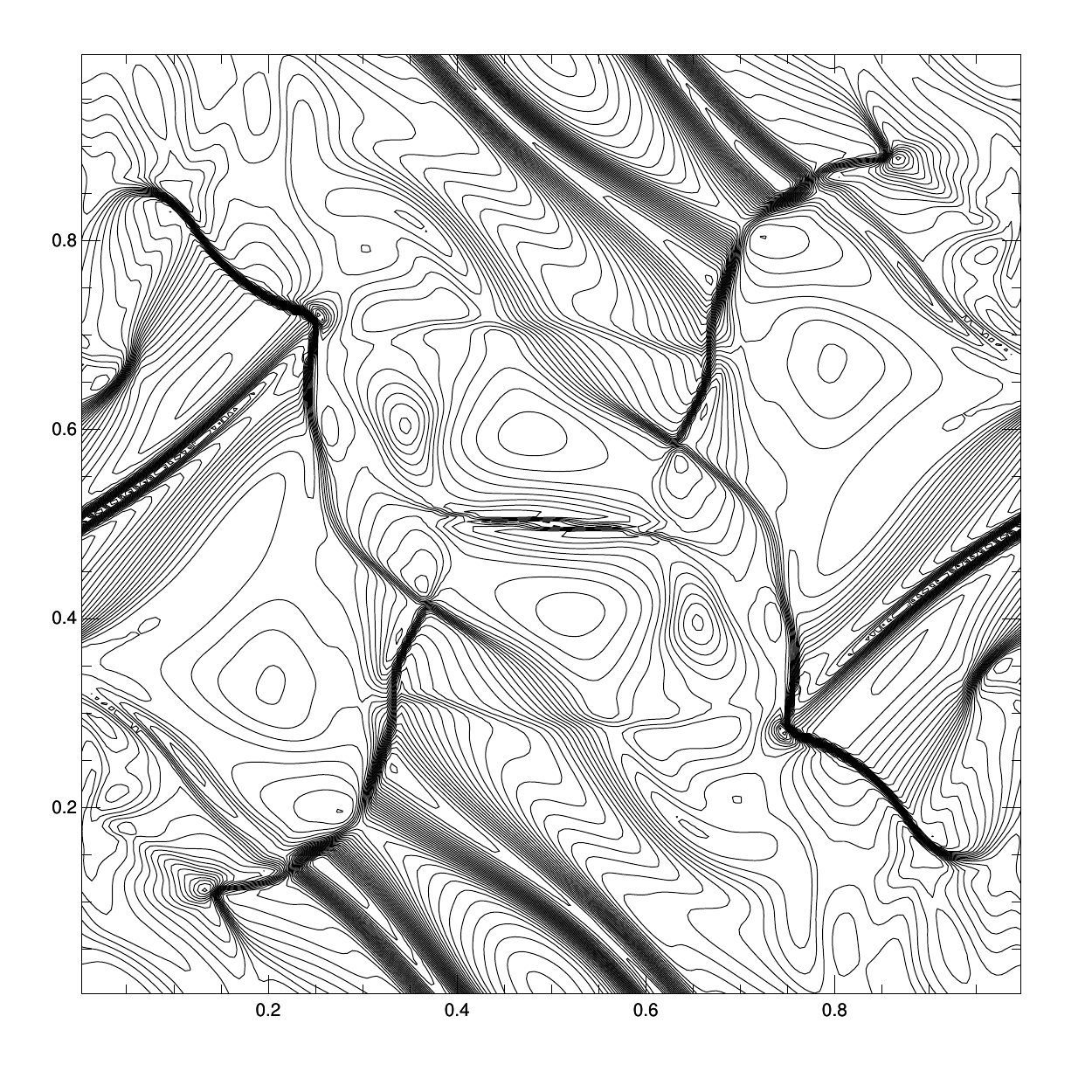}
\includegraphics[width=0.3\textwidth]
{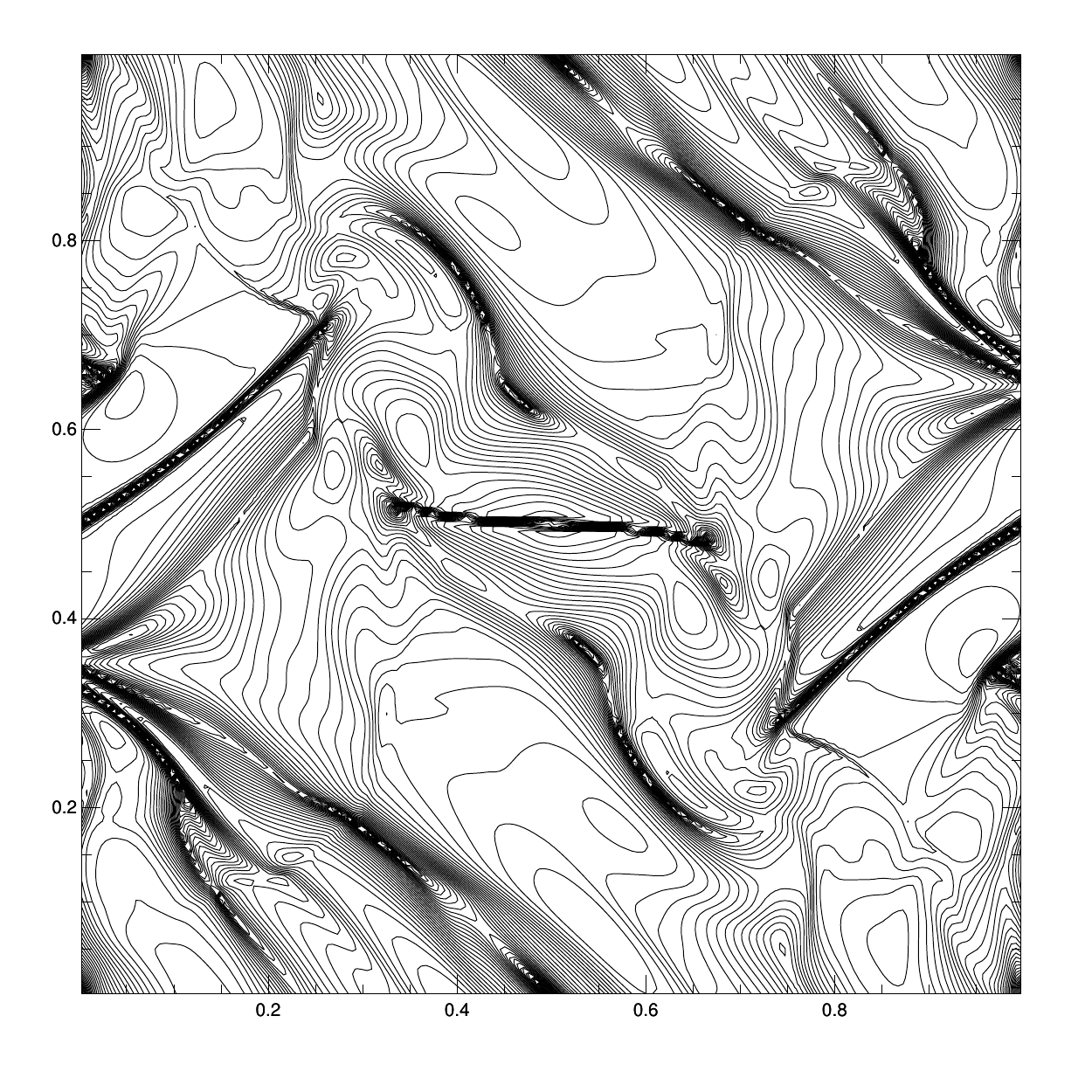}
\end{center}
\caption[]{Structure of the Orszag Tang vortex at $t=1.0$ on a $192^{2}$ grid. From left to right, the panels show contours of density, gas pressure and magnetic pressure. The panel showing the structure of
the density has $40$ contours arranged linearly over the range $5.0\times10^{-2}$--$5.0\times10^{-1}$, whilst the panels showing gas and magnetic pressure have $40$ contours arranged logarithmically covering the range $2.0\times10^{-2}$--$6.0\times10^{-1}$ and $1.0\times10^{-6}$--$1.0$ respectively.}
\label{OrszagTang} 
\end{figure}

\begin{figure}
\begin{center}
\includegraphics[width=0.5\textwidth]
{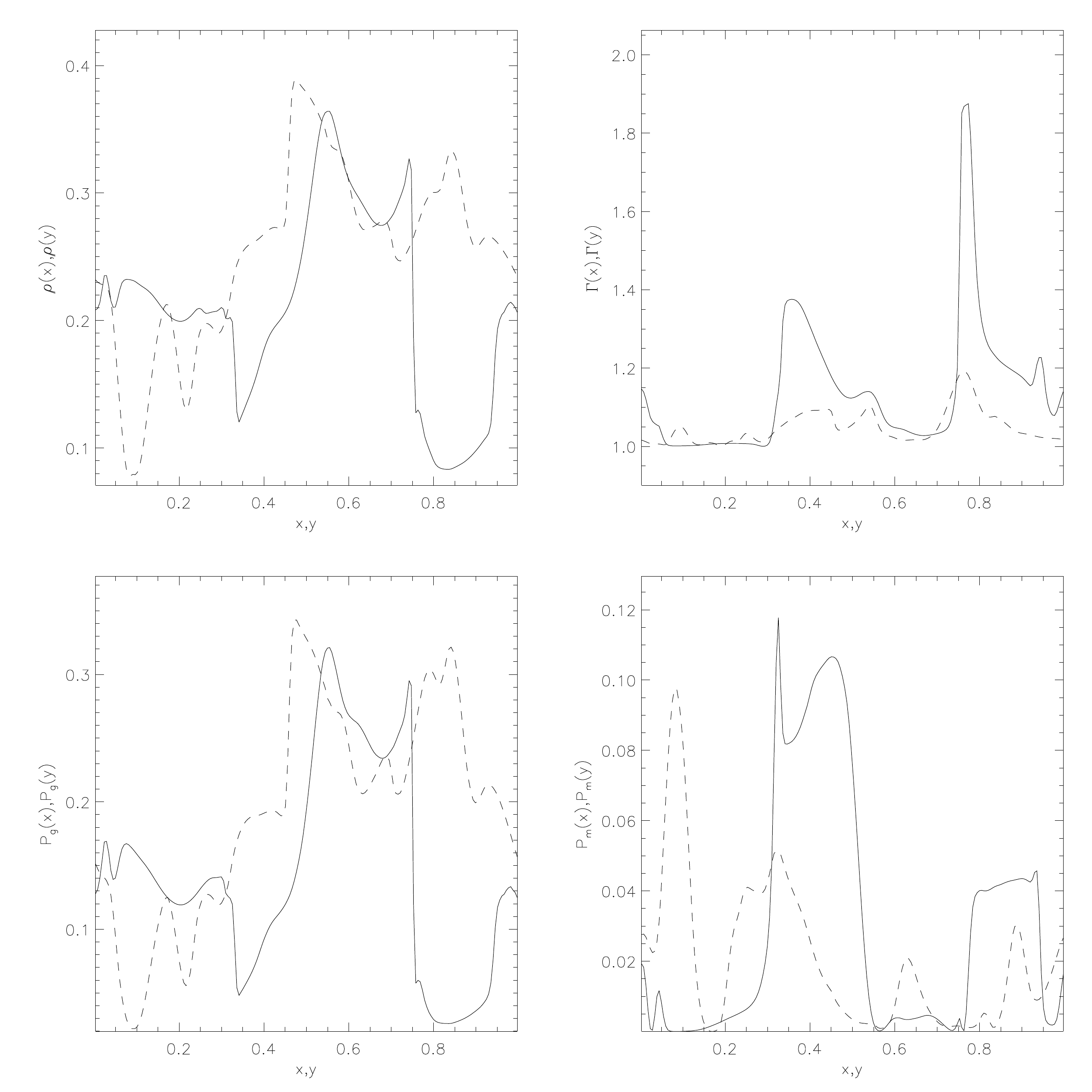}
\end{center}
\caption[]{Cuts through the Orszag-Tang vortex at $x=0.3125$ (solid lines), $y=0.3125$ (dashed lines) at $t=1.0$. Panels shows density (top left), Lorentz factor (top right), gas pressure (bottom left) and magnetic pressure (bottom right).}
\label{OrszagTangSlice1} 
\end{figure}

A useful test of the ability to maintain symmetry in complex flow is the
Orszag-Tang vortex \citep{Orszag:1979}. Our particular implementation
of this problem uses a square domain $0 \le x \le 1$; $0 \le y \le
1$ covered by $192\times192$ zones. The initial density and pressure
and uniform, with $\rho = 25/(36\pi)$, $P_g = 5/(12\pi)$ and $\gamma
= 5/3$. The velocity three-vector is initialized according to
$V^i = \left[ 0.5\sin(2\pi y), 0.5\sin(2\pi x)\right]$, whilst the
magnetic field is computed from the vector potential, $A_z = (B_0
/4\pi)\cos(4\pi x) + (B_0/2\pi)\cos(2\pi y)$ with $B_0 = 1/\sqrt{4\pi}$.
Distributions of density, gas pressure and magnetic pressure at
$t=1.0$ are shown in Figure \ref{OrszagTang}. In Newtonian MHD, the
VL+CT algorithm is able to maintain symmetry in this problem until
late times.  As can be seen in Figure \ref{OrszagTang}, in RMHD the
same is true using the HLLD solver, as well as both the HLLC and
HLLE solvers, although in the latter case the results are more
diffusive at a given resolution.  The most quantitative result from
this test is obtained from the horizontal and vertical slices shown
in Figure \ref{OrszagTangSlice1}.  Running
this test with $c=100$ to make it effectively non-relativistic as
in \cite{Gammie:2003} gives the same result as in \cite{Stone:2009}.
As far as we are aware, this is the first publication of a Orszag
Tang vortex for ideal RMHD (\citealt{Dumbser:2009} present a version
of this test in resistive RMHD).

\subsection{Multi-Dimensional Relativistic MHD Shocks}

One-dimensional shock tubes have long been a mainstay of numerical
algorithm development. We have found those presented in
\cite{Komissarov:1999,Balsara:2001,De-Villiers:2003,Mignone:2006}
particularly useful in testing the conservation properties of the
algorithm and the robustness of the primitive variable inversion
scheme. We have computed one-dimensional solutions to all of the
tests described by these authors and have found that our scheme is
able to obtain results comparable to those in the published literature.
We show an example below, as part of a
comparison to multidimensional versions of these tests.

\begin{figure}
\begin{center}
\includegraphics[width=\textwidth]{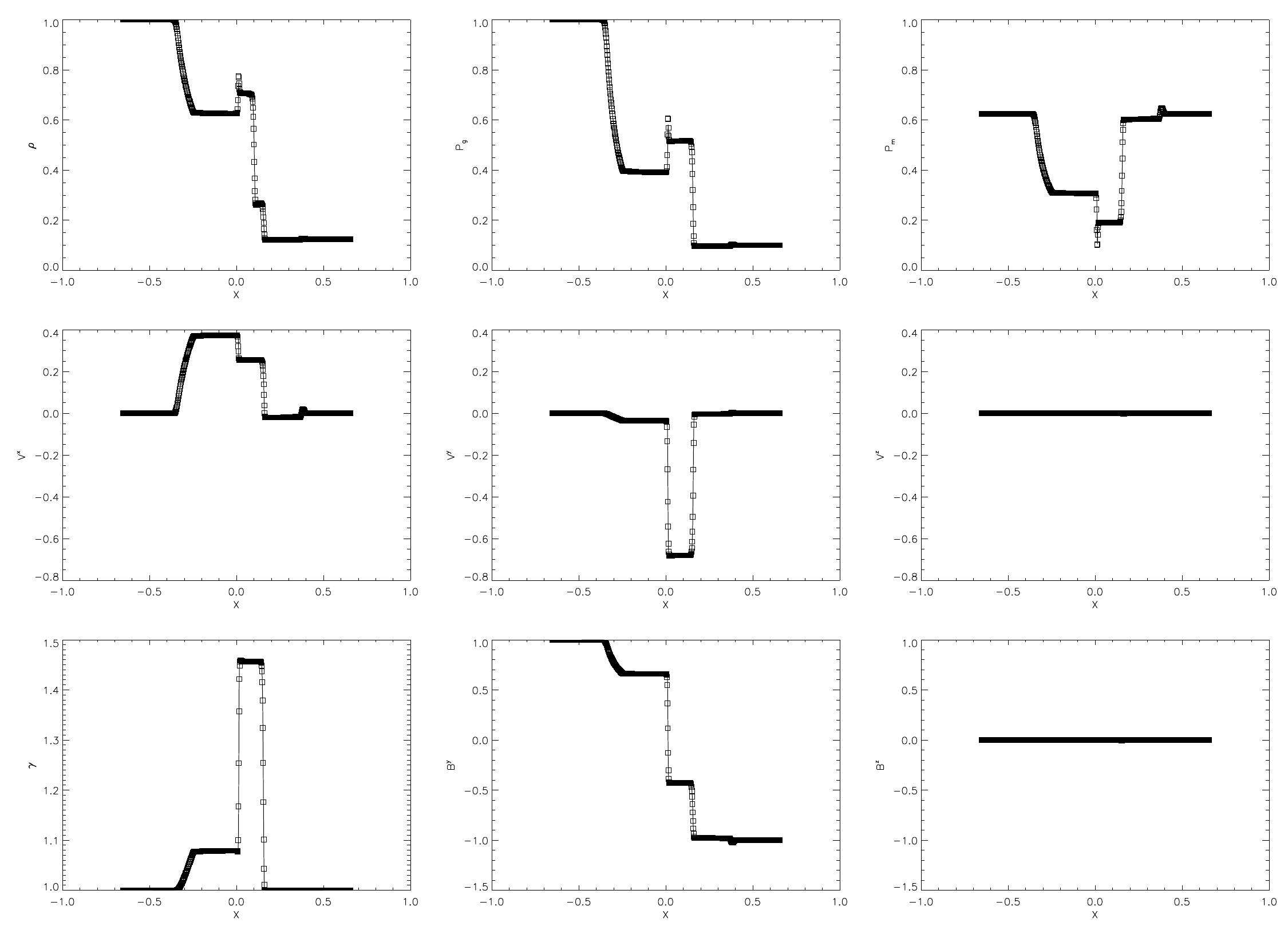}
\end{center}
\caption[]{$\gamma=2$ Brio-Wu shock at $t=0.4$. Solid lines indicate the one-dimensional solution, whilst squares indicate the three-dimensional solution.}
\label{SRmub1} 
\end{figure}

\begin{figure}
\begin{center}
\includegraphics[width=\columnwidth]{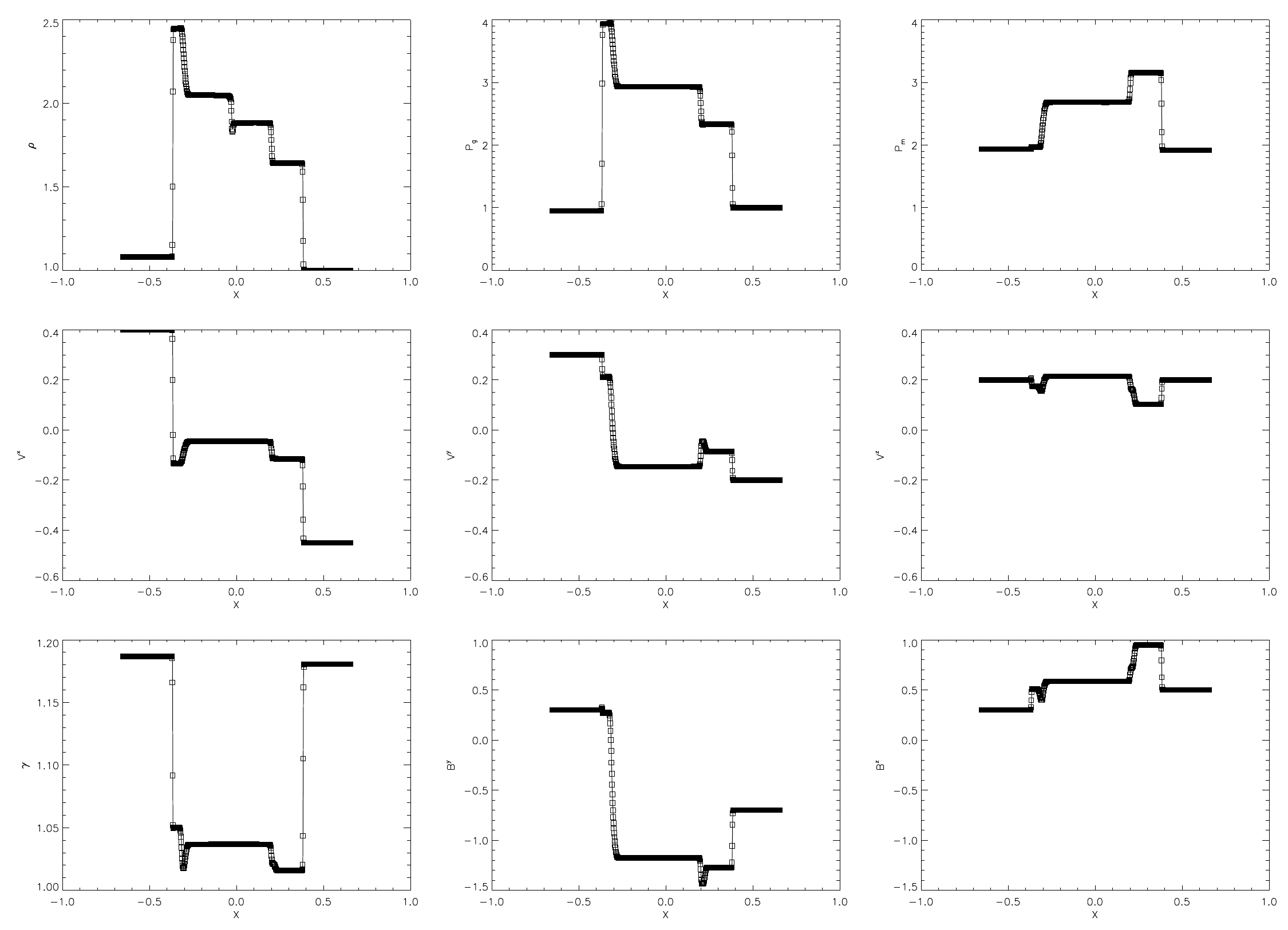}
\end{center}
\caption[]{Non-planar Riemann problem due to \cite{Balsara:2001} at $t=0.55$. Solid lines indicate the one-dimensional solution, whilst squares indicate the three-dimensional solution. }
\label{SRmub2} 
\end{figure}

One-dimensional shock tubes do not, however, reveal pathologies
associated with preservation of $\partial_i {\cal B}^i$, which can
result in jumps in the component of ${\cal B}$ normal to the shock
front \citep{Toth:2000}.  For this reason, we perform multidimensional
versions of these tests, with the initial discontinuity rotated so
that it is orientated obliquely to a three-dimensional grid following
the procedure described in \cite{Gardiner:2008}. The computational
grid has $768\times8\times8$ cells covering $-0.75\le x \le 0.75$,
$0\le y,z \le1/64$ such that it has $\delta x = \delta y = \delta
z = 1/512$.  Special boundary conditions are implemented in the
$y-$ and $z-$ directions to enforce periodicity \emph{parallel} to
the disconinuity (see \cite{Gardiner:2008}).   Figures \ref{SRmub1}
and \ref{SRmub2} show (appropriately rotated) multi-dimensional
solutions (denoted by squares) compared to one-dimensional solutions
run at equivalent resolution (lines) for the \cite{Brio:1988}
$\gamma=2$ shock tube at $t=0.4$ and the non-planar Riemann problem
due to \cite{Balsara:2001} at $t=0.55$.  The former is useful due
to its ubiquity as a test for schemes previously presented in the
literature. The latter is chosen for two reasons; firstly, the
Riemann fan contains three left-going waves (fast shock, Alfven
wave, and slow rarefaction), a contact discontinuity, and three
right-going waves (slow shock, Alfven wave and a fast shock);
secondly, the small relative velocities of the waves at the breakup
of the initial discontinuity lead to relatively fine structures,
which are hard to resolve \citep{Anton:2010}. As a result, comparison
of the results from this latter test in multi-dimensions with the
one-dimensional solution is very informative.  The data of Figures
\ref{SRmub1} and \ref{SRmub2} demonstrate excellent agreement between
the multi-dimensional and one-dimensional solutions; in neither
case are spurious magnetic fields generated normal to the shock
front. We conclude that the
integration scheme in combination with the HLLD Riemann solver does
an excellent job of resolving the structure of the Riemann fan in
multi-dimensions.

\subsection{Cylindrical Blast Wave}\label{blast}

A test problem that probes the ability of the code to evolve strong
multidimensional MHD shocks is the cylindrical blast wave. A popular
version of this test for relativistic MHD is originally due to
\cite{Komissarov:1999}; this problem has been considered subsequently
by \cite{Gammie:2003,Leismann:2005,Noble:2006,Mignone:2006,Del-Zanna:2007}.
In the original version of this problem, the blast wave is initialized
using a cylinder of over-pressured (by a factor of $3.33\times10^4$)
and over-dense (by a factor of $10^2$) gas, which expands into a
strongly magnetized ambient medium. \cite{Komissarov:1999} was able
to evolve such a setup by use of a strong artificial viscosity
(implemented in conservative form) and resistivity (implemented in
non-conservative form) for magnetizations ranging from $B_x = 0.01$
to $B_x=1.0$. The problem was subsequently reformulated by
\cite{Leismann:2005} such that the central cylinder was over-pressured
by a factor $2\times10^3$ compared to the ambient medium; this same
formulation was utilized by \cite{Del-Zanna:2007}. Both of these
authors ran what \cite{Komissarov:1999} described the ``moderately''
magnetized version of this test with $B_x = 0.1$. We have verified
that our integrator can execute this version of this test successfully,
along with the original \cite{Komissarov:1999} formulation at
moderate magnetization without the need for any artifical viscosity or
resistivity.

\begin{figure}
\begin{center}
\includegraphics[width=0.45\textwidth]
{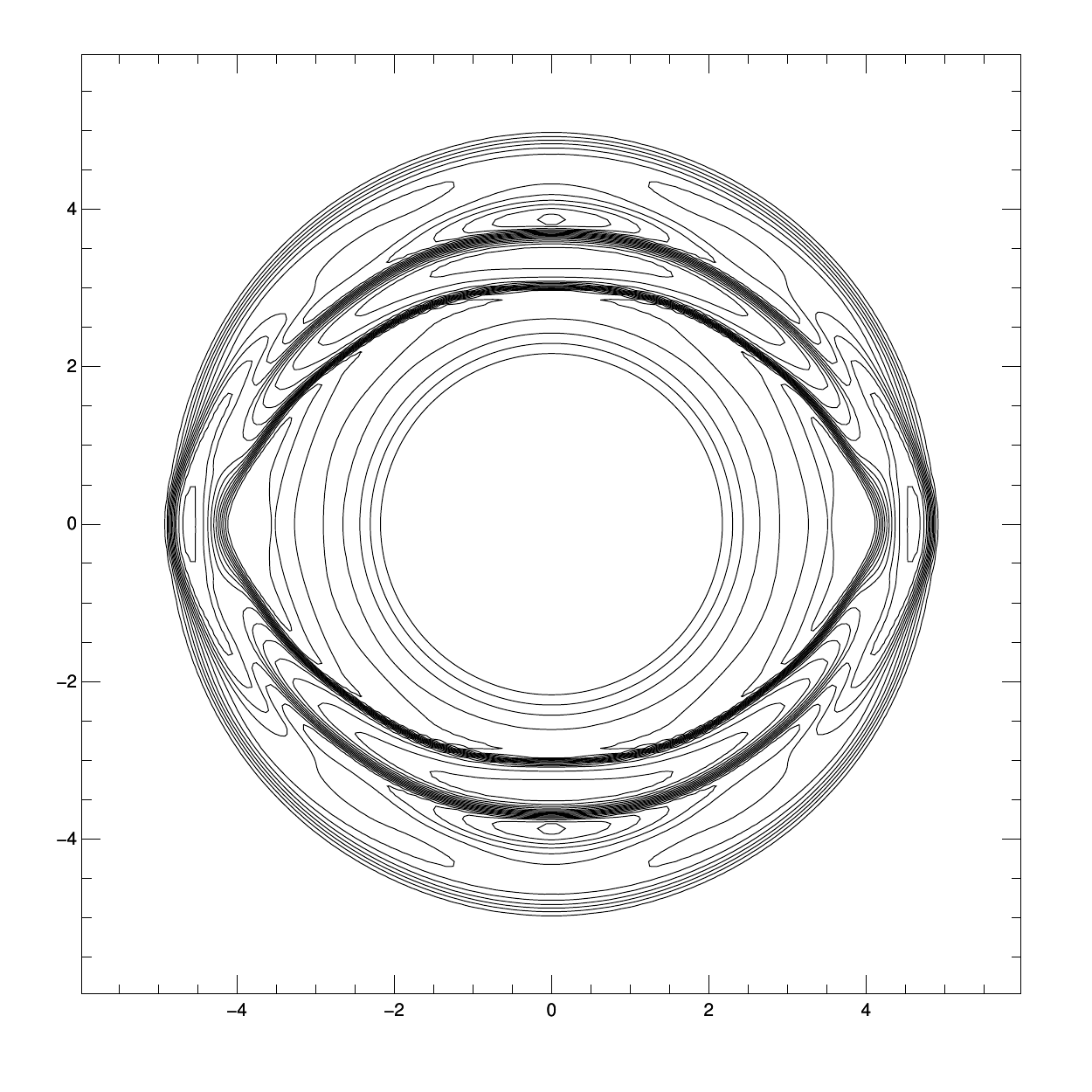}
\includegraphics[width=0.45\textwidth]
{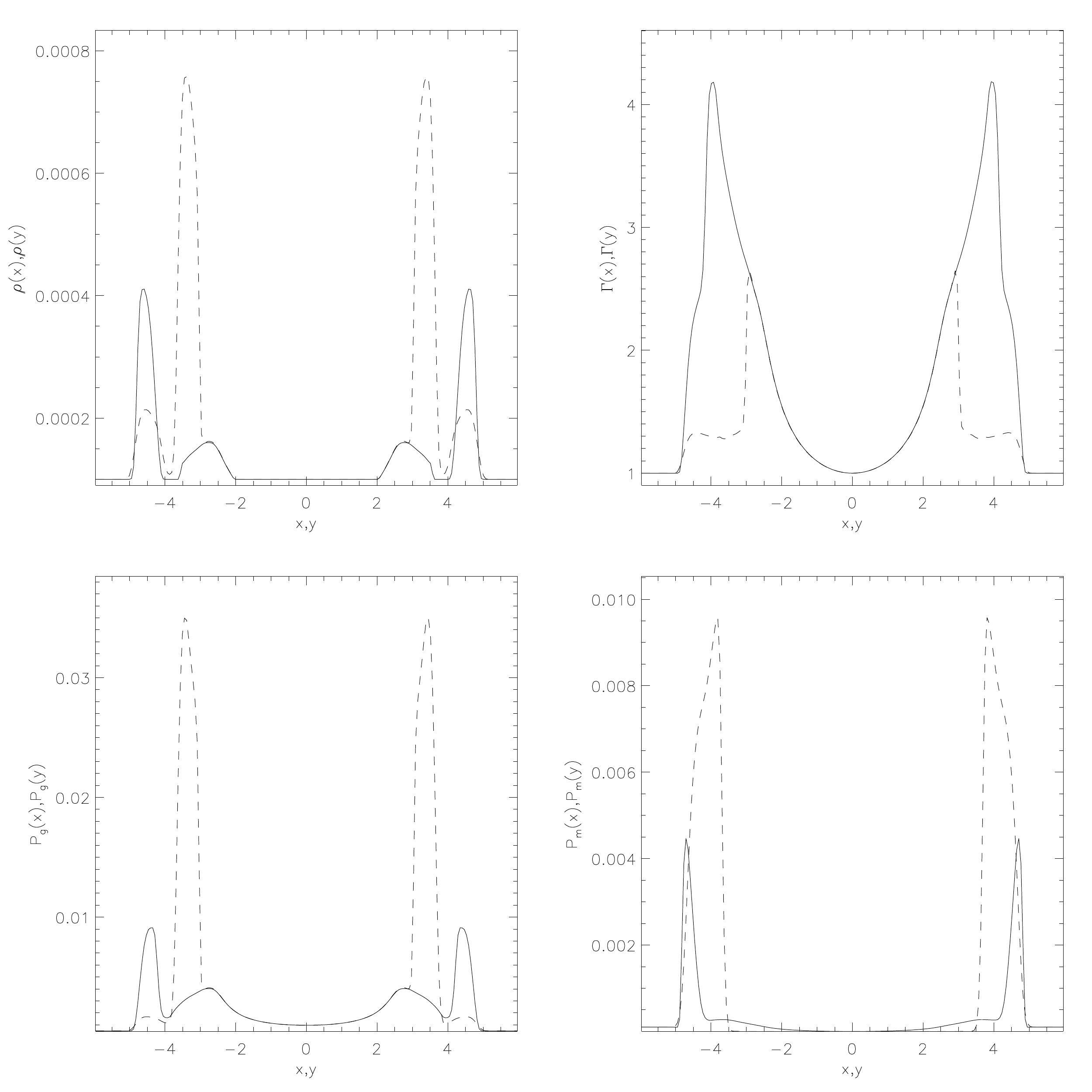}
\end{center}
\caption[]{Structure of the \cite{Leismann:2005} formulation of the \cite{Komissarov:1999} cylindrical blast wave in a moderately magnetized ($B=0.1$) medium at $t=4.0$. The left hand panel shows density using $40$ contours distributed logarithmically between $10^{-4}$ and $10^{-2}$. The right hand panels show one dimensional cuts along $y=0$ (solid lines) $x=0$ (dashed lines) for density, Lorentz factor, gas pressure and magnetic pressure.}
\label{DZ07Blast} 
\end{figure}

\begin{figure}
\begin{center}
\includegraphics[width=0.45\textwidth]
{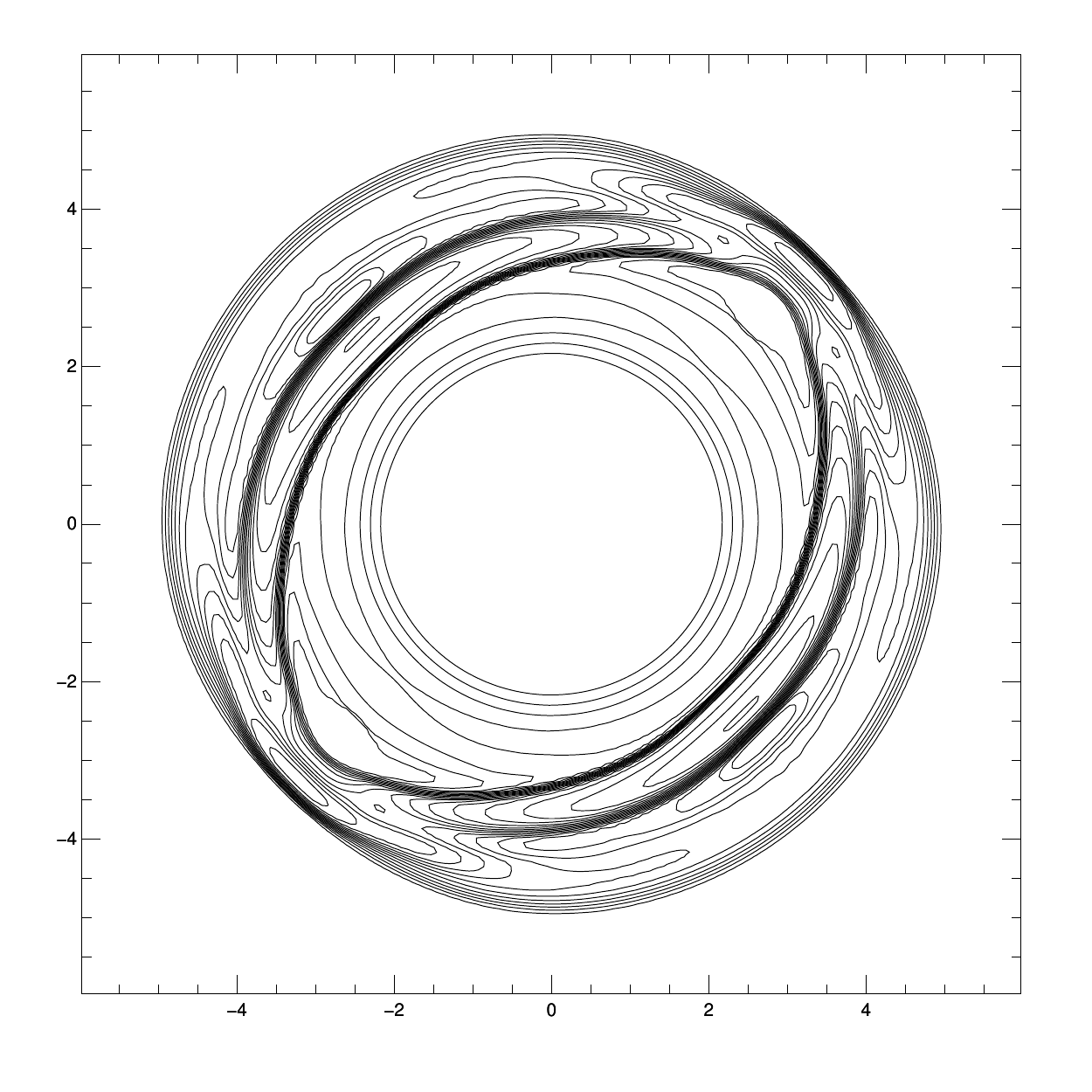}
\includegraphics[width=0.45\textwidth]
{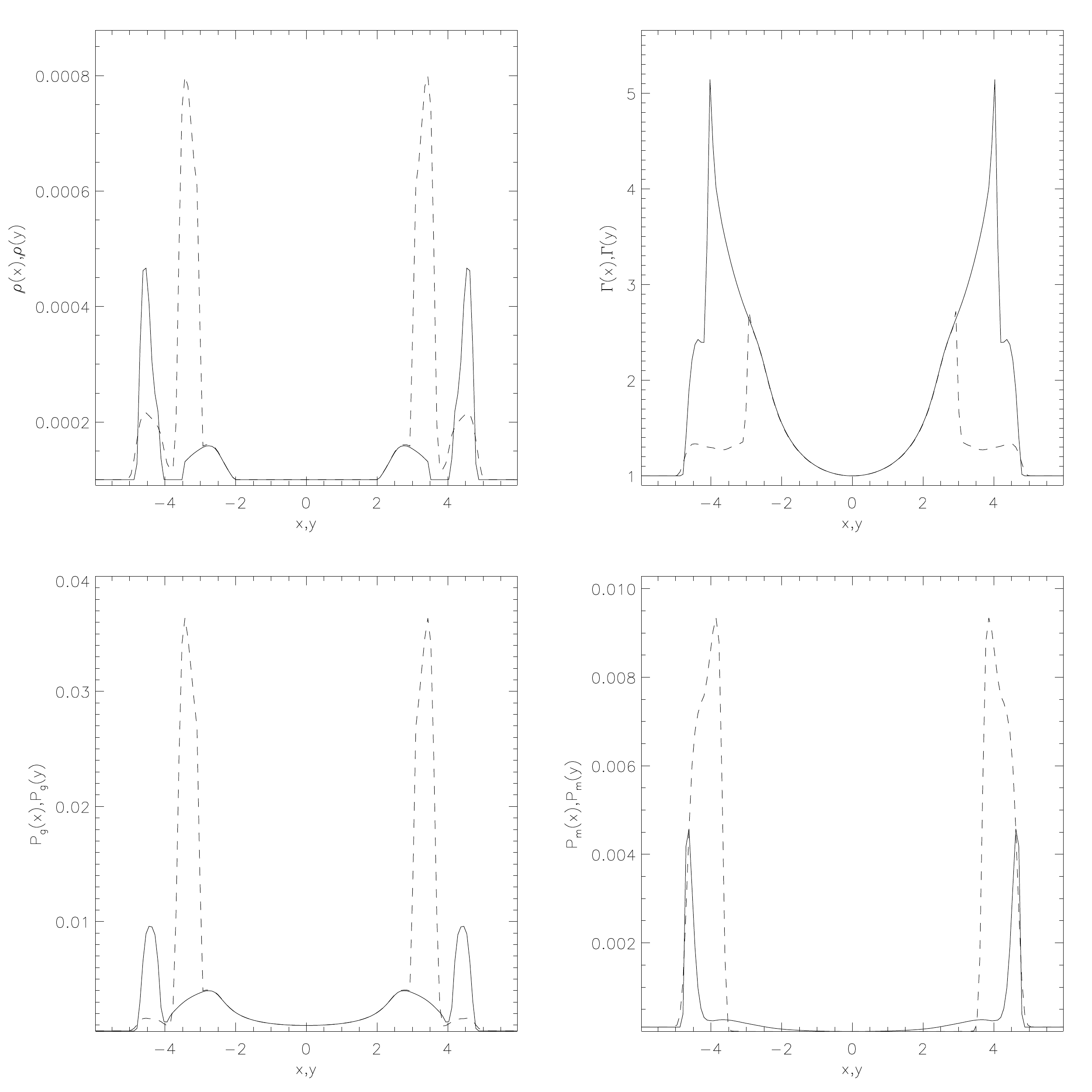}
\end{center}
\caption[]{As in Fig. \ref{DZ07Blast} for the magnetic field aligned at $\theta = 45^\circ$ to the grid. The right hand panels show slices along the grid diagonals.}
\label{DZ07Blast_rot} 
\end{figure}

We present results based on the \cite{Leismann:2005} version of the
test as \cite{Del-Zanna:2007} provide results that enable quantitative
comparison. The problem is run on a grid of $200\times200$ zones
covering a domain $-6.0 \le x \le 6.0$, $-6.0 \le y \le 6.0$ using
a Courant No. of $0.1$ \cite[as in][]{Noble:2006} and $\gamma =
4/3$. The ambient medium is filled with low density, $\rho = 10^{-4}$
and gas pressure, $P_g = 5.0\times10^{-4}$ with uniform magnetic
field, ${\cal B}^x = 0.1$, corresponding to an initial gas $\beta
= P_g / P_m = 10^{-1}$ and Alfven speed, $v_A = 0.91$ ($\Gamma_A =
(1-v^2_A)^{-1/2} = 2.4$) in the ambient medium. An over-pressured,
$P_g = 1.0$ and over-dense, $\rho = 10^{-2}$ cylinder of radius
$R=0.8$ is placed in the center of the grid. The structure of the
blast wave at $t=4.0$ is shown in Figure \ref{DZ07Blast}. The blast
wave is top-bottom and left-right symmetric in all of the variables
shown. To make this test as quantitative as possible, we show slices
along the lines $x=0$, $y=0$ for density, lorentz factor, gas
pressure and magnetic pressure.  Comparison with the results of
\cite{Del-Zanna:2007} (who ran this problem using a fifth order
scheme) suggest that we have obtained results of comparable accuracy.
We have also verified that we can run this test with the magnetic
field oblique to the grid, without the development of significant
grid related artifacts; Figure \ref{DZ07Blast_rot} shows the structure of
the blast wave at $t=4.0$ for the case where the magnetic field is placed at a $45^\circ$ angle to the grid. While the overall structure of the blast
wave is similar between the aligned and rotated cases, there are differences
particularly in the maximum Lorentz factor of the blast wave parallel
to the field lines($\Gamma = 4.0$ in the aligned case vs. $\Gamma = 5.0$ in the oblique case), which we attribute to different levels of numerical diffusivity being produced by the Riemann solver for obliquely aligned magnetic fields.

The discussion of \cite{Del-Zanna:2007} suggests that
this test at higher magnetizations is extremely challenging due to
independent reconstruction errors in flow variables along with
imbalances in terms in the energy equation for flows with fluid or
Alfven velocities close to the speed light.  \cite{Komissarov:1999}
and \cite{Mignone:2006} avoid these problems by breaking total
energy conservation and, in the case of \cite{Mignone:2006} by
applying shock-limiting techniques. However, the need to resort to such
strategies limits the usefulness of this particular
formulation of the test.  This has led us to consider
a new variant of the cylindrical blast wave test at very high magnetizations
which most schemes (including ours) should be able to evolve without resorting to ad-hoc changes to the algorithm.

\begin{figure}
\begin{center}
\includegraphics[width=0.45\textwidth]
{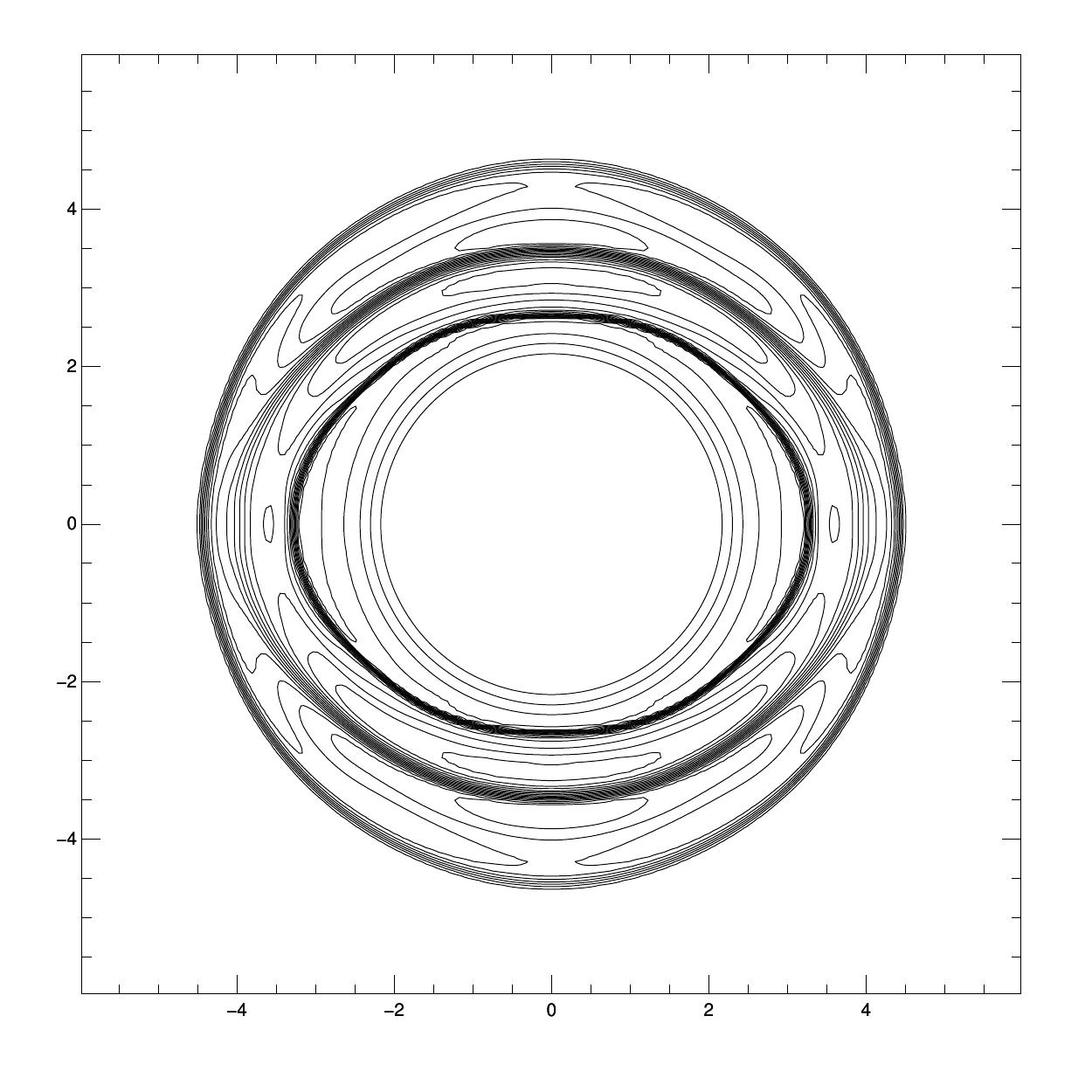}
\includegraphics[width=0.45\textwidth]
{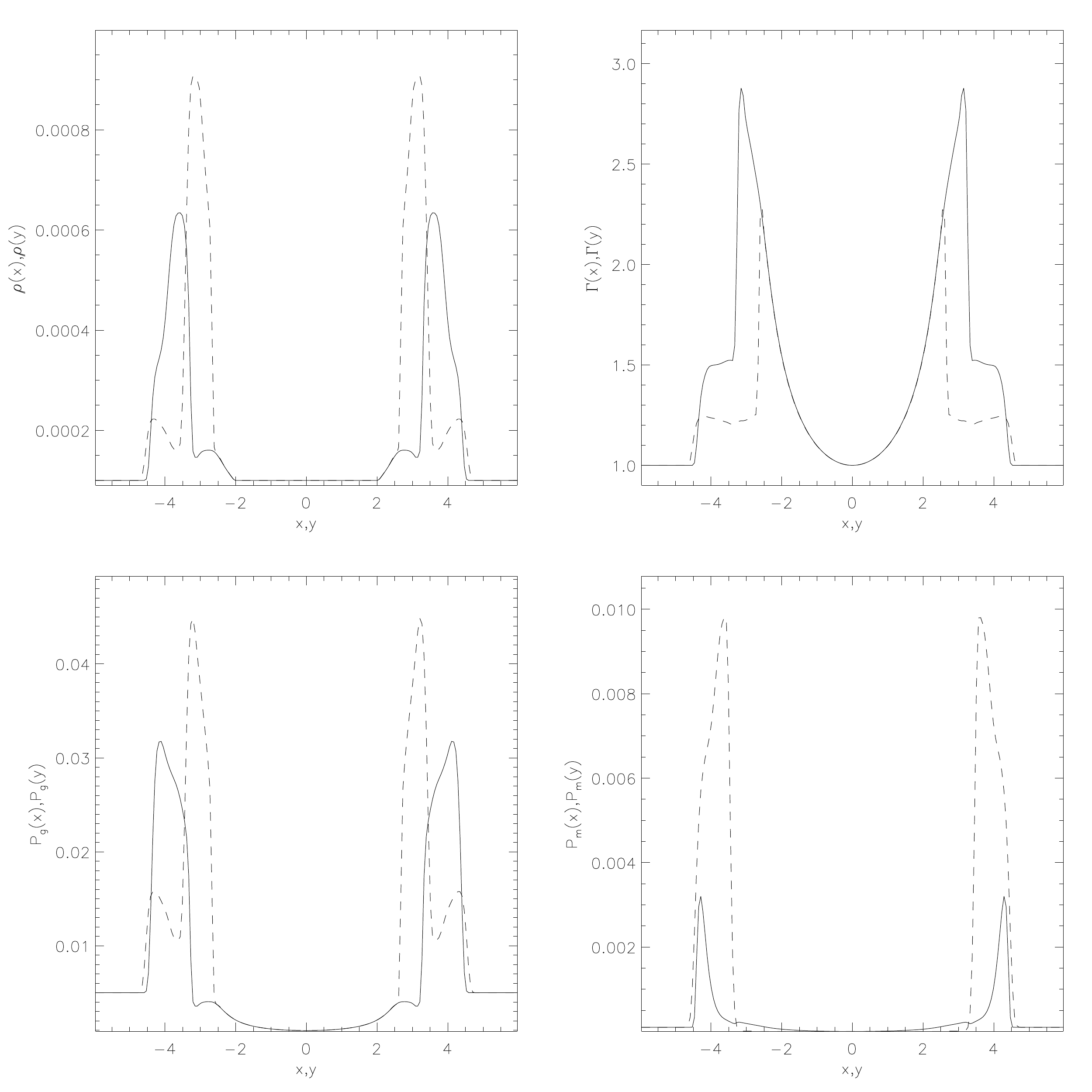}
\end{center}
\caption[]{Structure of cylindrical blast wave in a weakly magnetized ($B=0.1$) medium at $t=4.0$. The left hand panel shows density using $40$ contours distributed logarithmically between $10^{-4}$ and $10^{-2}$. The right hand panels show one dimensional cuts along $y=0$ (solid lines) $x=0$ (dashed lines) for density, Lorentz factor, gas pressure and magnetic pressure.}
\label{NewBlast_B01} 
\end{figure}

\begin{figure}
\begin{center}
\includegraphics[width=0.45\textwidth]
{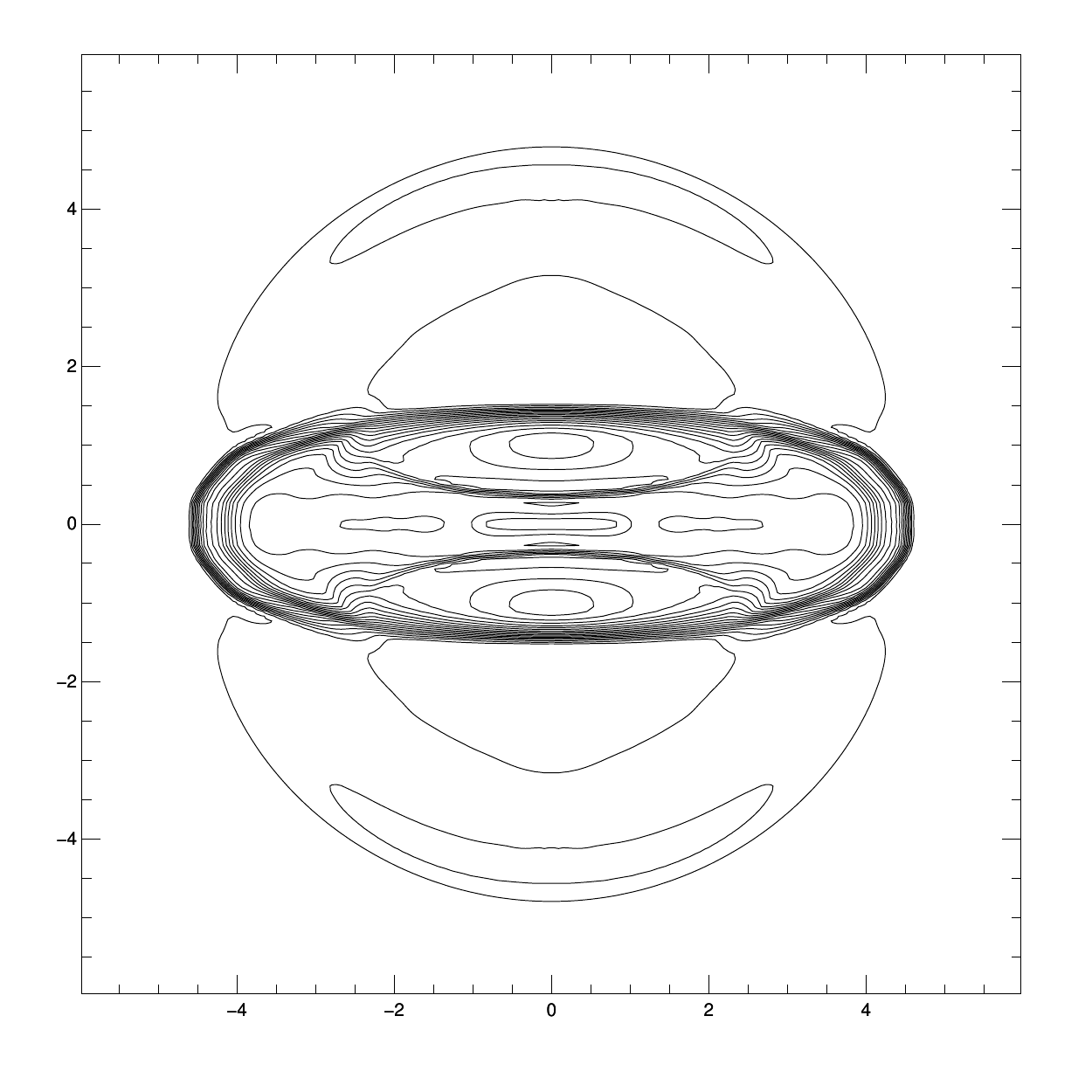}
\includegraphics[width=0.45\textwidth]
{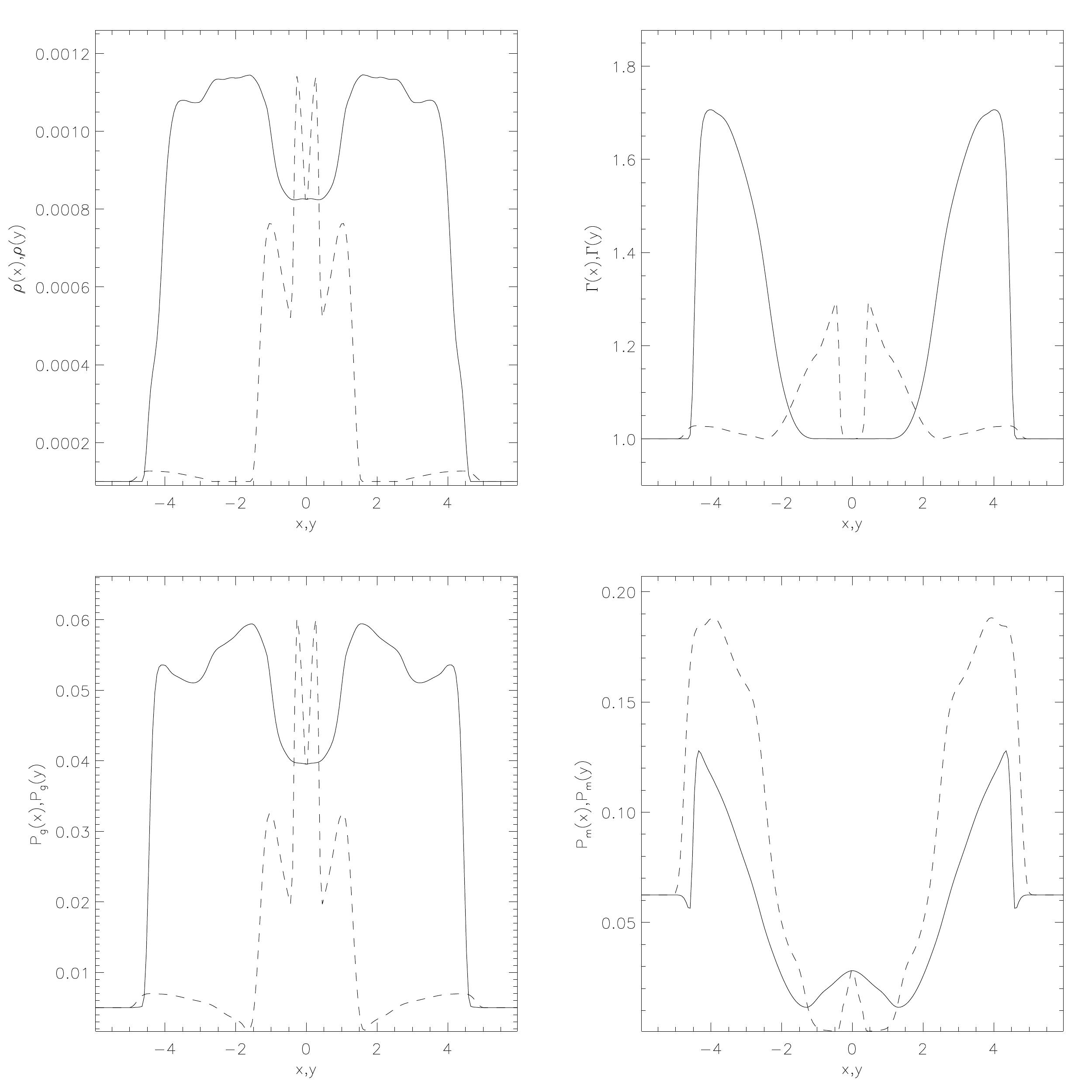}
\end{center}
\caption[]{Structure of cylindrical blast wave in a moderately magnetized ($B=0.5$) medium at $t=4.0$. The left hand panel shows density using $40$ contours distributed logarithmically between $10^{-4}$ and $10^{-2}$. The right hand panels show one dimensional cuts along $y=0$ (solid lines) $x=0$ (dashed lines) for density, Lorentz factor, gas pressure and magnetic pressure.}
\label{NewBlast_B05} 
\end{figure}

\begin{figure}
\begin{center}
\includegraphics[width=0.45\textwidth]
{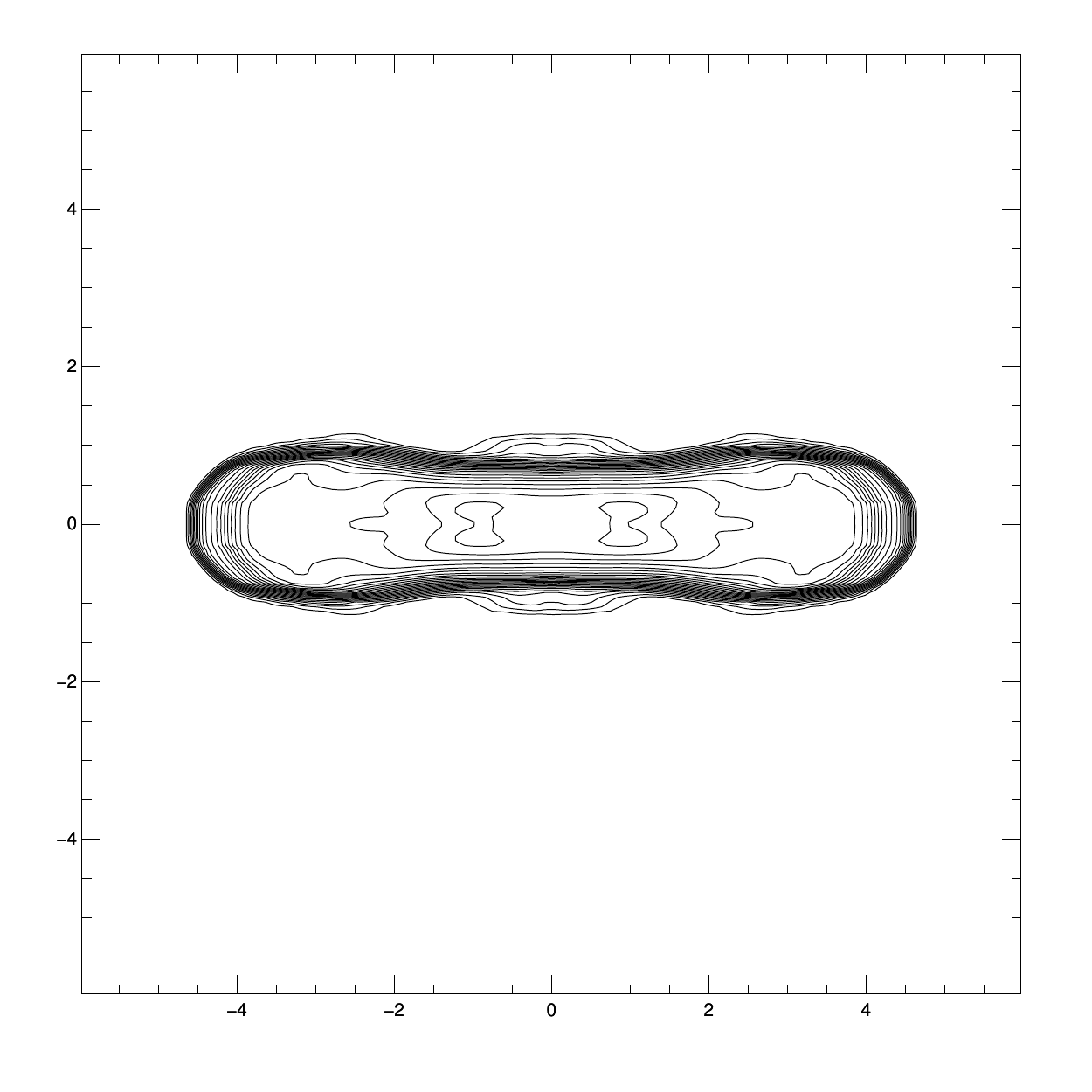}
\includegraphics[width=0.45\textwidth]
{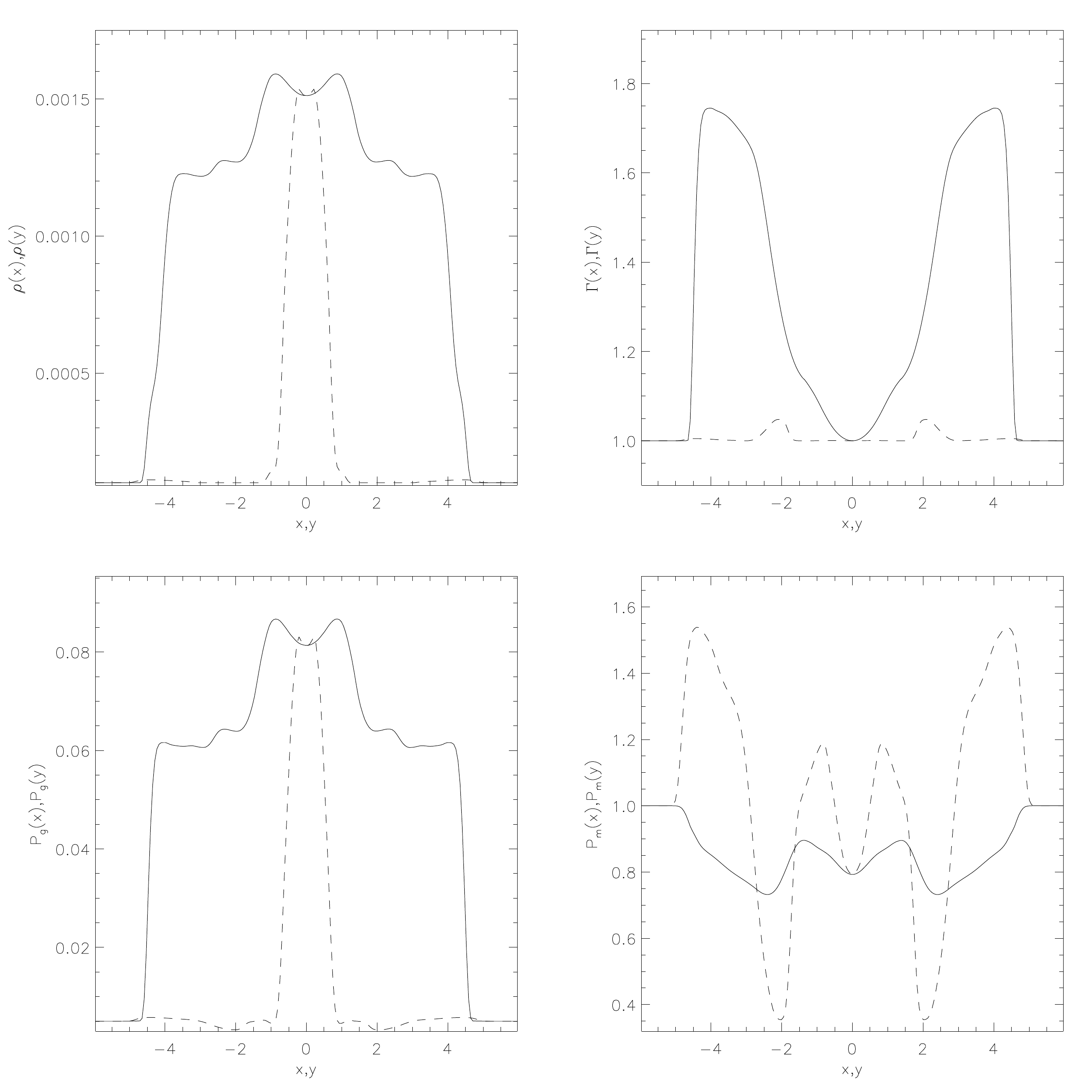}
\end{center}
\caption[]{Structure of cylindrical blast wave in a strongly magnetized ($B=1.0$) medium at $t=4.0$. The left hand panel shows density using $40$ contours distributed logarithmically between $10^{-4}$ and $10^{-2}$. The right hand panels show one dimensional cuts along $y=0$ (solid lines) $x=0$ (dashed lines) for density, Lorentz factor, gas pressure and magnetic pressure.}
\label{NewBlast_B10} 
\end{figure}

Our modified version of this test adopts
a gas pressure $P_g = 5\times10^{-3}$ in the ambient
medium and $P_g = 1.0$ in the over-pressured,
over-dense cylinder.  This modification allows
us to probe to up to ${\cal B}^x=1.0$, corresponding to an initial
gas $\beta = P_g / P_m = 10^{-2}$ and Alfven speed, $v_A = 0.98$
($\Gamma_A = (1-v^2_A)^{-1/2} = 5.0$) in the ambient medium.
We emphasize that by the former measure, the maximum
magnetization obtained in this test is an order of magnitude stronger
than in the \cite{Leismann:2005} test. The rest of the parameters
remain the same as in the original formulation by \cite{Komissarov:1999}.
The results of this test at $t=4.0$ are shown in Figure \ref{NewBlast_B01}
for the weakly magnetized, ${\cal B}^x = 0.1$ case; Figure
\ref{NewBlast_B05} for the moderately magnetized, ${\cal B}^x =
0.5$ case and Figure \ref{NewBlast_B10} for the strongly magnetized,
${\cal B}^x = 1.0$ case. All of the tests reveal a high degree of
symmetry and conform to expectations based on previous cylindrical
blast wave simulations; i.e. as the magnetization is increased, the
blast wave is confined to propagate along the magnetic field lines,
creating a structure elongated in the $x$-direction. We have found
that we are able to successfully evolve the weakly- and moderately-
magnetized version of the test with the magnetic field obliquely
aligned to the grid; grid related artifacts are produced for the
strongly magnetized version of the test when executed at second
order using the HLLD solvers. These issues are absent for the HLLE
solver due to the higher numerical diffusion applied to the solution
in that case.

Our ability to execute the reformulated version of the blast wave
test at high magnetization (${\cal B}^x=1.0$, $\beta = P_g / P_m =
10^{-2}$ ,$\Gamma_A = (1-v^2_A)^{-1/2} = 5.0$) led to us to investigate
the maximum magnetization that the code is capable of evolving. At
very high Alfven speeds ($\Gamma_a > 5.0$), the fast magnetosonic
wave and Alfven waves become increasingly degenerate, which we have
found to cause stability problems within the HLLD Riemann solver.
Experiments using the HLLE solver and third-order reconstruction
of primitive variables have shown that we are able to successfully
evolve configurations with at least ${\cal B}^x=10^3$, corresponding
to an initial gas $\beta = P_g / P_m = 10^{-8}$ and Alfven speeds
equivalent to $\Gamma_A = (1-v^2_A)^{-1/2} \sim 5\times10^{3}$,
i.e. highly relativistic Alfven speeds. Clearly,
the origin of the numerical issues in evolving strongly magnetized
blast waves does not lie in reconstruction errors, or imbalances
in the energy equation for Alfven velocities close to the speed of
light, as was suggested by \cite{Del-Zanna:2007}. Instead, our
results suggest that the origin of the difficulties in evolving
strongly magnetized versions of the \cite{Komissarov:1999,Leismann:2005}
blast wave problem lie in the initial conditions of the hydrodynamic
variables, i.e. the blast wave itself. To investigate this possibility,
we have compared the properties of the \cite{Leismann:2005} blast
wave executed with ${\cal B}^x=1.0$ with our reformulation at this
same magnetization. We find that the maximum Lorentz factor of the
blast wave in the former case is $\Gamma \sim 4.0$, while in the
latter it is $\Gamma\sim1.8$. We note, however, that while the strength of these two blast waves differ by a factor of two (by the measure of the relative Lorentz factor), they result in similar amplitude variations
in the magnetic field, $\delta |{\cal B}|^2 / |{\cal B}|^2$, which in turn
suggests that problems in evolving the \cite{Leismann:2005} formulation
do not result solely from the increased magnetization.
In addition, we find that the structure of the density  is similar
between the two evolutions, except that the \cite{Leismann:2005}
blast wave exhibits grid scale artifacts which are absent in the new
formulation. Executing the strongly magnetized version of the
\cite{Leismann:2005} blast wave in Newtonian physics removes these
artifacts. Finally, if we execute an unmagnetized version of the blast
wave in relativistic physics, we find that the algorithm exhibits a similar
failure mode (grid scale artefacts in the density) when the
blast wave exceeds $\Gamma = 8$.
Taken together, these results suggest that the origin of the problems in executing strongly magnetized blast waves with
$\Gamma\ge4.0$ lies in the primitive variable inversion scheme,
rather than in reconstruction operations or imbalances in the energy
equation. This is not surprising, as for high Lorentz factor, high
magnetization flows, the recovered primitive variables are likely
to be dominated by roundoff error. There are several possible
resolutions to this issue; one can switch to a primitive variable
inversion scheme that evolves the difference of the total energy
and the density \citep{Mignone:2007}; alternatively, one could
utilize the $2D$ scheme of \cite{Noble:2006,Del-Zanna:2007}, where
one treats $|V|^2$ and $\rho h \Gamma^2$ as independent variables.
We leave detailed investigation of these issues to future work.

\subsection{Kelvin-Helmholtz Instability}\label{kh2d}

As a final test of our integration scheme, we present calculations
of the linear growth phase of the two-dimensional Kelvin-Helmholtz
instability (KHI).  \cite{Mignone:2009} presented a convergence
study of the linear phase of the two-dimensional KHI using both the
HLLE and HLLD Riemann solvers, finding an order of magnitude increase
in the total power in velocity fluctuations transverse to the shear
layer for the latter of these two Riemann solvers compared to the
former, even in the case where the solution was converged. These
authors suggested that the origin of this difference is
the ability of the HLLD Riemann solver to resolve small scale
structures within turbulence generated in the nonlinear regime,
leading to an enhancement in the effective resolution compared to
the HLLE case. We therefore focus our discussion on the convergence
of the simulations computed with each of the HLLE, HLLC and HLLD
Riemann solvers and the power spectrum of nonlinear RMHD turbulence
driven by the KHI in each simulation.

The initial conditions for this test consist of a combination of those
described in \cite{Mignone:2009} and \cite{Zhang:2009}.
The shear velocity profile is given by
\begin{equation}
V^x = \left\{ \begin{array}{cc}
V_{\mathrm{shear}} \tanh \left(\frac{y-0.5}{a}\right) & \mathrm{if} \; y > 0.0 \\
-V_{\mathrm{shear}} \tanh \left(\frac{y+0.5}{a}\right) & \mathrm{if} \;  y \le 0.0
\end{array}\right.
\end{equation}
Here, $a = 0.01$ is the characteristic thickness of the shear layer,
$V_{shear} = 0.5$, corresponding to a relative Lorentz factor of
$2.29$. The instability is seeded by application of a single mode
perturbation of the form
\begin{equation}
V^y = \left\{ \begin{array}{cc}
A_0 V_{\mathrm{shear}} \sin \left( 2 \pi x \right) \exp \left[ -\left(\frac{y+0.5}{\sigma}\right)^2 \right] & \mathrm{if} \;y > 0.0 \\
- A_0 V_{\mathrm{shear}} \sin \left( 2 \pi x \right) \exp \left[ -\left(\frac{y+0.5}{\sigma}\right)^2 \right]  & \mathrm{if} \; y \le0.0\\
\end{array}\right.
\end{equation}
Here, $A_0 = 0.1$ is the perturbation amplitude and $\sigma = 0.1$
describes the characteristic length scale over which the perturbation
amplitude decreases by a factor $e$. Symmetry is broken by applying
$1\%$ Gaussian perturbations modulated by the same exponential
distribution as used above to the $x,y$ components of the initial
velocity field. The initial pressure distribution is uniform with
$P_g = 1.0$ and $\gamma = 4/3$. The density distribution is initialized
using the same profile used to define the shear velocity,
with $\rho = 1.0$ in regions with $V^x = 0.5$ and $\rho =
10^{-2}$ in regions with $V^x = -0.5$. The magnetic field was aligned
with the $x$ direction and initialized with ${\cal B}^x =10^{-3}$.
Finally, periodic boundary conditions were applied in all directions.

\begin{figure}
\begin{center}
\includegraphics[width=0.45\textwidth]{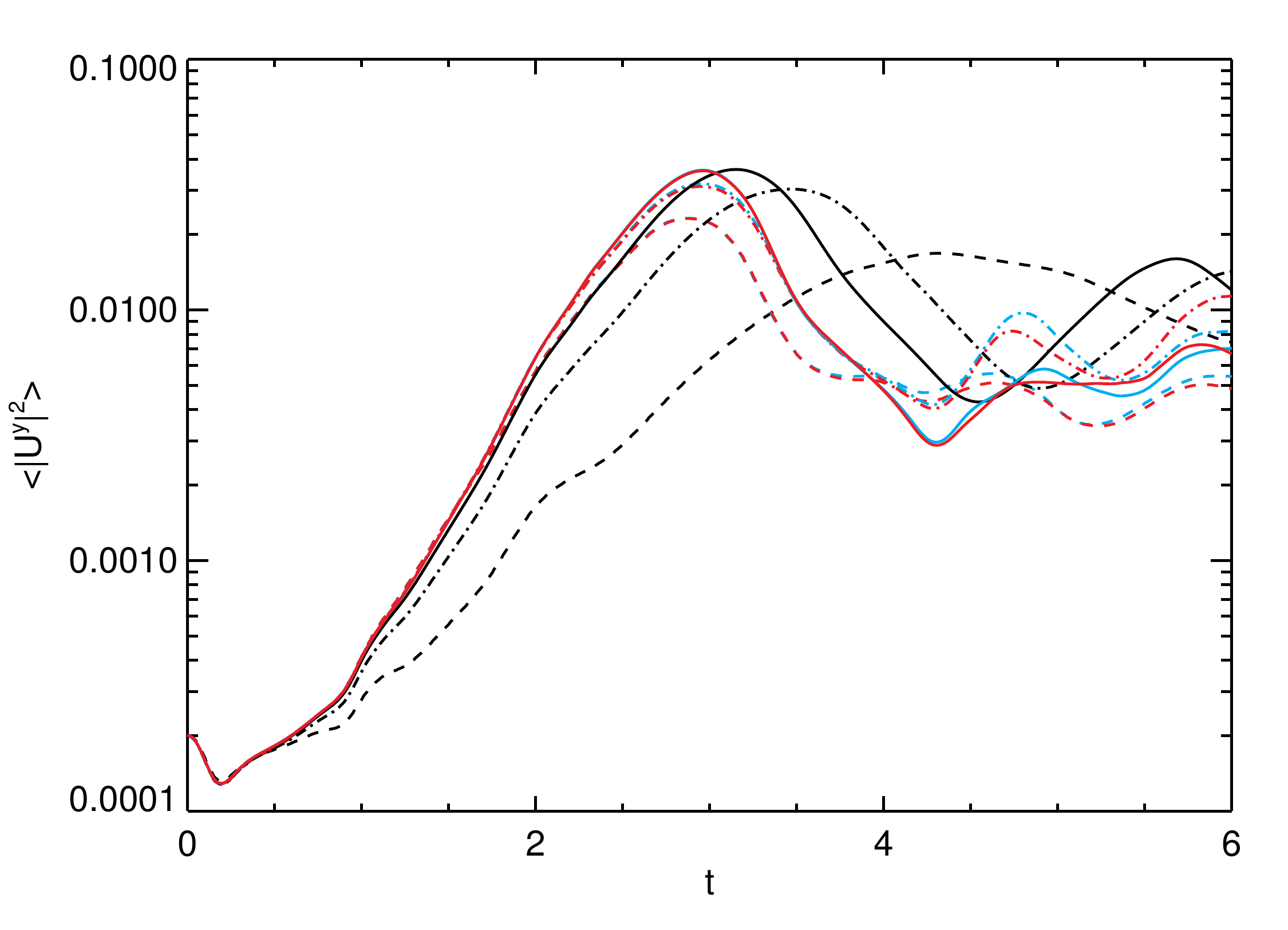}
\end{center}
\caption[]{Area averaged four-velocity transverse to the shear layer, $\left< |U^y|^2 \right>$ during the linear growth phase of the two-dimensional Kelvin-Helmholtz test problem. Black lines show results obtained with the HLLE Riemann solver, blue lines results obtained with the HLLC Riemann solver and red lines results obtained with the HLLD Riemann solver. Dashed lines denote results from low resolution simulations ($128\times256$ zones); dash-dot lines denote results from medium resolution simulations ($256\times512$ zones) and solid lines results from high resolution simulations ($512\times1024$ zones). Note that results for the HLLC and HLLD Riemann solver (blue and red lines) are essentially indistinguishable during the linear growth phase.}
\label{kh2d_ky} 
\end{figure}

The simulations are run on a domain covering $-0.5\le x \le 0.5$,
$-1.0\le y \le 1.0$ using $128\times256$ (low resolution),
$256\times512$ (medium resolution) and $512\times1024$ (high
resolution) zones with the HLLE, HLLC and HLLD Riemann
solvers. We assess the convergence using the area
averaged four-velocity transverse to the shear layer,
$\left<|U^y|^2\right>$, during the linear growth stage of the
instability (see Figure \ref{kh2d_ky}). \cite{Mignone:2009} found
that in simulations computed with the HLLD solver, the linear growth
rate displayed converged behavior even at low resolutions; while
the linear growth rate in simulations computed with the HLLE solver
increased with increasing resolution, tending to the growth rate
measured in the HLLD based simulations. The data of Figure \ref{kh2d_ky}
display the same behavior as described by \cite{Mignone:2009} with
the additional result that the HLLC and HLLD Riemann solvers display
identical linear growth rates. This leads us to the conclusion that
it is the inclusion of the contact discontinuity within the Riemann
fan for these two solvers that leads to this behavior, which is not
surprising given the density variation across the shear layer in the
initial conditions.  We note that
the maximum amplitude of $\left<|U^y|^2\right>$, which marks the
termination of the linear growth phase, occurs at $t=3.0$ for
simulations computed using the HLLC and HLLD Riemann solvers and
that by this measure, simulations computed using these two Riemann
solvers exhibit converged behavior at a resolution of $256\times512$
zones, corresponding to the characteristic thickness of the shear
layer, $a$, being resolved by $2$ zones.

\begin{figure}
\begin{center}
\includegraphics[width=0.1\textwidth, viewport=0 400 100 600,clip]{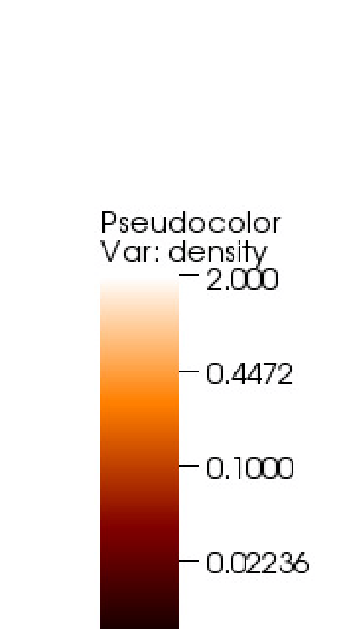}
\includegraphics[width=0.29\textwidth, viewport=140 10 430 560,clip]
{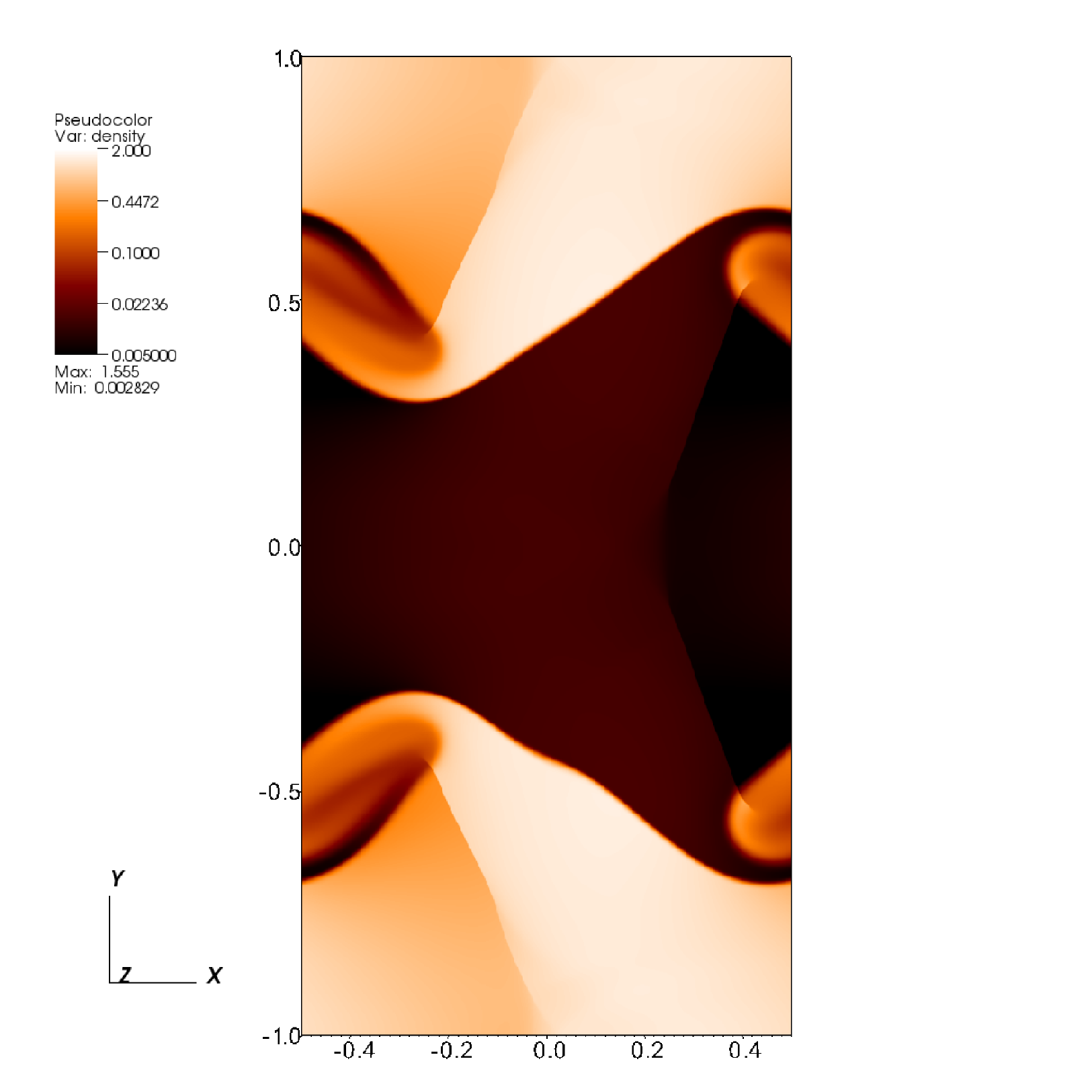}
\includegraphics[width=0.29\textwidth, viewport=140 10 430 560,clip]
{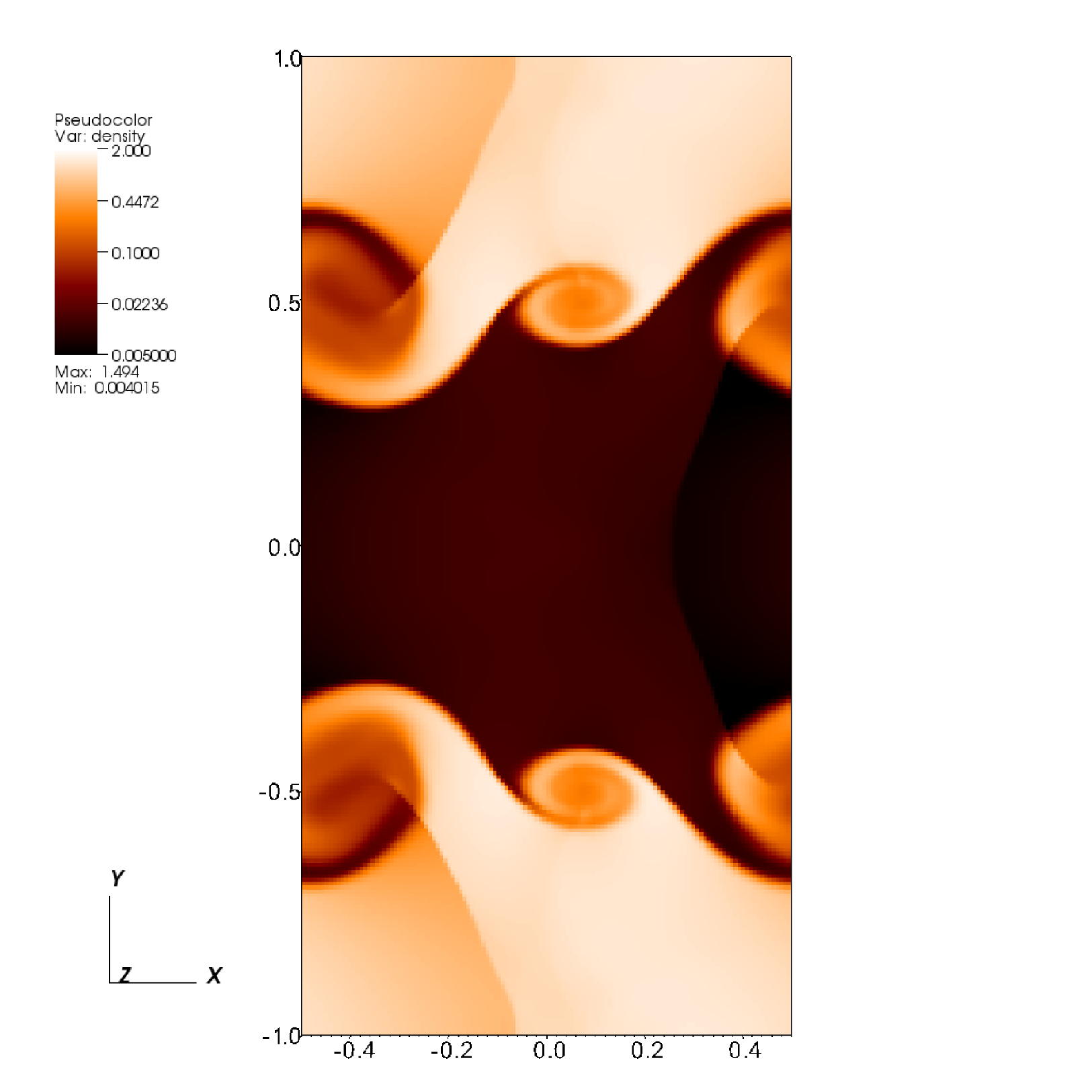}
\includegraphics[width=0.29\textwidth, viewport=140 10 430 560,clip]
{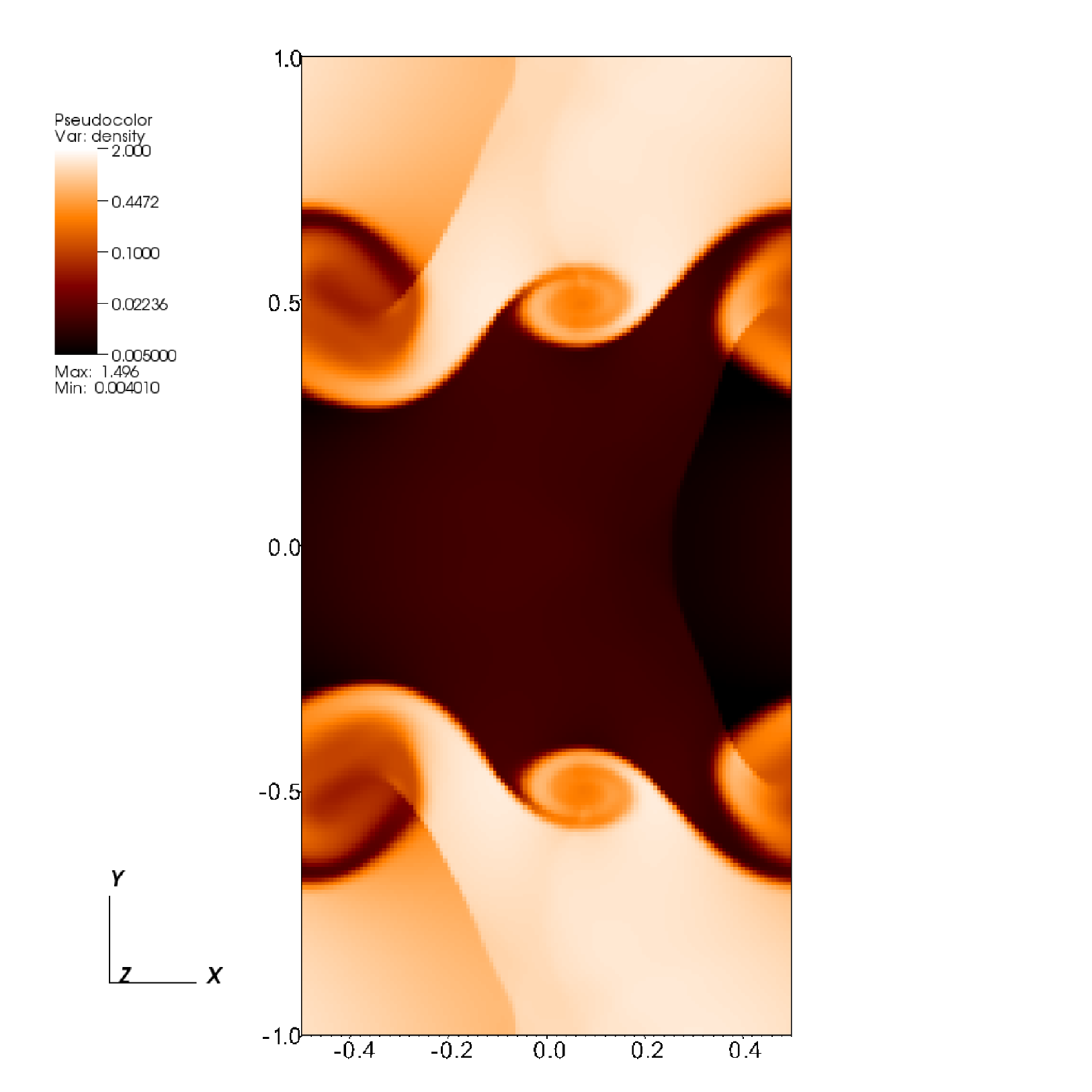}
\end{center}
\caption[]{Comparison of density distributions at $t=3.0$ for the two-dimensional version of the Kelvin-Helmholtz instability. The left panel shows results obtained using the HLLE solver at $512\times1024$ zones, the center panel results obtained using the HLLC solver at $128\times256$ zones and the right panel results obtained using the HLLD solver at $128\times256$. Note the secondary vortex visible at $x=|0.5|,y=0.0$ in results obtained from the HLLC and HLLD Rieman solvers at low resolution which is absent in results obtained from the HLLE solver even at high resolution.}
\label{kh2d_den} 
\end{figure}

In Figure \ref{kh2d_den}, we compare the density distribution
measured at $t=3.0$ (chosen to correspond with the termination of
the linear growth phase) in the high resolution simulation computed
using the HLLE Riemann solver with that measured in the low resolution
simulations computed using the HLLC and HLLD Riemann solvers.
Immediately apparent is the absence of the secondary vortex in the
former case, even though the resolution employed in this case is a
factor of four greater than that for the other solvers. We have further verified that this secondary vortex does not appear in simulations using up to $4096\times8192$ zones (a factor $32^2$ increase in resolution) and the HLLE solver.  Clearly, the absence of the contact discontinuity in the Riemann fan of the HLLE solver has a substantial impact on the structure of the instability even at the highest resolution computed here.

\begin{figure}
\begin{center}
\includegraphics[width=0.32\textwidth]{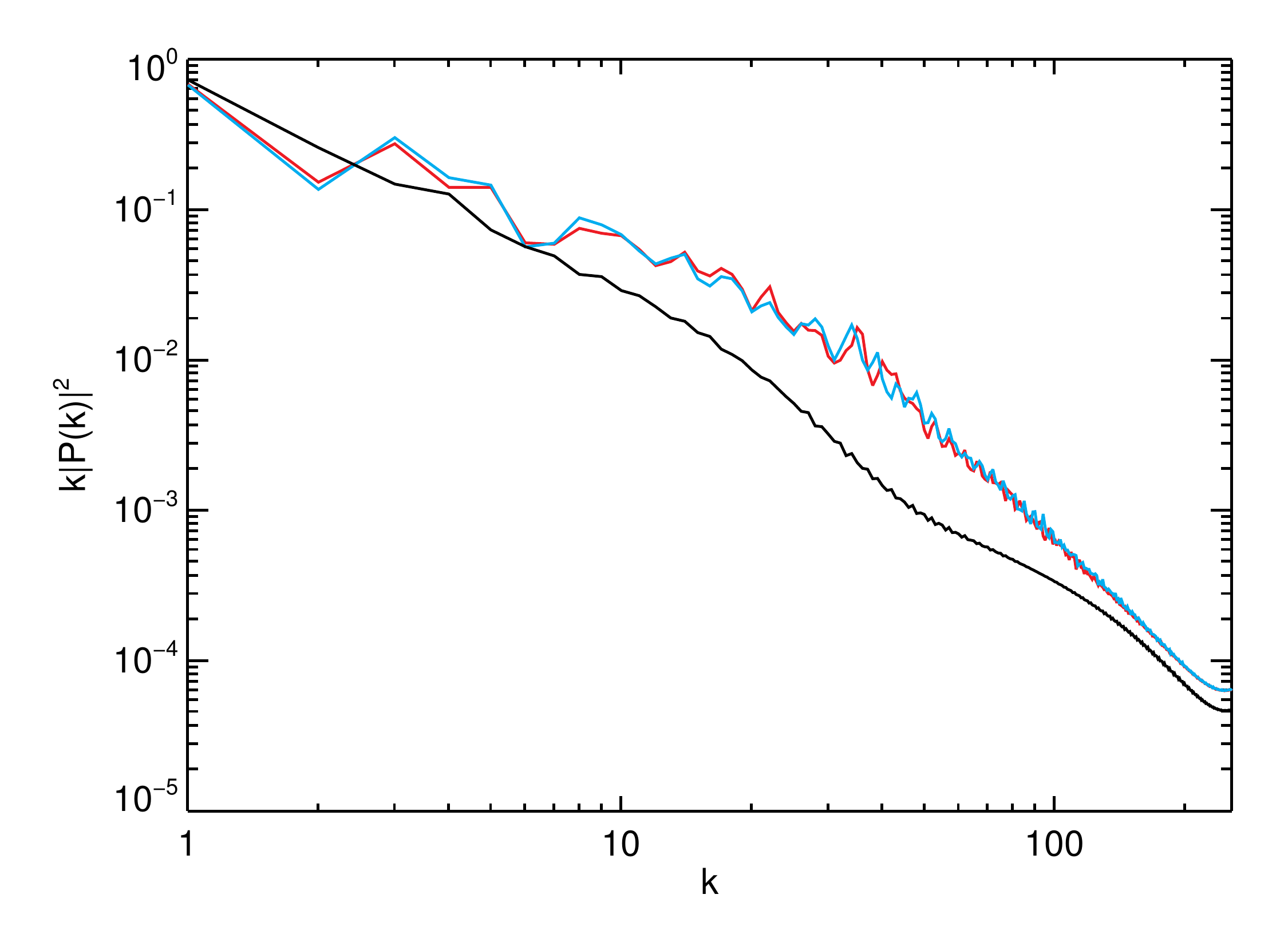}
\includegraphics[width=0.32\textwidth]{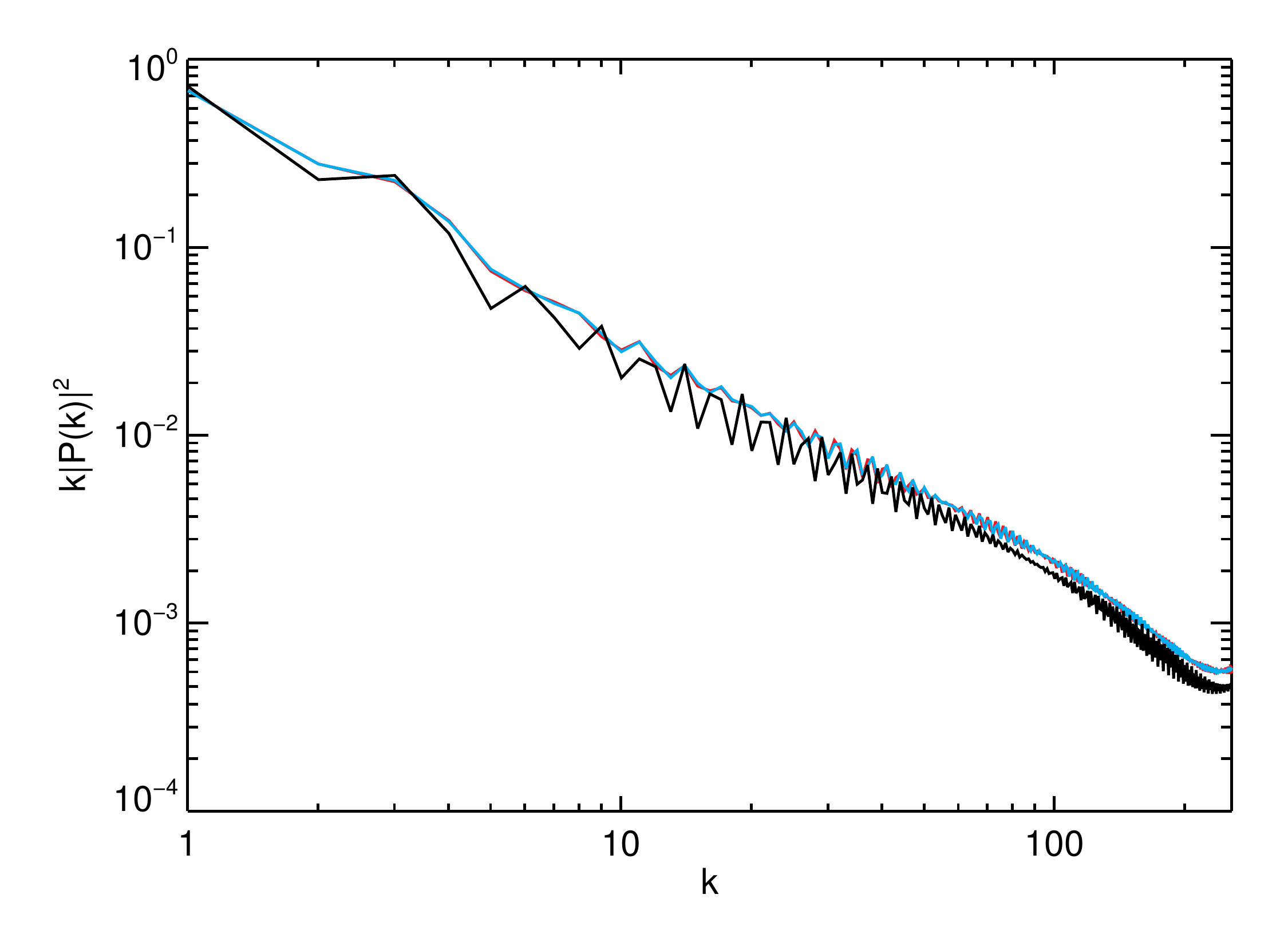}
\includegraphics[width=0.32\textwidth]{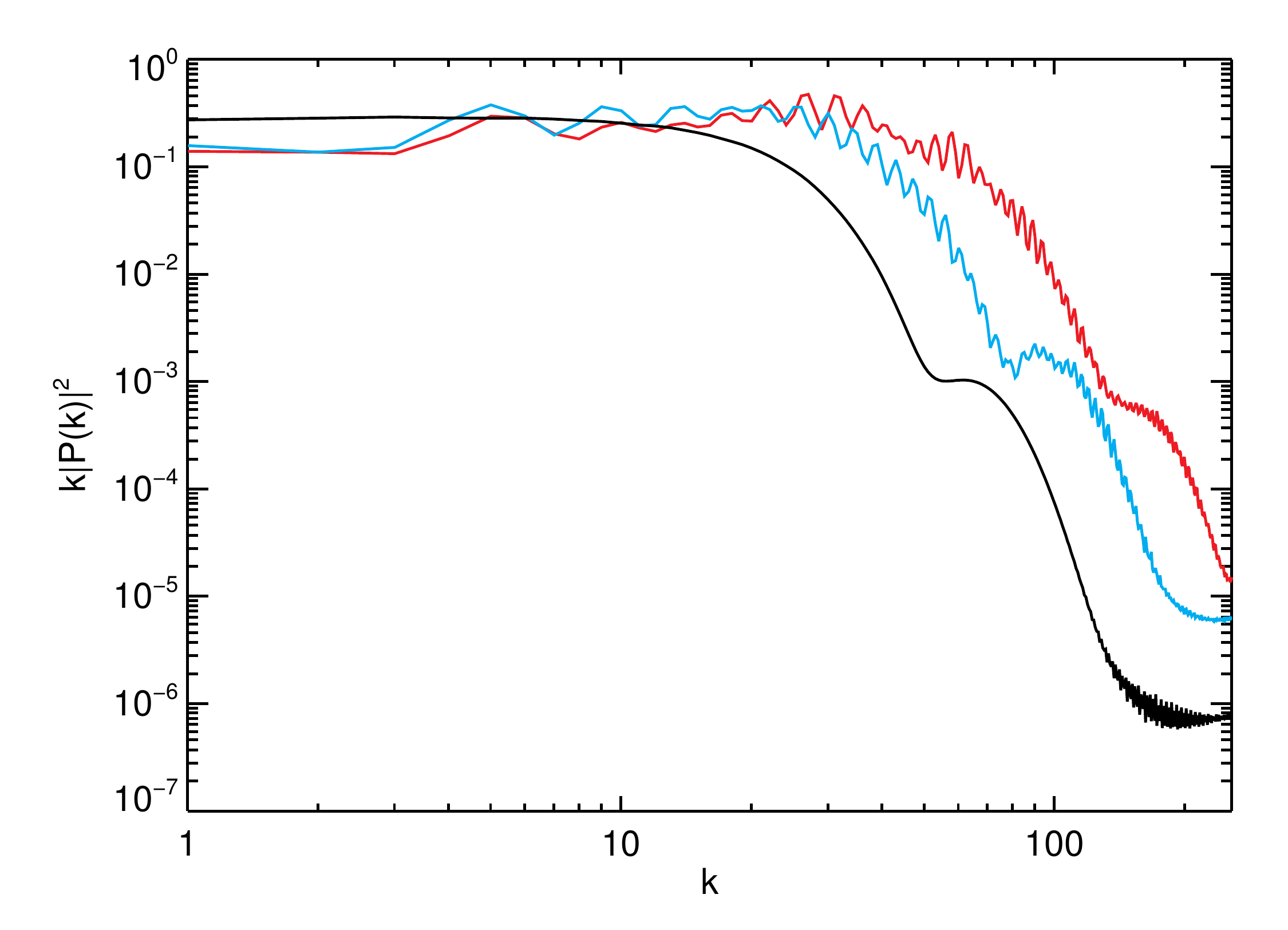}
\end{center}
\caption[]{Power spectra in density (left panel), lorentz factor (center panel) and magnetic pressure (right panel) for high resolution simulations computed with the HLLE (black lines), HLLC (blue lines) and HLLD (red lines) Riemann solvers. Each power spectrum is normalized such that $\int^{k_{s}}_{1} |P(k)|^2 dk = 1$ and plotted as $k |P(k)|^2$}
\label{kh2d_fft} 
\end{figure}

A more
quantitative comparison can be obtained through study of the
integrated power spectrum, $|P(k)|^2$
\begin{equation}
|P(k)|^2 = \int^{y_{\mathrm{max}}}_{y_{\mathrm{min}}} |p(k,y)|^2 dy
\end{equation}
where $p(k,y)$ is the one-dimensional discrete Fourier transform of the 
quantity $q(x,y)$ along the $x$-direction
\begin{equation}
p(k,y) =
\frac{1}{N} \sum^{N-1}_{x=0} {q(x,y) \exp \left( -\frac{ 2\pi i }{N} k x \right)}
\end{equation}
Figure \ref{kh2d_fft} compares the volume-averaged power spectrum
of density, $|\rho(k)|^2$, Lorentz factor, $|\Gamma(k)|^2$ and
magnetic pressure, $|P_m (k)|^2$ for the high resolution simulations
computed with each Riemann solver. Each of these power spectra are
normalized such that $\int^{k_s}_{1} |P(k)|^2 dk = 1$, where $k_s$
is the Nyquist critical frequency.  We consider the power spectrum
of $|\Gamma(k)|^2$ rather than $|V^{y}(k)|^2$ as this diagnostic
enables to make contact with the three-dimensional simulations
described in the next section; we have also examined the power spectrum in
$|V^{y}(k)|^2$ and found that the same qualitative conclusions
apply.  The power spectra of the density and Lorentz factor are
indistinguishable for the simulations computed with the HLLC and
HLLD Riemann solvers; in the former case, the power spectrum consists
of a broken power law, with $k |\rho(k)|^2 \propto k^{-1}$ for $k
\le 30$ and $k |\rho(k)|^2 \propto k^{-8/3}$ for $k > 30$, whilst
in the latter the power spectrum is well described by a single power
for $5<k<100$, $k |\Gamma(k)|^2 \propto k^{-7/6}$, steepening
slightly at larger scales and softening at smaller scales. In this
latter case, the power spectrum for the simulation computed with
the HLLE solver is identical to the HLLC and HLLD cases. For the
density power spectrum, $|\rho(k)|^2$, we find reduced power at
intermediate scales ($10 \le k \le 100$) for the  simulation computed
using the HLLE solver compared to those computed HLLC and HLLD
Riemann solvers, as suggested by the data of Figure \ref{kh2d_den}.
The power spectrum for the magnetic pressure, $k |P_m (k)|^2$ is
more complex; at large scales, we find that $k |P_m (k)|^2 \sim
\mathrm{cons.}$ up to some frequency, $k_{break}$. Above this
frequency, the power spectrum is described by two power laws,
separated by the plateau. Despite this complexity, we find that $k
|P_m (k)|^2$ computed from the HLLE case can be transformed into
that computed in the HLLC and HLLD cases by increasing $k$ (whilst
leaving the normalized $k |P_m (k)|^2$ fixed) by a factor of $1.6$
and $1.6^2$ respectively.

\begin{figure}
\begin{center}
\includegraphics[width=0.45\textwidth]{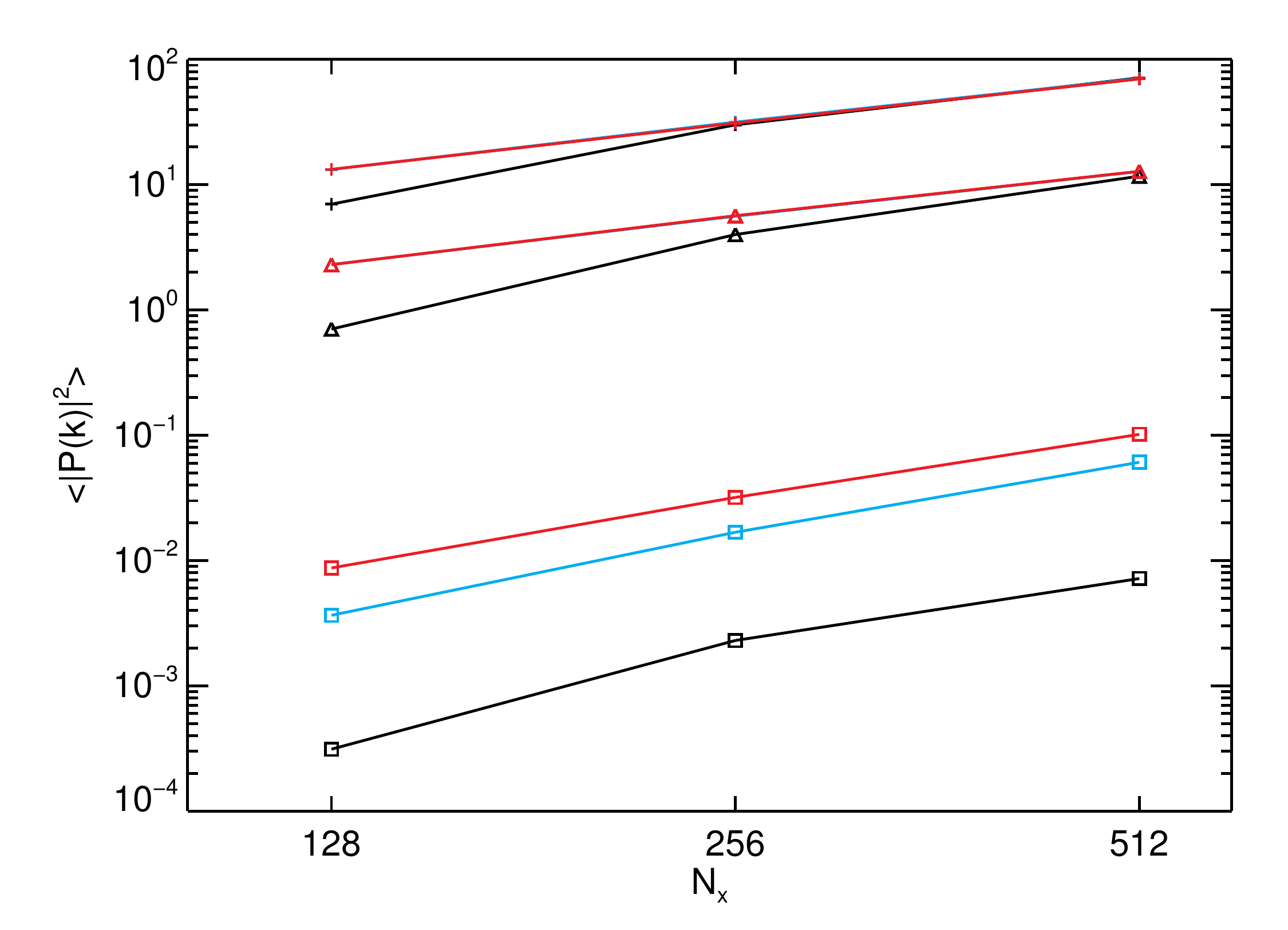}
\end{center}
\caption[]{Total power, $\left< |P(k)|^2 \right> = \int^{k_{max}}_{k_{min}} |P(k)|^2 dk$ for $|\rho(k)|^2$ (crosses); $|\Gamma(k)|^2$ (triangles) and $|P_m(k)|^2$ (squares) for simulations computed with the HLLE (black lines), HLLC (blue lines) and HLLD (red lines) Riemann solvers.}
\label{kh2d_ptot} 
\end{figure}

As a final comparison for these two-dimensional simulations, Figure
\ref{kh2d_ptot} shows the total power, $\left< |P(k)|^2 \right>$
for each of the power spectra shown in Figure \ref{kh2d_fft}, where
\begin{equation}
\left< |P(k)|^2 \right> = \int^{k_{s}}_1 |P(k)|^2 dk
\end{equation}
and $k_s$ is the Nyquist frequency. As expected from the
results presented above, $\left< |\rho(k)|^2 \right>$ and $\left<
|\Gamma(k)|^2 \right>$ are identical between the HLLC and HLLD
Riemann solvers. We further find that the HLLE solver produces the
same  $\left< |\rho(k)|^2 \right>$ and $\left< |\Gamma(k)|^2 \right>$
at the highest resolution, where the linear growth stage of the
instability is converged for this solver. As above, the data for
the magnetic pressure, $\left< |P_m(k)|^2 \right>$ exhibits different
behavior. At the highest resolution resolution, the simulation
computed using the HLLE Riemann solver has $\left< |P_m(k)|^2
\right>$ a factor $14$ less than the simulation computed at the
same resolution using the HLLD solver; at this resolution, $\left<
|P_m(k)|^2 \right>$ for the HLLE solver is still less than this
quantity computed using the HLLD solver at the \emph{lowest}
resolution considered here. The difference in $\left< |P_m(k)|^2
\right>$ between simulations computed using the the HLLC and HLLD
solvers is markedly less pronounced, at the highest resolution
$\left< |P_m(k)|^2 \right>$ computed in the former case is at most
a factor of $1.7$ smaller than in the latter. Based on these results,
we therefore conclude that inclusion of the contact discontinuity
in the Riemann fan has a profound influence on the evolution of the
linear growth stage of the KHI when the density varies across the
shear layer.  Even when the evolution of the
instability appears converged in simulations utilizing the simple
HLLE Riemann solver, the shape of the power spectrum in the density
is fundamentally different to solvers that include this discontinuity
in the Riemann fan, even when the total integrated power is similar.
The lack of the contact discontinuity in the Riemann fan lowers the
effective spectral resolution (by a factor $1.6$) and total
power (by a factor $8.5$) in the magnetic pressure. The presence
of rotational discontinuities (Alfven waves) in the Riemann fan
increases the spectral resolution and total power in the
magnetic pressure by factors $1.6$ and $1.7$, respectively; i.e.
we see a similar increase in spectral resolution due to the presence
of this discontinuity as was the case for the inclusion of the
contact discontinuity, but a markedly smaller increase in overall
power.
\section{Example Applications}\label{apps}

The two-dimensional Kelvin-Helmholtz instability (KHI) test presented
in the previous section suggested that choice of Riemann solver can
play an important role in determining the overall spectral resolution
of a given integration scheme. In particular, we demonstrated that
solutions computed using the HLLE approximate Riemann solver converged
to the wrong solution during the linear growth phase of the KHI,
due to the absence of the contact discontinuity in the Riemann fan.
In this section, we examine the impact of these results on two
popular applications of RMHD codes, dynamo amplification
of magnetic fields within three-dimensional turbulence driven by the KHI
\cite[see e.g.][]{Zhang:2009},
and the propagation of three-dimensional relativistic
jets \cite[see .e.g][]{Mignone:2010}. In this study it is important to remember that as we are performing computations in ideal relativistic MHD, none of the
solutions in the non-linear regime can be regarded as ``converged''.
For convergence, a physical dissipation scale (provided by
either e.g. a Navier-Stokes viscosity or Ohmic resistivity) would
have to be included in the problem. Computations using physics
beyond ideal RMHD are, however, extremely challenging and are well
beyond the scope of the work presented here. In addition, for many
applications, such as turbulence within magnetized accretion disks
close to the black hole event horizon, the physical dissipation
scale is many orders of magnitude smaller than smallest scales that
can currently be probed by state of the art numerical studies, e.g.
\cite{Noble:2010,Penna:2010}. For this reason, it is important to
assess the role played by the Riemann solver in determining the
properties of fully three-dimensional non-linear problems in
RMHD without explicit dissipation.

\subsection{Dynamo Amplification of Magnetic Fields in Three-Dimensional Simulations of the KHI}\label{kh3d}

\begin{figure}
\begin{center}
\includegraphics[width=0.45\textwidth]{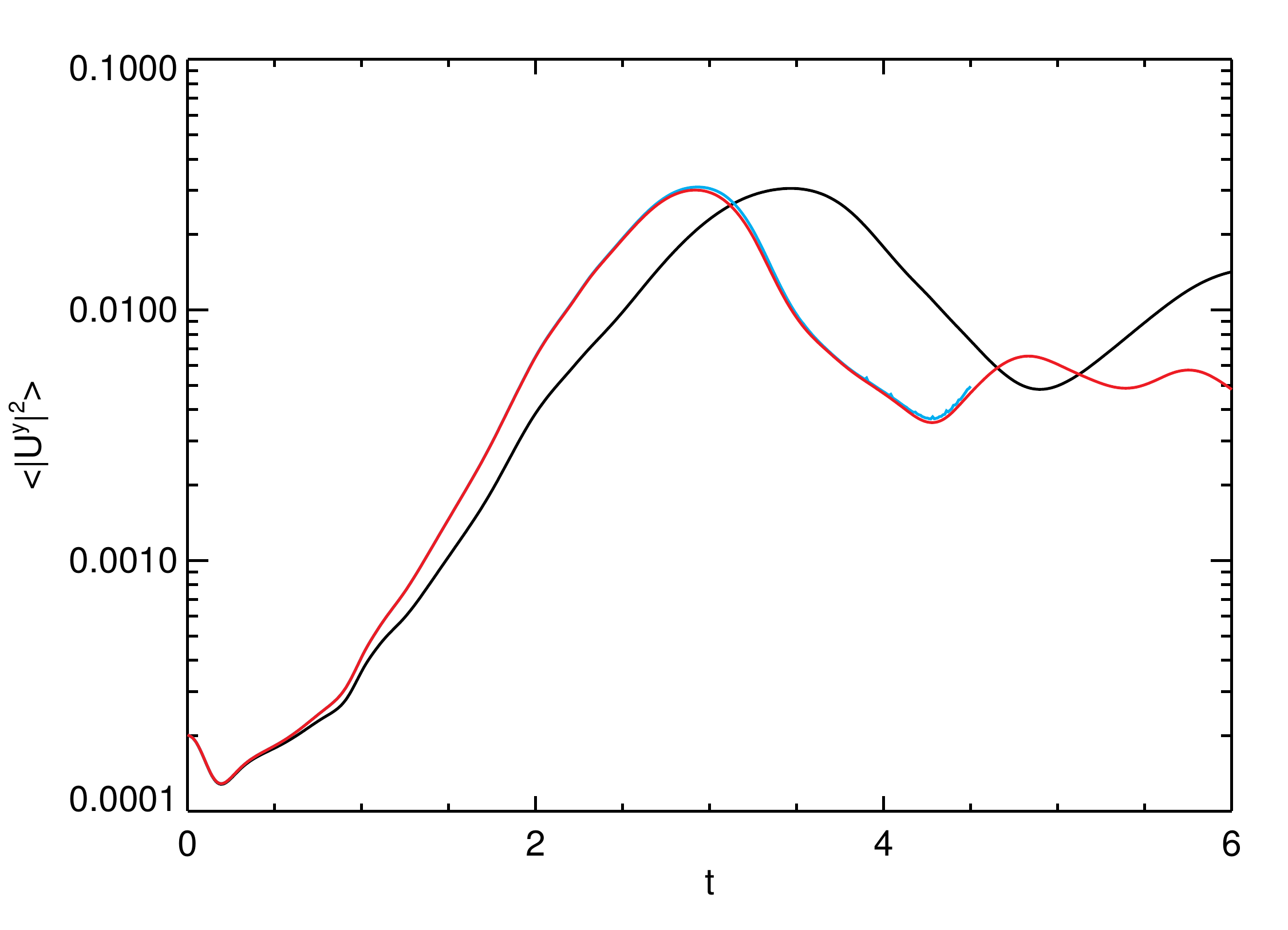}
\end{center}
\caption[]{Volume averaged four-velocity transverse to the shear layer, $\left< |U^y|^2 \right>$ during the linear growth phase of the three-dimensional Kelvin-Helmholtz instability. Black lines show results obtained with the HLLE Riemann solver, blue lines results obtained with the HLLC Riemann solver and red lines results obtained with the HLLD Riemann solver. Note that results for the HLLC and HLLD Riemann solver (blue and red lines) are essentially indistinguishable during the linear growth phase.}
\label{kh3d_ky} 
\end{figure}

We begin by building directly on the results of \S\ref{kh2d} and
study dynamo amplification of magnetic fields in three-dimensional
simulations of the Kelvin-Helmholtz instability (KHI).  Whilst the
two-dimensional simulations presented previously are useful for
probing the linear growth phase, they cannot probe the dynamo
\citep{Moffatt:1978}. To do so, we extend the calculations presented
in \S\ref{kh2d} to three-dimensions. The initial conditions for the
simulations were identical to those used in \S\ref{kh2d}, with the
addition of $1\%$ Gaussian perturbations modulated by an exponential
distribution to the $z$-component of the three-velocity in order
to break symmetry along the $z$-axis.  We present three calculations,
one for each Riemann solver at a resolution corresponding to the
medium resolution case for the two-dimensional simulations, i.e. a
domain covering $-0.5\le x \le 0.5$, $-1.0\le y \le 1.0$, $-0.5\le
z \le 0.5$ using $256\times512\times256$ zones.  As discussed above,
we found that at this resolution, the linear growth stage had
converged for simulations utilizing the HLLC and HLLD Riemann
solvers. Figure \ref{kh3d_ky} shows the evolution of the volume-averaged
four-velocity transverse to the shear layer, $\left< |U^y|^2 \right>$
during the linear growth stage of the instability. The evolution
of this quantity is similar to the two dimensional case, both in
terms of growth rate and maximum amplitude of $\left< |U^y|^2
\right>$. We note however, that the simulation computed using the
HLLC solver failed at $t=4.5$, likely due to the pathologies described
in \S\ref{HLLC-solver} (hence the truncation of the corresponding
line in this plot). We therefore concentrate on results obtained
using the HLLE and HLLD solvers in the remainder of this section.

\begin{figure}
\begin{center}
\includegraphics[width=0.66\textwidth]{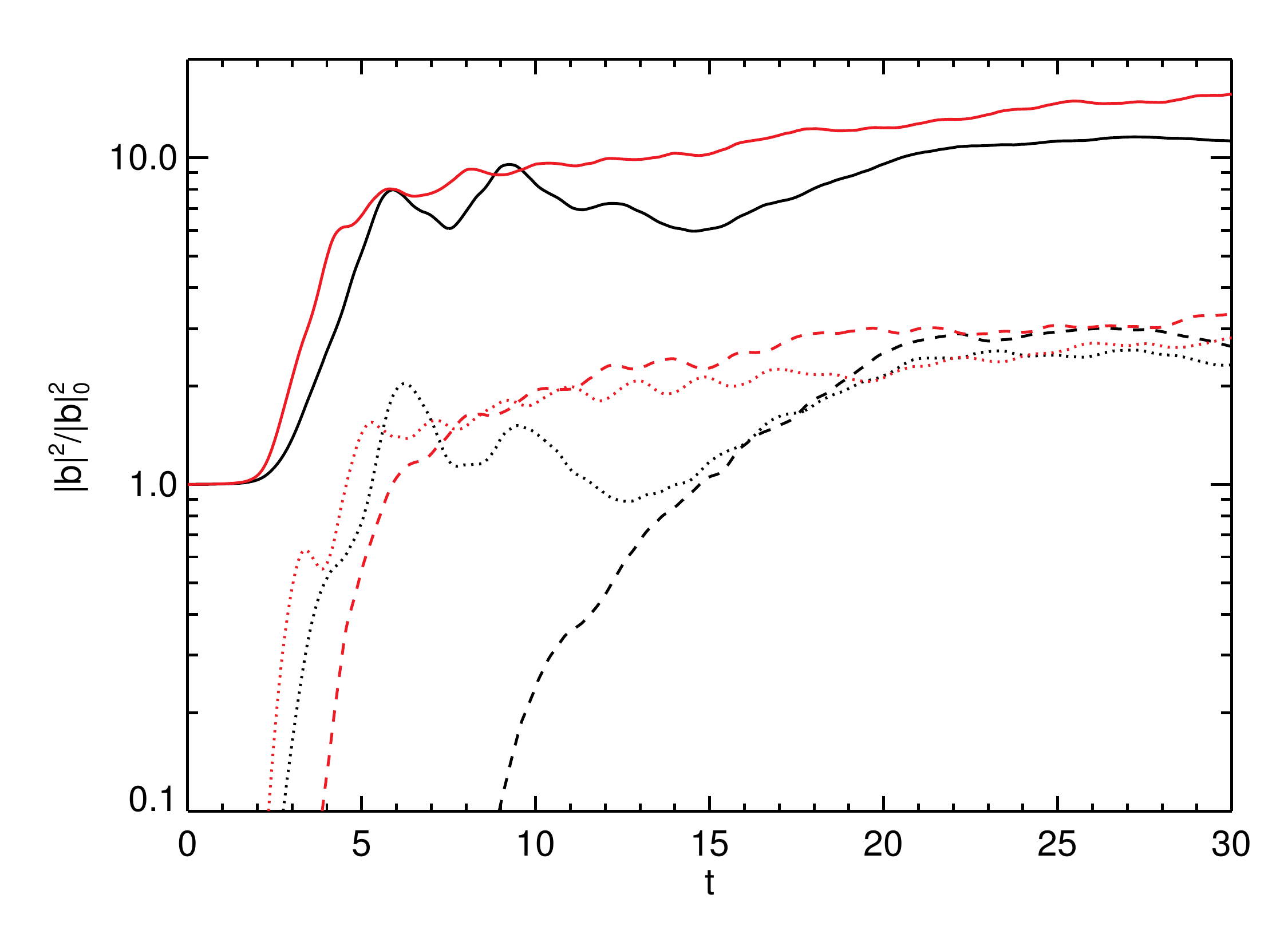}
\end{center}
\caption[]{Time history of volume averaged magnetic field strength for the three-dimensional Kelvin-Helmholtz instability test. Results for the HLLE solver are denoted using black lines, results for the HLLD solver using red lines. Solid lines denote $|b|^2$, dotted lines $|b^y|^2$ and dash lines $|b^z|^2$.}
\label{khhist3d} 
\end{figure}

Figure \ref{khhist3d} shows the time histories of the volume averaged
magnetic field strength
for simulations conducted with the HLLE and HLLD Riemann solvers.
Marked differences are found in the \emph{non-linear} phase of the
evolution ($t > 3.0$), in particular the development of 
magnetic energies in the $z$-direction. Growth of 
this energy occurs roughly a factor of two later for the HLLE simulation compared to the HLLD simulation
at $t=9$ and $t=4$ respectively.
We further find that it takes a greater
amount of time for the turbulence to enter a steady state (characterized
by approximately constant volume averaged
$\left<|b^y|^2\right>$, $\left<|b^z|^2\right>$); in the simulation computed using the HLLE solver, the turbulence enters an approximate steady state at $t=25$, whilst in the HLLD case, this approximate steady state occurs at
$t=15$. We note that in this steady state,
$\left<|b^y|^2\right>$, $\left<|b^z|^2\right>$
are comparable between the two simulations, whilst the simulation
conducted using the HLLD solver has $\left<|b|^2\right>$ that is
a factor $\sim1.8$ greater that in the HLLE case.

\begin{figure}
\begin{center}
\includegraphics[width=0.32\textwidth]{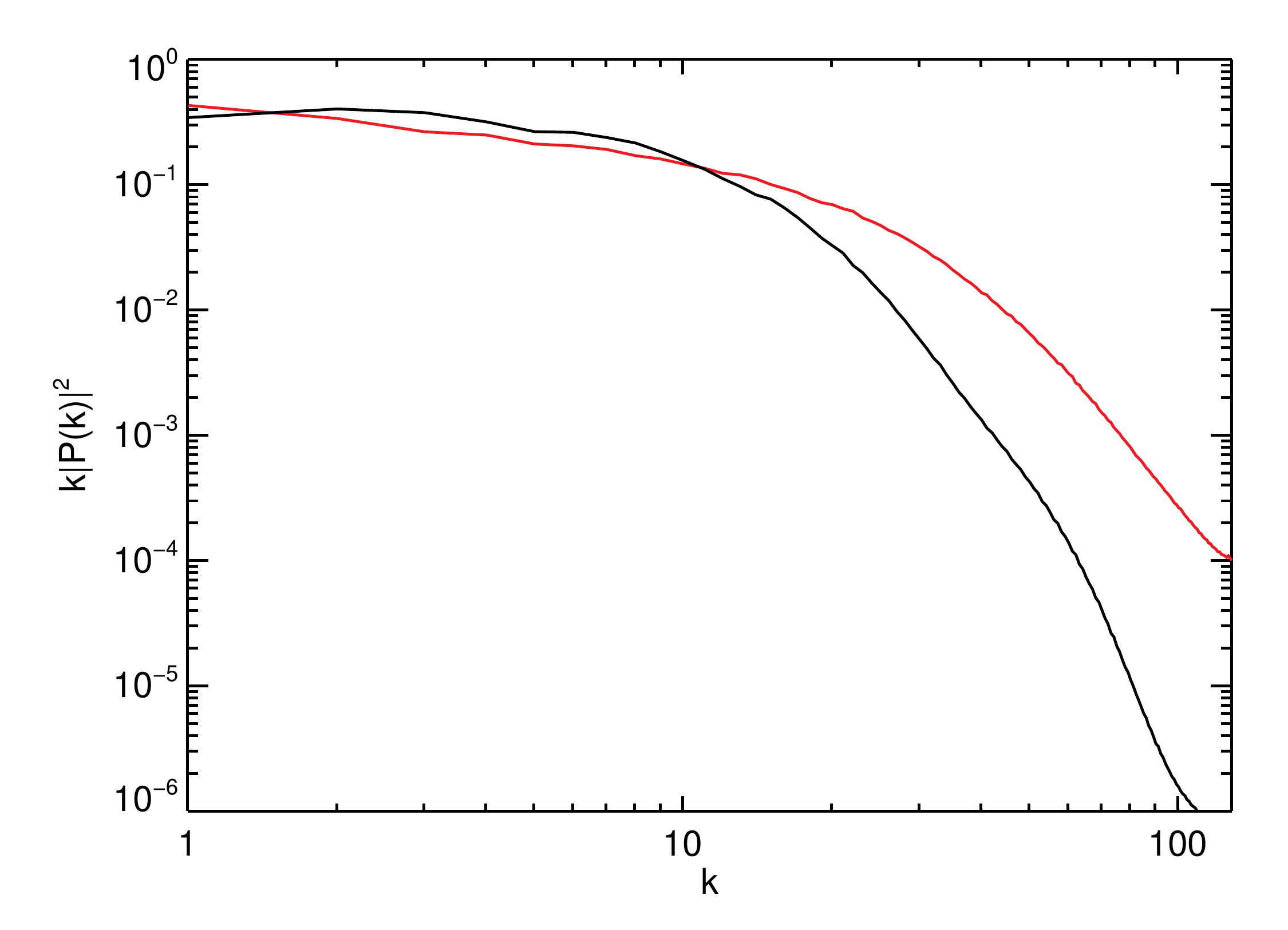}
\includegraphics[width=0.32\textwidth]{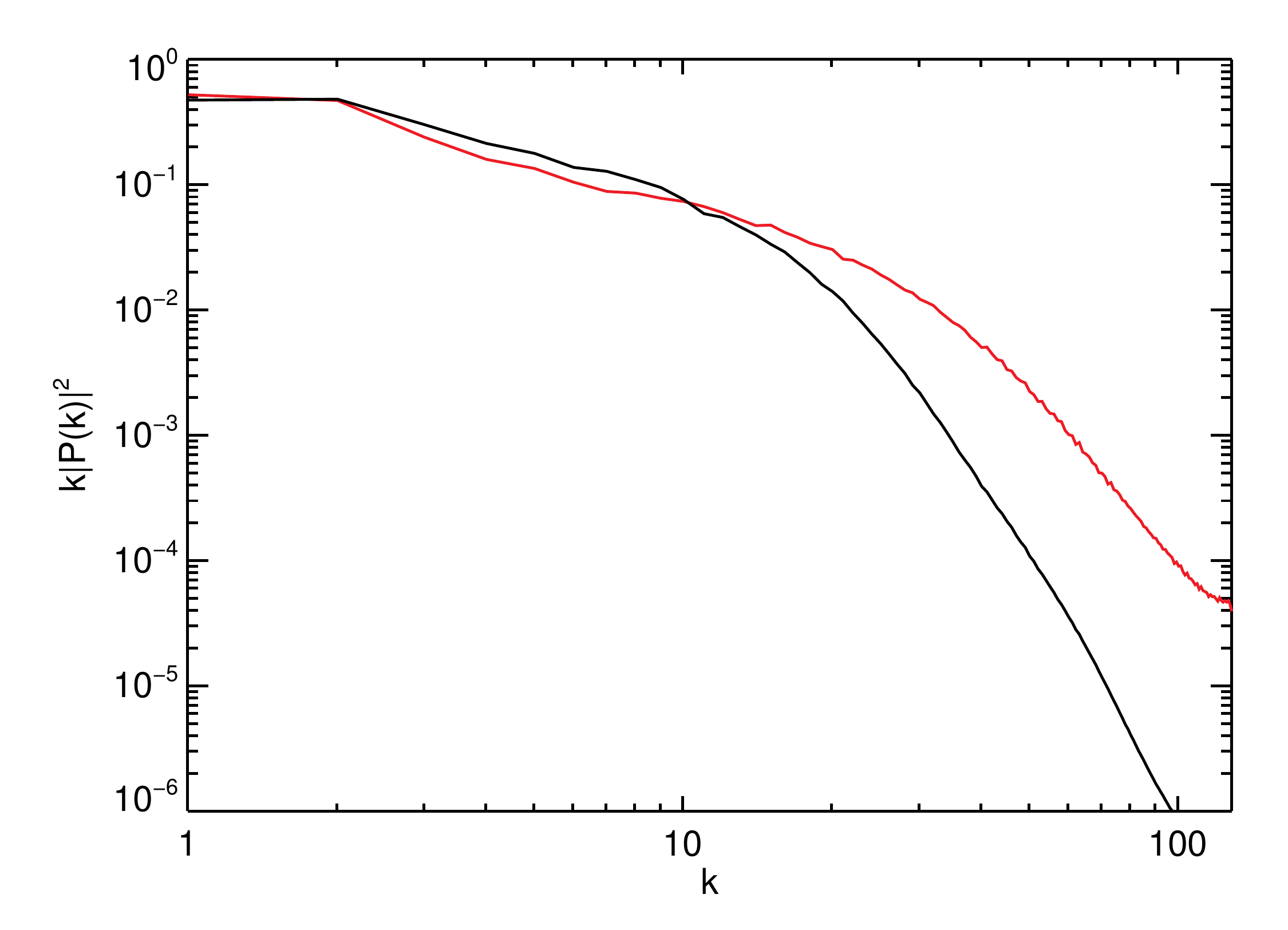}
\includegraphics[width=0.32\textwidth]{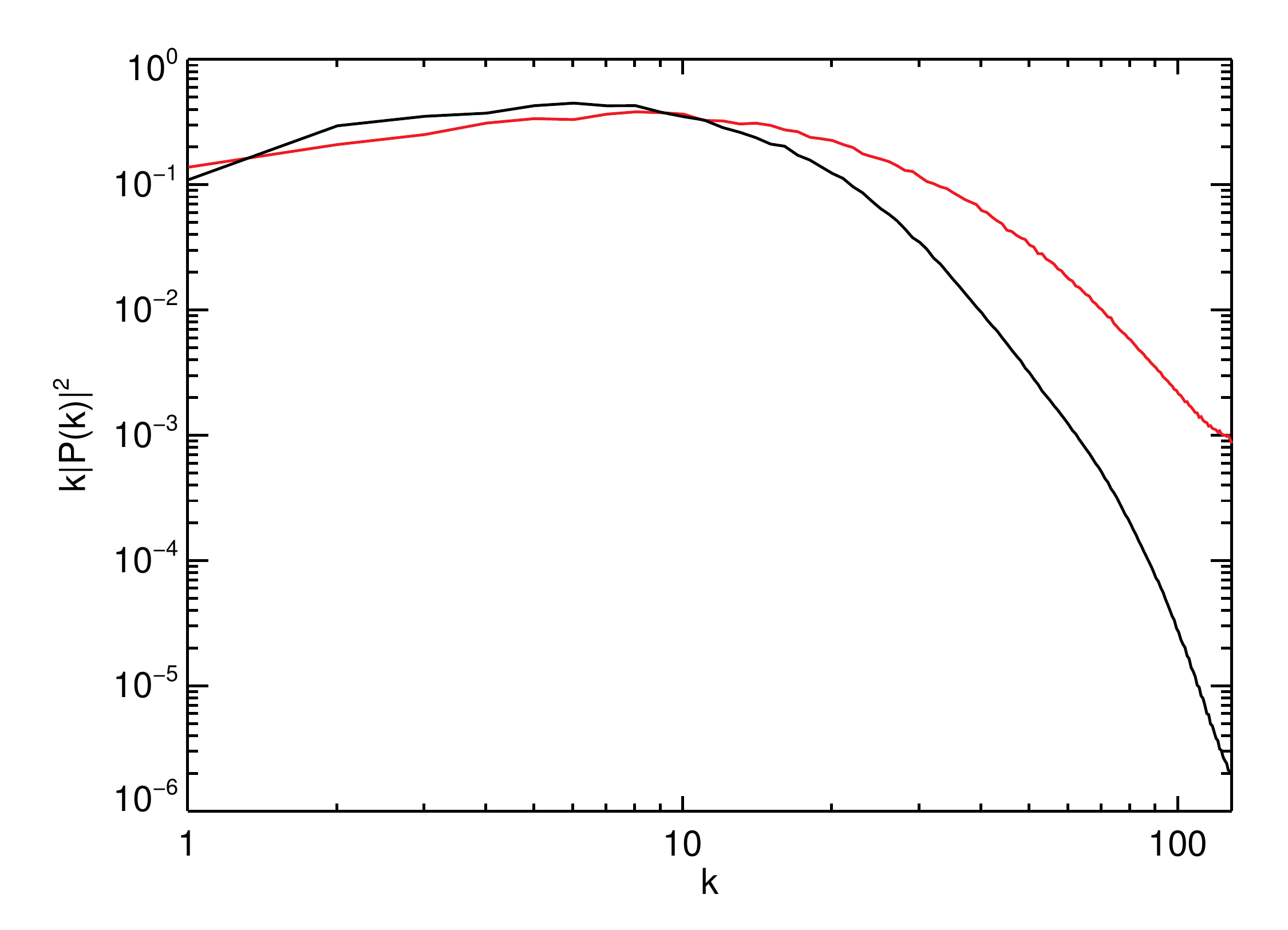}
\end{center}
\caption[]{Power spectra in density (left panel), lorentz factor (center panel) and magnetic pressure (right panel) for high resolution simulations computed with the HLLE (black lines) and HLLD (red lines) Riemann solvers at $t=30$. Each power spectrum is normalized such that $\int^{k_{s}}_{1} |P(k)|^2 dk = 1$ and plotted as $k |P(k)|^2$}
\label{kh3d_fft_3000} 
\end{figure}

Figures \ref{kh3d_rho} and \ref{kh3d_bsq} show three-dimensional
volumetric renderings of the density and magnetic field strength
distributions at $t=10$ and $30$. Quantitative comparison of the structure
of the turbulence in these two simulations is made in
Figure \ref{kh3d_fft_3000} using shell-integrated
power spectra of the density, $k |\rho(k)|^2$, Lorentz factor $k
|\Gamma(k)|^2$ and magnetic field strength $k |b^2(k)|^2$. We compute
$|P(k)|^2$ by first computing the two-dimensional power spectrum
on slices of constant $y$
\begin{equation}
p(k_x,y,k_z) =
\frac{1}{N_x N_z} \sum^{N_x-1}_{x=0} \sum^{N_z-1}_{z=0}
{q(x,y,z) \exp \left[ -\frac{ 2\pi i }{N} (k_x x + k_z z) \right]}
\end{equation}
From this, we compute the integrated two-dimensional power 
spectrum $|p(k_x,k_z)|^2$
\begin{equation}
|p(k_x,k_z)|^2 = \int^{y_{max}}_{y_{min}} |p(k_x,y,k_z)|^2  dy
\end{equation}
The shell-integrated power spectrum is then $|P(k)|^2 = 2 \pi k^2
dk^2 |p(k)|^2$. Here $|p(k)|^2$ denotes the average of $|p(k_x,k_z)|^2$
over shells of constant $k = (k^2_x + k^2_z)^{1/2}$. Finally, we
normalize $|P(k)|^2$ such that $2 \pi \int^{k_s}_1 |p(k)|^2 k dk =
1$. The data presented in these Figures demonstrates marked differences
between the simulations; in particular, data for the HLLE simulation
at $t=10$ reveals that the fluid possesses little in the way of
structure along the $z$-direction, by contrast, the data from the
HLLD simulation shows significant three-dimensional structure around
the shear layers at this time.
By $t=30$, three-dimensional MHD turbulence has
filled the entire volume of both simulations. At this time, each
power spectra shows an excess of power at large scales ($k<10$) in
the simulation computed with the HLLE solver compared to the HLLD
solver at the expense of power at intermediate-to-small scales
($k>10$). Examining the total power, $\left< |P(k)|^2 \right> =
\pi \int^{k_s}_1 |p(k)|^2 k dk$ in each quantity at $t=30$, we find
that $\left< |P(k)|^2 \right>$ for each quantity in the HLLE
simulation is approximately half that computed from the HLLD
simulation. Overall, these data confirm the results from study of
the linear growth phase of the KHI for three-dimensional RMHD
turbulence arising from this instability. The effective spectral
resolution of the HLLD solver is approximately a factor of two
higher than that of the HLLE solver, which affects both the shape
and amplitude of the power spectrum in the turbulent steady state.
We note that as a result, the development of fully three-dimensional
MHD turbulence is delayed by around a factor of two in the HLLE
simulation compared to the HLLD simulation. Limited computational
resources mean that we have not been able to compute three-dimensional
simulations at the highest resolution computed in the two-dimensional
case. However, the similarities in the linear growth phase
in the three-dimensional case compared to the two-dimensional case
gives us confidence that a factor two increase in resolution will
not significantly alter our conclusions.
Finally, we remind the reader that the non-linear
phase of these simulations cannot be regarded as converged due to
the absence of physical dissipation in these simulations; for this
reason, we do not regard the three-dimensional simulations presented
here as a quantitative test of the code (for such a test, we refer
the reader to the linear growth phase of the KHI presented in
\S\ref{kh2d}); rather, these simulations serve as a qualitative
demonstration of the pitfalls of using overly simple Riemann solvers
in the study of non-linear flows.

\subsection{Propagation of Relativistic Jets using SMR}\label{srjet}

As a further example application of the code, we present the
propagation of a three-dimensional relativistic jet using SMR.
Understanding the structure and evolution of these
systems is a compelling area of astrophysical research from both a
theoretical and observational standpoint.  From purely dynamical
arguments, it is expected that unbound outflows from astrophysical
systems are both launched and collimated magnetically \cite[see
e.g.][]{Blandford:1977,Blandford:1982}. The cylindrical MHD equilibria
that describe such flows are expected to be unstable to a variety
of reflection, Kelvin-Helmholtz, current-driven and kink modes
\cite[see e.g.][]{Begelman:1998}. Observationally (i.e. at large
radius), astrophysical jets are found to be \emph{stable} objects
dominated by kinetic, rather than magnetic energy \cite[see
e.g.][]{Sikora:2005}. One possible resolution for this apparent
contradiction is that whilst astrophysical jets could be
electromagnetically dominated at their origin, the aforementioned
instabilities could act to dissipate electromagnetic energy so that
the jet is kinetic energy dominated at large radius \cite[again
see][]{Sikora:2005}. Understanding the circumstances in which these
such processes can operate is therefore an important area of
astrophysical research. Due to the fundamental multidimensional
nature of the problem, numerical investigation is an important tool
in this study. The data of the preceding section suggest that choices
regarding the complexity of the Riemann solver can play a non-linear
role in determining the effective resolution of the code for
multi-dimensional problems, particularly for the magnetic field.
In this section, we therefore present a series of three-dimensional
simulations of relativistic jets designed to investigate this issue.
Since we are primarily interested in the dissipation of magnetic
field, we focus our attention on the role played by the rotational
discontinuity in this process by comparing results from simulations
performed with the HLLC (which includes only the contact discontinuity
within the Riemann fan) and HLLD (which includes both the contact
discontinuity and rotational discontinuities within the Riemann fan)
Riemann solvers.

For this study, we utilize a variant of the configuration described
by \cite{Mignone:2006}. In this setup, the simulation domain is
filled with an ambient medium of constant gas pressure, density and
magnetic field with
\begin{equation}
\rho_a = 1 ; \;\;
\left(P_g\right)_a = \frac{\eta |V_b|^2}
{\gamma (\gamma -1) M^2_s - \gamma |V_b|^2}; \;\;
{\cal B}^x = \sqrt{2 P_g / \beta_b}
\end{equation}
where $M_s = |V_b| / c^2_s = 4$, $\eta=10^{-2} = \rho_b / \rho_a$
is the density ratio between the jet beam and the ambient medium
and $\gamma = 5/3$. The jet is injected with $\Gamma = 7$ (corresponding
to $|V_b|=0.99$) through a circular nozzle on the $(y,z)$-plane of
radius $r_{jet} = 1$, centered on $x=y=z=0$.  The jet has the same
gas and magnetic pressure as the ambient medium, whilst the density
is a factor of $\eta$ lower. Inside the nozzle, boundary values are
kept fixed, whilst outside the nozzle, we apply a standard conducting
boundary condition. We adopt outflow boundary conditions on the
remainder of the boundaries.  Note that we do not perturb the jet at the
nozzle, which means that the structures produced in the simulations
are driven by physical instabilties seeded by grid noise.  More quantitative
tests would require both explicit dissipation to produce converged solutions,
as well as physical perturbations to seed instabilities.

For each Riemann solver, this problem was run with
three different resolutions.
The problem domain covers $0.0\le x \le 51.2$, $-25.6\le
y \le 25.6$, $-25.6\le z \le 25.6$ using $256^3$ (low resolution,
$5$ zones per jet radius) and $512^3$ (medium resolution, $10$ zones
per jet radius) zones. The highest resolution simulation was run with SMR
(see \S\ref{smr}) using
three levels of refinement.
The coarsest level covered a domain $0.0\le x \le
51.2$, $-25.6\le y \le 25.6$, $-25.6\le z \le 25.6$ using $(256)^3$
zones, corresponding to a resolution of $5$ zones per jet radius.
The intermediate level covered a domain $0.0\le x \le
51.2$, $-12.8\le y \le 12.8$, $-12.8\le z \le 12.8$ using $512$
zones in the $x$-direction and $256$ zones in the $y,z$-directions,
corresponding to a resolution of $10$ zones per jet radius. The
finest level covered a domain $0.0\le x \le 51.2$,
$-6.4\le y \le 6.4$, $-12.8\le z \le 12.8$ using $1024$ zones in
the $x$-direction and $256$ zones in the $y,z$-directions, corresponding
to a resolution of $20$ zones per jet radius. Note that to achieve
the finest resolution across the entire domain would require $1024^3$
zones, a factor of $10$ increase in computational cost. The simulations
utilizing the HLLC Riemann solver were run with $C=0.4$,
whilst the simulations utilizing the HLLD Riemann solver
were run with $C=0.2$. In this latter case,
use of a Courant number $> 0.2$ resulted in the formation of strongly
magnetized current sheets close to the jet axis, which eventually
destroyed the evolution in a similar fashion to that described in
\S\ref{current_sheet}.

\begin{figure}
\begin{center}
\includegraphics[width=0.32\textwidth]{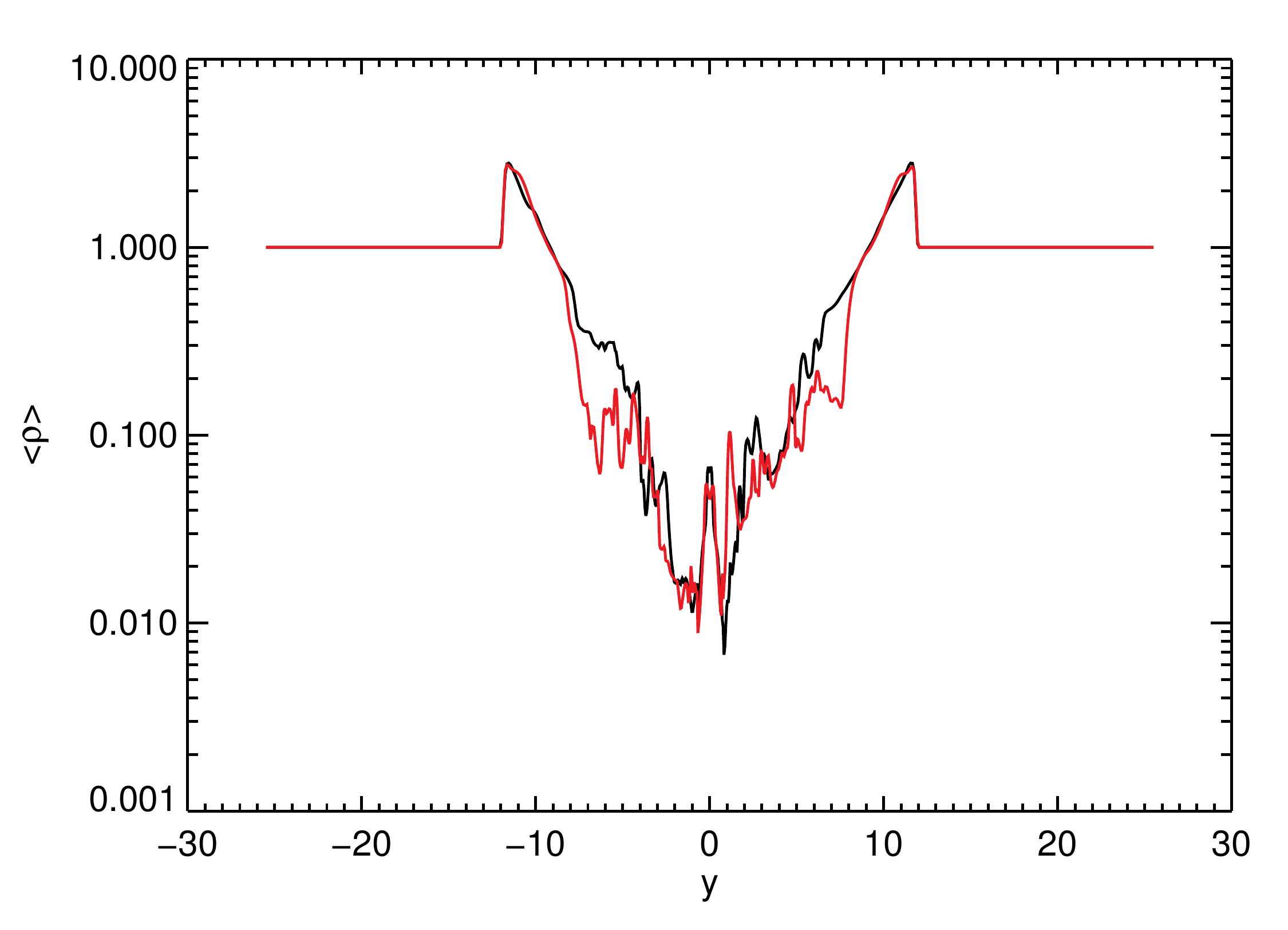}
\includegraphics[width=0.32\textwidth]{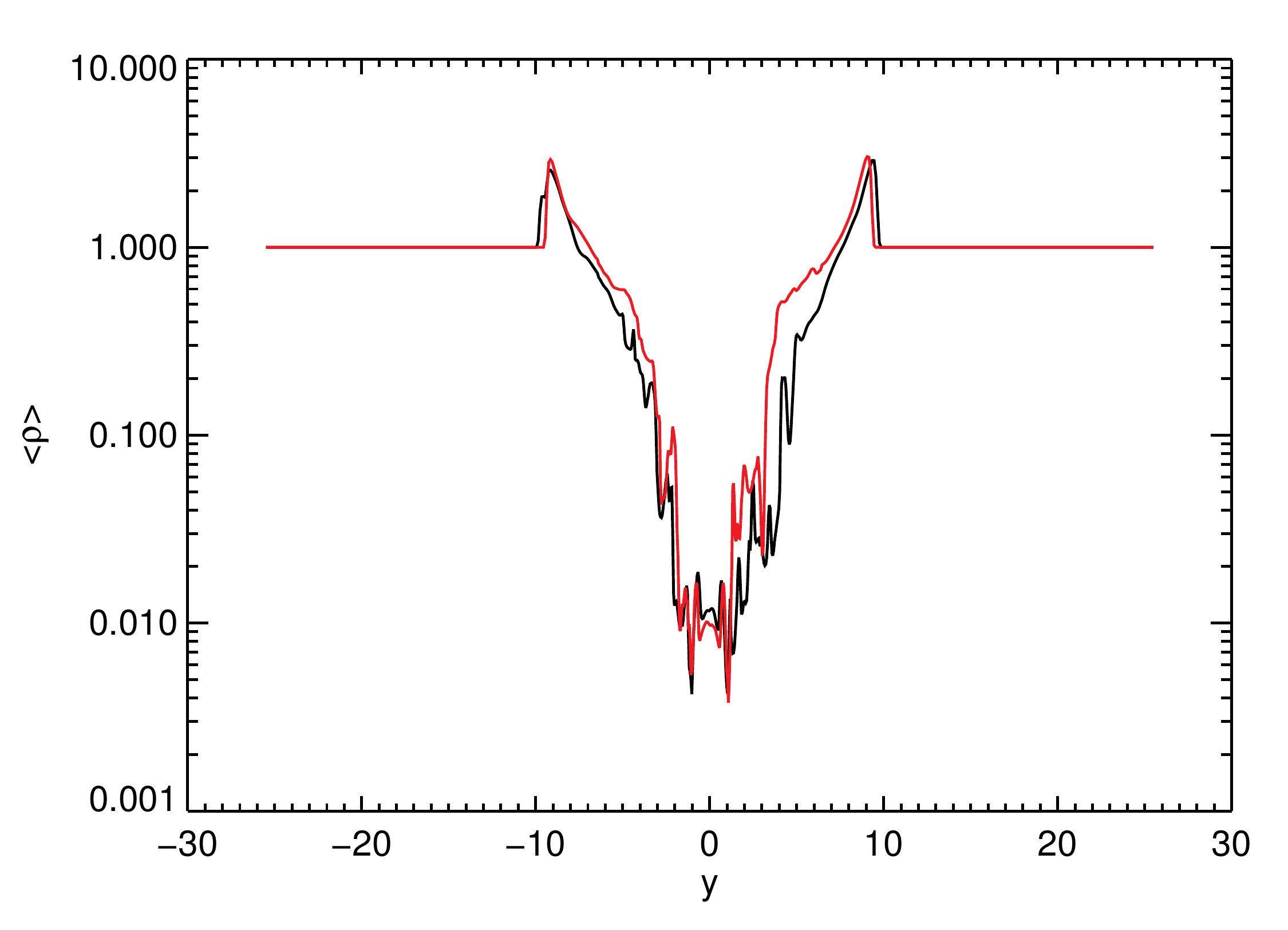}
\includegraphics[width=0.32\textwidth]{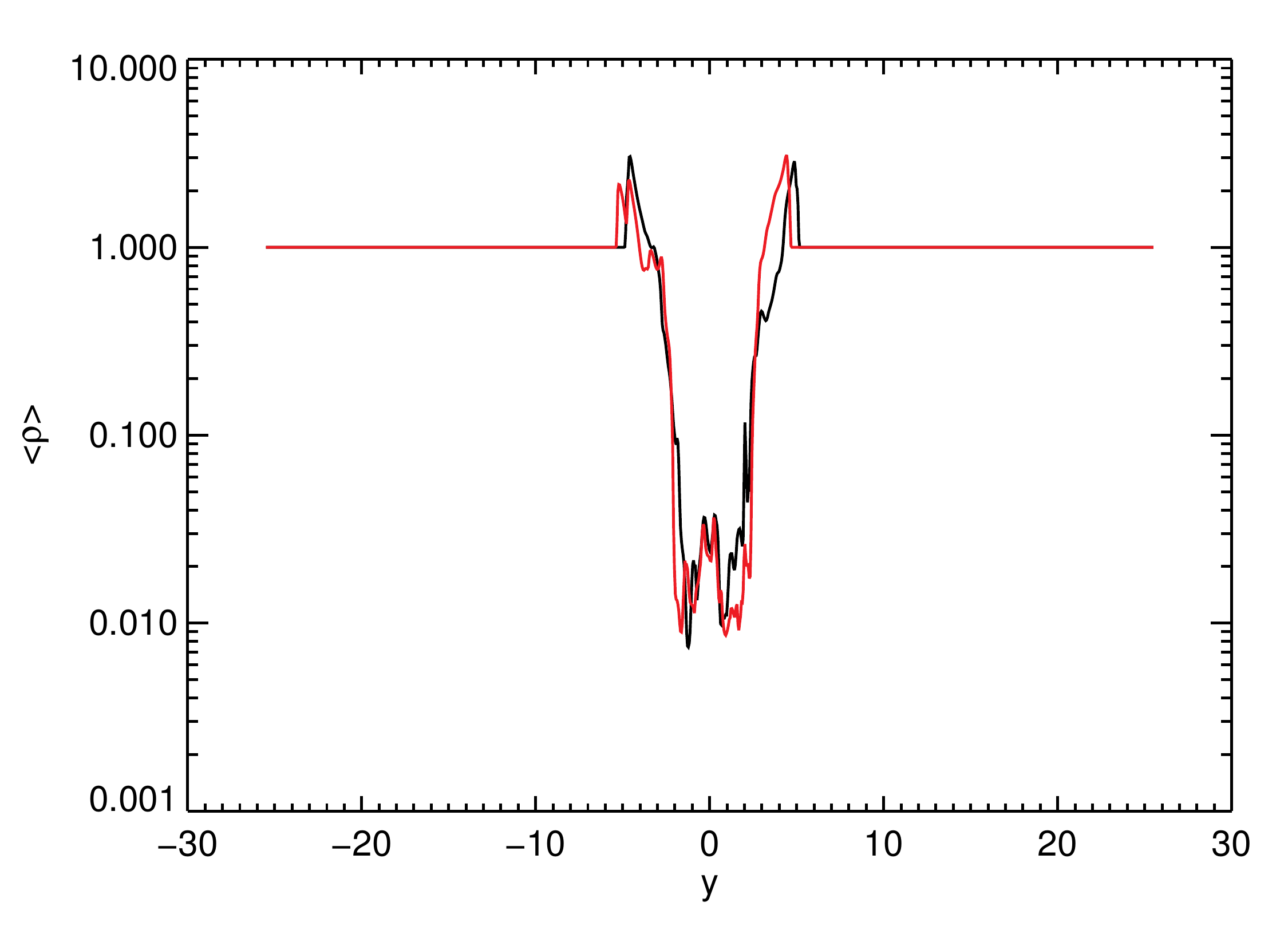}
\end{center}
\caption[]{One-dimensional profiles of quantities transverse to the jet axis, $\rho_{avg} (x) = 0.5[\rho(x,y=0,z) + \rho(x,y,z=0)]$ at $x=12.8$ (left panel), $x=25.6$ (center panel) and $x=38.4$ (right panel) for the HLLC (black lines) and HLLD (red lines) solvers calculated at $t=100$.}
\label{SRjet_rho_1d} 
\end{figure}

\begin{figure}
\begin{center}
\includegraphics[width=0.32\textwidth]{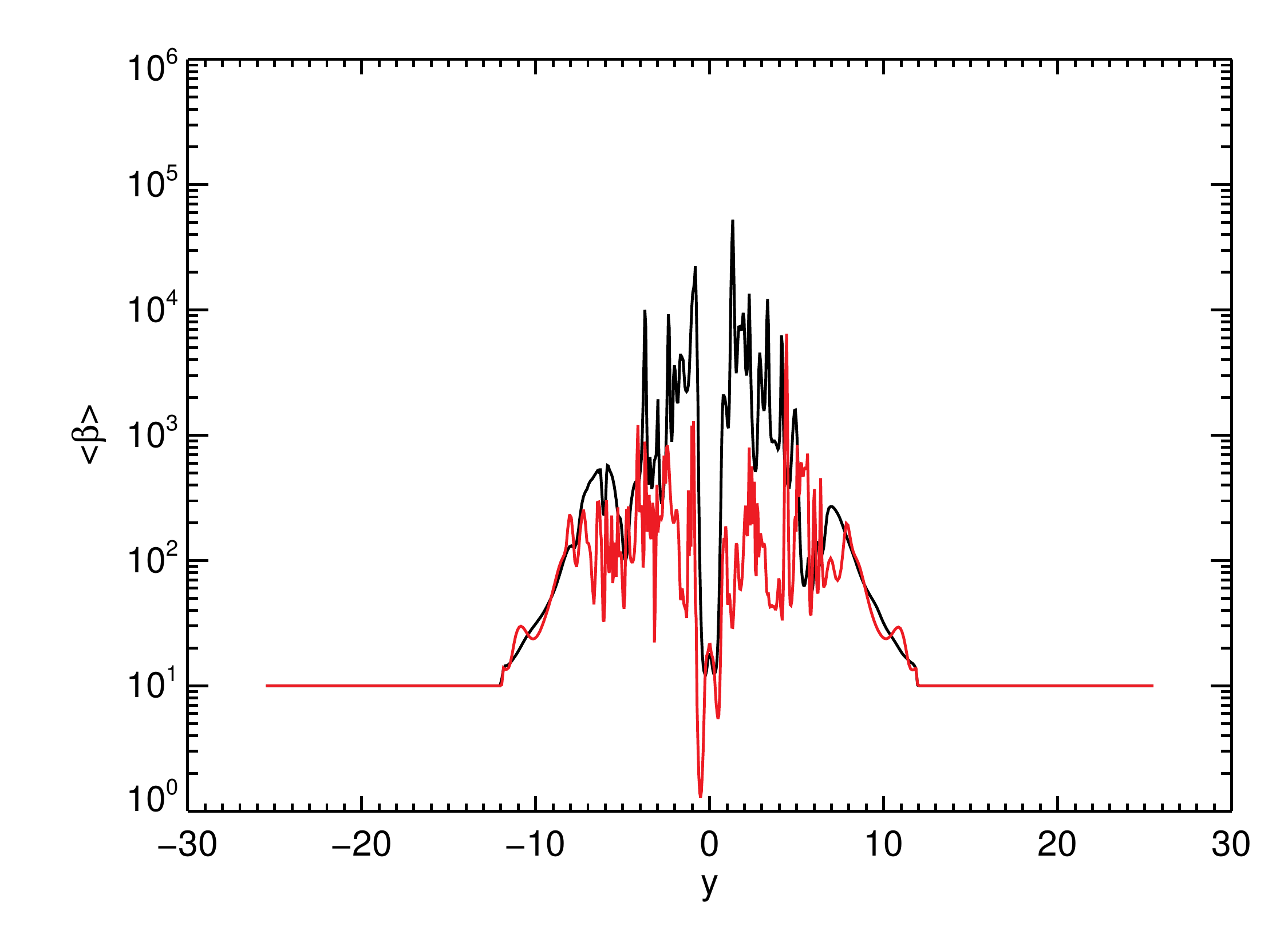}
\includegraphics[width=0.32\textwidth]{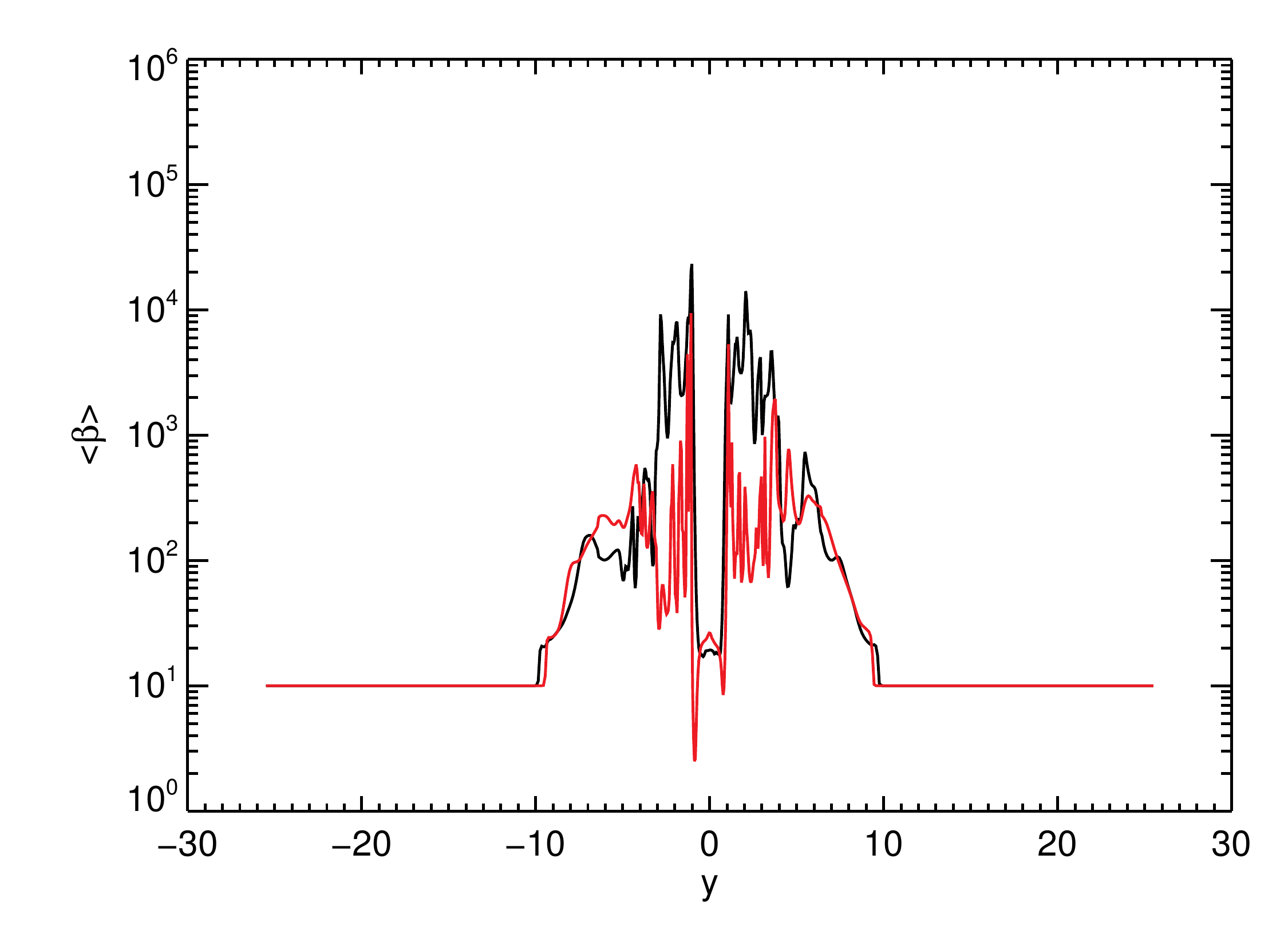}
\includegraphics[width=0.32\textwidth]{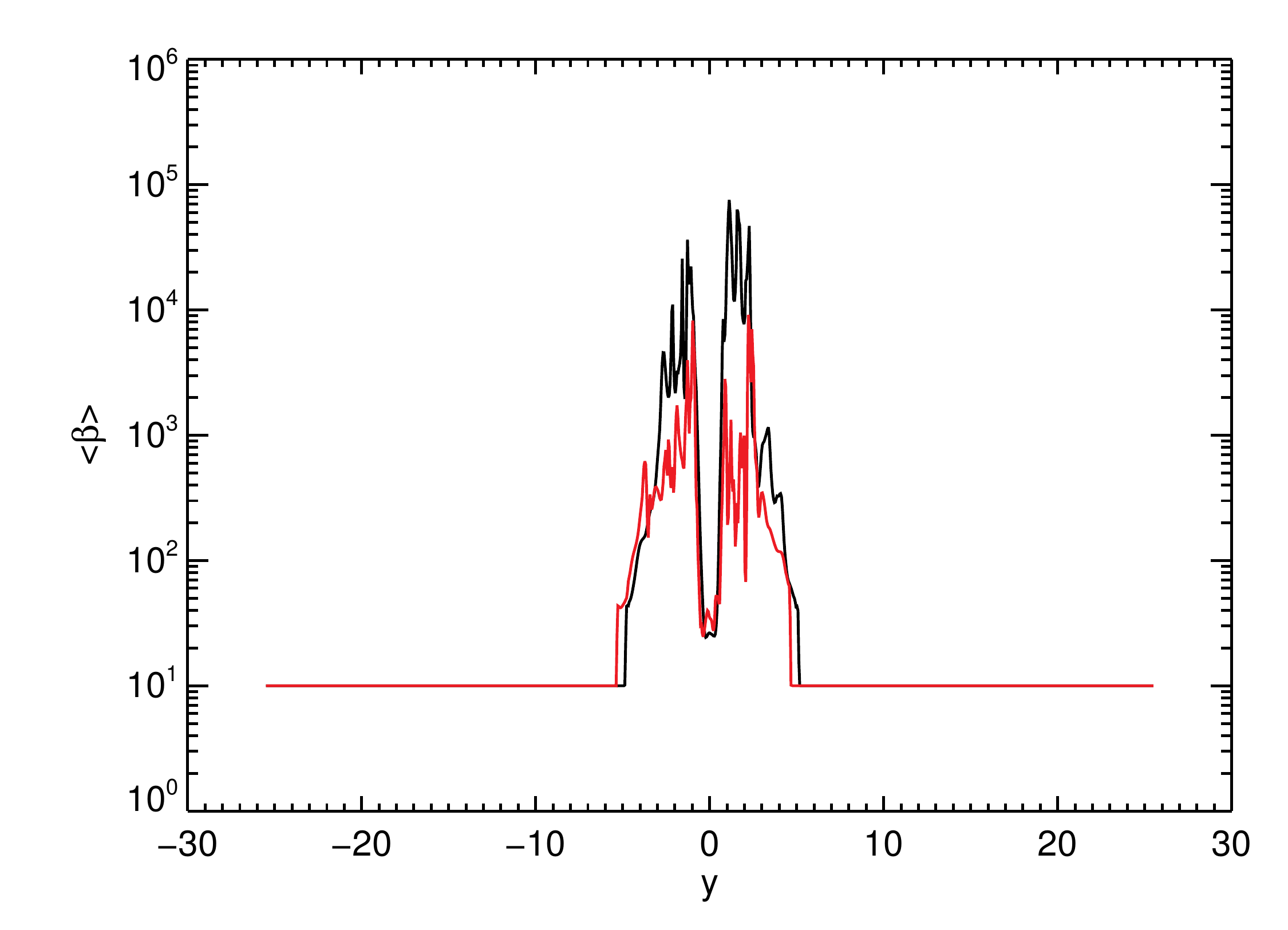}
\end{center}
\caption[]{As in Figure \ref{SRjet_rho_1d} for $\beta_{avg}(x) = 2 (P_g)_{avg}(x) / |b|^2_{avg}(x)$.}
\label{SRjet_beta_1d} 
\end{figure}

Figures \ref{SRjet_rho_3d} and \ref{SRjet_bsq_3d} compare the
distributions of density and magnetic field strength for the two
high resolution simulations via volumetric renderings. Figures
\ref{SRjet_rho_1d}--\ref{SRjet_beta_1d} compare these distributions
via one-dimensional profiles transverse to the jet axis. This latter
diagnostic is computed via
\begin{equation}
q_{avg}(x) = \frac{Q(x,y=0,z) + Q(x,y,z=0)}{2}
\end{equation}
where $Q$ is one of $\rho,|b|^2,P_g$ and $\beta_{avg}(x) =
2(P_g)_{avg}(x) / |b|^2_{avg}(x)$.  As can be seen from the
figures, the overall structure of the jet is similar between
these two simulations. Since many of the details of the jet structure
have been described at length by previous authors \cite[see
e.g.][]{Komissarov:1999b,Leismann:2005,Mignone:2006}, we instead
focus our attention on the differences between the two simulations.
Examination of Figures \ref{SRjet_rho_3d} and
\ref{SRjet_rho_1d} reveals that the nose of the jet is narrower and
has propagated slightly further in the HLLC simulation compared to
the HLLD simulation, whilst in the latter case, the outgoing bow
shock is offset slightly from the jet axis at $x=38.4$.  The greatest
contrast between the high resolution simulations is found in the
structure of the magnetic field in the jet cocoon. Inspection of
Figures \ref{SRjet_bsq_3d} and
\ref{SRjet_beta_1d}  suggest that in the HLLD simulation, the jet
cocoon is filled with turbulent magnetic fields approximately an
order of magnitude greater in strength than found in the HLLC
simulation. In the former case, we also found that the jet core is
surrounded by a (strongly) magnetized sheath with $\beta \sim 1$,
which is almost entirely absent in the simulation computed with the
HLLC solver. That stronger magnetic fields are observed in the HLLD
based simulation is not a surprise; the results of \S\ref{kh2d}
emphasize the importance of inclusion of rotational discontinuities
within the Riemann fan for studies of MHD turbulence. In these
simulations, turbulent amplification
of magnetic fields occurs within the shear layer between the jet
cocoon and the ambient medium \citep{Mignone:2009} and so we expect
the results of \S\ref{kh3d} to apply here as well. What is surprising
is that the inclusion of the rotational discontinuities within the
Riemann fan can make such a \emph{substantial} difference within
the structure of the jet; in results presented thus far, we have
observed factor $\sim2$ enhancements in overall resolution between
simulations that include rotational discontinuities within the
Riemann fan compared to those without, whereas in the simulation
presented here, the differences are closer to an order of magnitude.

\begin{figure}
\begin{center}
\includegraphics[width=0.45\textwidth]{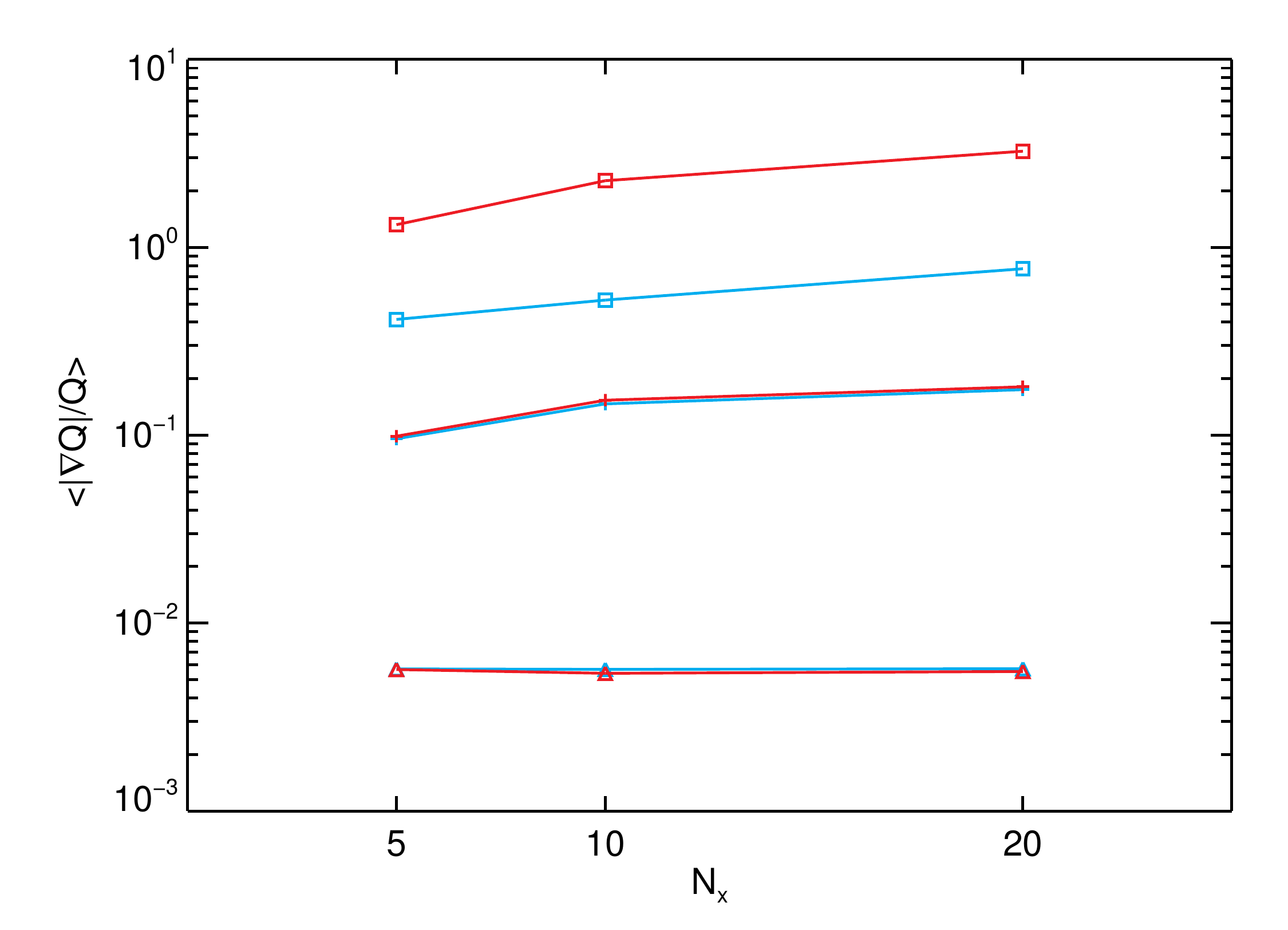}
\end{center}
\caption[]{Volume averaged normalized gradient, $\left<|\nabla Q|/Q\right>$ for density, $\rho$ (crosses), Lorentz factor $\Gamma$ (triangles) and magnetic field strength, $|b|^2$ (squares) for simulations computed with the HLLC  (blue lines) and HLLD (red lines) Riemann solvers for simulations with $5,10,20$ zones per jet radius.}
\label{SRjet_grad} 
\end{figure}

An alternative measure of the effective resolution of these jet
simulations is through the normalized gradient of a quantity,
$|\nabla Q|/Q$. \cite{Mignone:2009} compared the evolution of the
gradients of the poloidal magnetic field $| \nabla B^2_p | / B^2_p$
for axisymmetric simulations computed at a range of resolutions for
the HLLE and HLLD solvers, finding a factor of two increase in the
effective resolution by this measure. Figure \ref{SRjet_grad} shows
the volume-average of the normalized gradient of $Q$, $\left<|\nabla
Q|/Q \right>$ for $Q=\rho, \Gamma, |b|^2$ at $t=100$ for each of
the simulations described above. According to the measures
$\left<|\nabla \rho|/\rho \right>$ and $\left<|\nabla \Gamma|/\Gamma
\right>$, simulations computed using the HLLC and HLLD solvers have
the same effective resolution. The measure $\left<|\nabla |b|^2|/|b|^2
\right>$ indicates that the effective resolution of the HLLD solver
is a factor of $3-4$ greater than that of the HLLC solver; we also
find that $\left<|\nabla |b|^2|/|b|^2 \right>$ for the HLLC solver
at the highest resolution is still smaller than $\left<|\nabla
|b|^2|/|b|^2 \right>$ for the HLLD solver at the lowest resolution.
Examination of the profile of  $\left<|\partial_y |b|^2 + \partial_z
|b|^2|/|b|^2 \right>$ along the jet axis (where the averaging is
now performed on surfaces of constant $x$) suggests that $\left<|\nabla
|b|^2|/|b|^2 \right>$ is enhanced in the HLLD simulations throughout
the jet, rather than being concentrated in one region; i.e. in the
HLLD simulations, the entirety of the jet cocoon is filled with
magnetic fields that are both stronger (by up to an order of
magnitude) and possess steeper gradients (by up to a factor of four)
than simulations conducted with the HLLC solver at the same resolution,
further demonstrating the importance of the rotational discontinuity.

Overall, these simulations demonstrate that, for the second order
integration scheme presented here, use of more complex Riemann
solvers for three-dimensional evolutions of relativistic jets is
essential in order to correctly capture the dynamics of the magnetic
field.  However, we once again caution that without explicit
dissipation and physical perturbations, the solutions shown here
are not converged.  Therefore, as in the previous section, these
simulations serve as a qualitative demonstration of the importance
of utilizing advanced Riemann solvers in studying multi-dimensional
non-linear flows.

\section{Summary}\label{conclusion}

We have described a new, second order accurate Godunov scheme for RMHD. This scheme is distinguished from previous work in two important respects. Firstly, we utilize the staggered, face-centered field version of
the constrained transport (CT) algorithm
with the method of \cite{Gardiner:2005,Gardiner:2008}
to compute the electric fields at cell edges, which keeps the cell-centered, volume averaged discretization of the divergence to
be kept zero to machine precision. Secondly, we make use of a dimensionally unsplit integrator \citep{Stone:2009} which preserves the conservative form of the RMHD equations without the need for characteristic decomposition of the equations of motion in the primitive variables. Because of the tight coupling between the conserved variables in RMHD, maintaining both the divergence condition and the conservative form of the equations offers clear advantages over either divergence cleaning methods \citep{Anninos:2005,Mignone:2010} or dimensionally split methods \citep{Powell:1999}.

We documented four additional parts of the algorithm that we have found important; a scheme for computing the primitive variables from conserved quantities, which we base on the $1D_W$ scheme described by \cite{Noble:2006}, as modified by \cite{Mignone:2007}; the method for calculating the wavespeeds in RMHD, which amounts to finding the roots of a quartic polynomial; a variety of approximate Riemann solvers used to compute fluxes in RMHD; a hierarcy of correction steps designed to correct errors corresponding to unphysical primitive variables. We have made the resulting numerical scheme publicly available as part of the Athena code \cite{Stone:2008}

We presented a variety of multi-dimensional numerical tests
which build on those previously available in the literature for
both relativistic and Newtonian MHD. The solutions to these tests
are designed to highlight important properties of the numerical
method, such as it's ability to hold symmetry, or to test that the solenoidal
constraint is preserved on the correct numerical stencil. Of these tests,
we have found that the large amplitude circularly polarized Alfven wave
test of \cite{Del-Zanna:2007} and the field loop advection test to be
particularly revealing. The former of these revealed failures in the HLLC
Riemann solver due to \cite{Mignone:2006} due to a flux-singularity that
exists for multi-dimensional MHD problems; the latter test probes both
the codes ability to maintain the solenoidal constraint on the correct
numerical stencil \citep{Gardiner:2005,Gardiner:2008} and the ability
of the primitive variable inversion scheme to maintain a uniform, non-zero velocity field when fluid momentum, energy, density and magnetic field
are advected obliquely to the grid.

We demonstrated that the integration scheme is able to evolve strong blast waves in strongly magnetized media via a modified version of the relativistic
blast wave originally due to \cite{Komissarov:1999}. In this modification,
the blast wave achieves a terminal Lorentz factor of $\Gamma = 1.8$, while allowing the evolution of magnetic fields with strengths correspond to $\beta = 10^{-8}$ in the ambient medium. Comparing results between blast waves of different strengths in media with different degrees of magnetization using both relativistic and non-relativistic physics led us to the conclusion that problems in previously reported blast wave tests arise not due to imbalances in the energy equation \cite[see e.g.][]{Del-Zanna:2007}, but due to errors in the primitive variable inversion scheme.

We applied the integration scheme to two interesting problems in computational relativistic astrophysics; the development of the Kelvin-Helmholtz Instability (KHI) and the propagation of a relativistic magnetized jet. Simulations of the KHI were computed in both two- and three-dimensions using a variety of approximate Riemann solvers. The development of and the turbulence arising from the instability was found to be strongly affected by the choice of Riemann solver. The most diffusive solver, the HLLE approximate Riemann solver, produced converged solutions at the end of the linear growth phase of the instability that lacked secondary vortices clearly present at the same time in solutions computed at a factor $32^2$ lower in resolution with the HLLC and HLLD approximate Riemann solvers, which differ by from the HLLE solver by the presence of contact (HLLC/D) and rotational discontinuities (HLLD). In three-dimensions, these lacks strongly affected the structure of the non-linear turbulence that arose from the instability. Since the scheme is stable and the HLLE solver is consistent, this result suggests that solutions computed using the HLLE solver converge to a different weak solution to the conservation law than those computed using the HLLC and HLLD solvers \citep{LeVeque:2002}.

The final problem that we considered was that of the evolution of
a relativistic, magnetized jet, the highest resolution simulations
of which were computed using SMR. These simulations
were used to probe the impact of rotational discontinuities on the
structure of the magnetic field within the jet. Simulations computed
with Riemann solvers that contained these discontinuities (HLLD)
exhibited magnetic fields within the jet cocoon
that were an order of magnitude stronger and contained a factor
$3-4$ more structure than simulation computed with Riemann solvers
that did not (HLLE).  Furthermore, simulations using the former
exhibited strongly magnetized sheath that surrounded
jet core, which was absent in simulations using the latter. The large scale
dynamics of astrophysical jets are thought to be intimately tied
to the mechanisms through which magnetic fields close to the launch
sites are dissipated and as such our results demonstrate the
importance of utilizing more accurate Riemann solvers for such studies.

The algorithms described here are but a first step in extending the
Athena code to RMHD. Future algorithmic projects will include extending the
integrator described here to generalized curvilinear coordinates
(i.e. GRMHD), extending the primitive variable inversion scheme to other equations of state \cite[e.g. the Synge gas][]{Mignone:2007a} and possibly an extension of the ``CTU+CT'' integrator of \cite{Gardiner:2005,Gardiner:2008} to relativistic fluids, utilizing the work of \cite{Anton:2010}. Finally, we will apply the algorithms described here to a full investigation of the relativistic Kelvin-Helmholtz instability and current-driven instabilities in relativistic magnetized jets in future work.


\acknowledgements{This work was supported by NASA under grant NNX09AG02G
from the Astrophysics Theory and Fundamental Physics program, under
NNX09AB90G from the Origins of Solar Systems Program and by the NSF under
grants AST-0807471 and AST-0908269. The authors acknowledge the Texas
Advanced Computing Center (TACC) at The University of Texas at Austin,
and the Princeton Institute of Computational Science and Engineering,
for providing HPC and visualization resources.  We thank Phil Armitage,
Mitch Begelman, Alexander Tchekhovskoy, Jon McKinney and Andrea Mignone for useful discussions and advice. Finally, we thank Phil Armitage,
Charles Gammie, Alexander Tchekhovskoy and an anonymous referee for careful reading of the manuscript and making several suggestions that greatly improved earlier versions of this work.}


\newpage

\clearpage

\clearpage

\begin{figure}
\begin{center}
\subfloat[][$t=10$, HLLE approximate Riemann solver]{
\includegraphics[width=\textwidth, viewport=0 0 500 600,clip]
{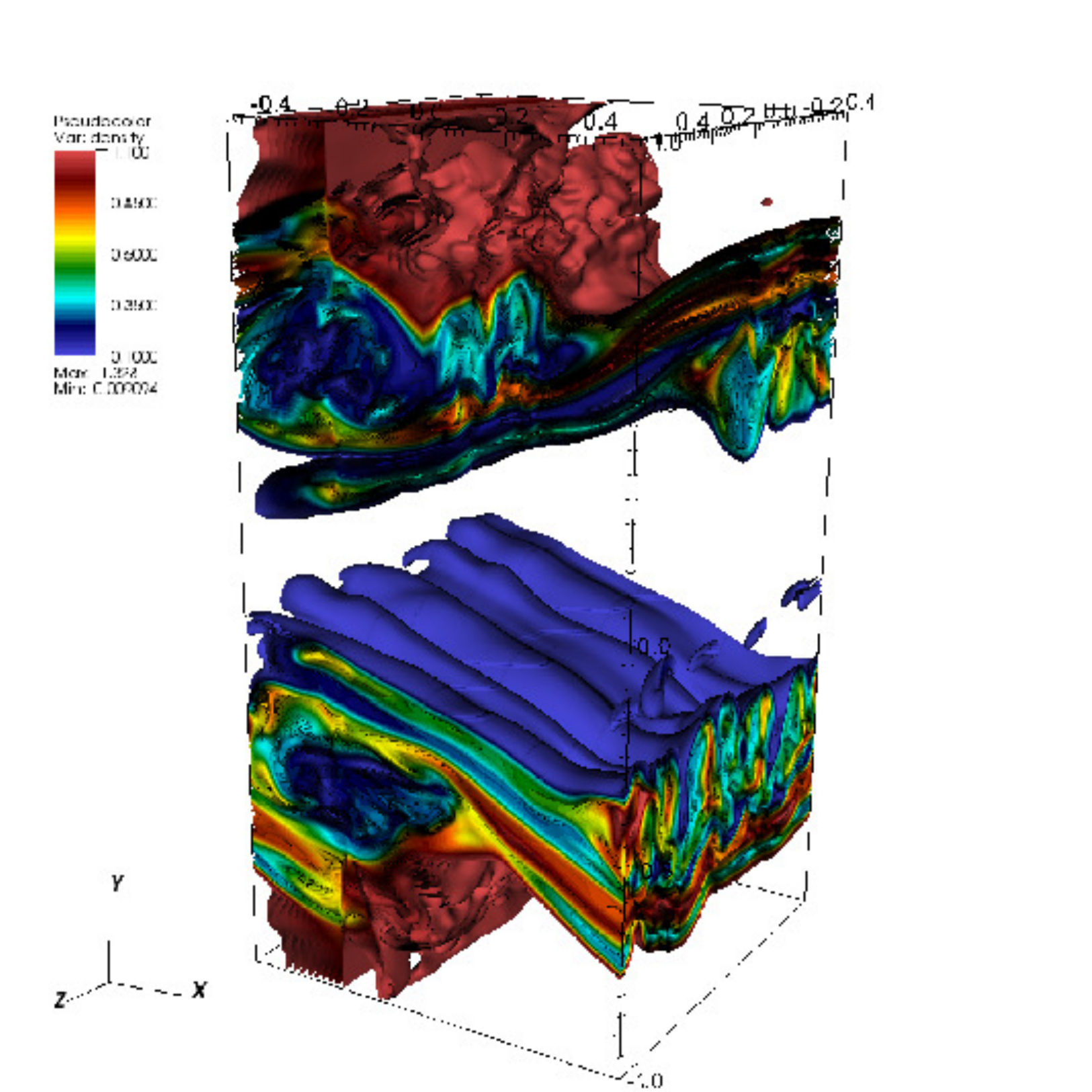}
\label{kh3d_rho_a} 
}
\end{center}
\caption[]{Volumetric rendering of the density distribution for the three-dimensional Kelvin-Helmholtz instability test. The time and Riemann solver used is given on each panel.}
\label{kh3d_rho} 
\end{figure}

\begin{figure}
\ContinuedFloat
\begin{center}
\subfloat[][$t=10$, HLLD approximate Riemann solver]{
\includegraphics[width=\textwidth, viewport=0 0 500 600,clip]
{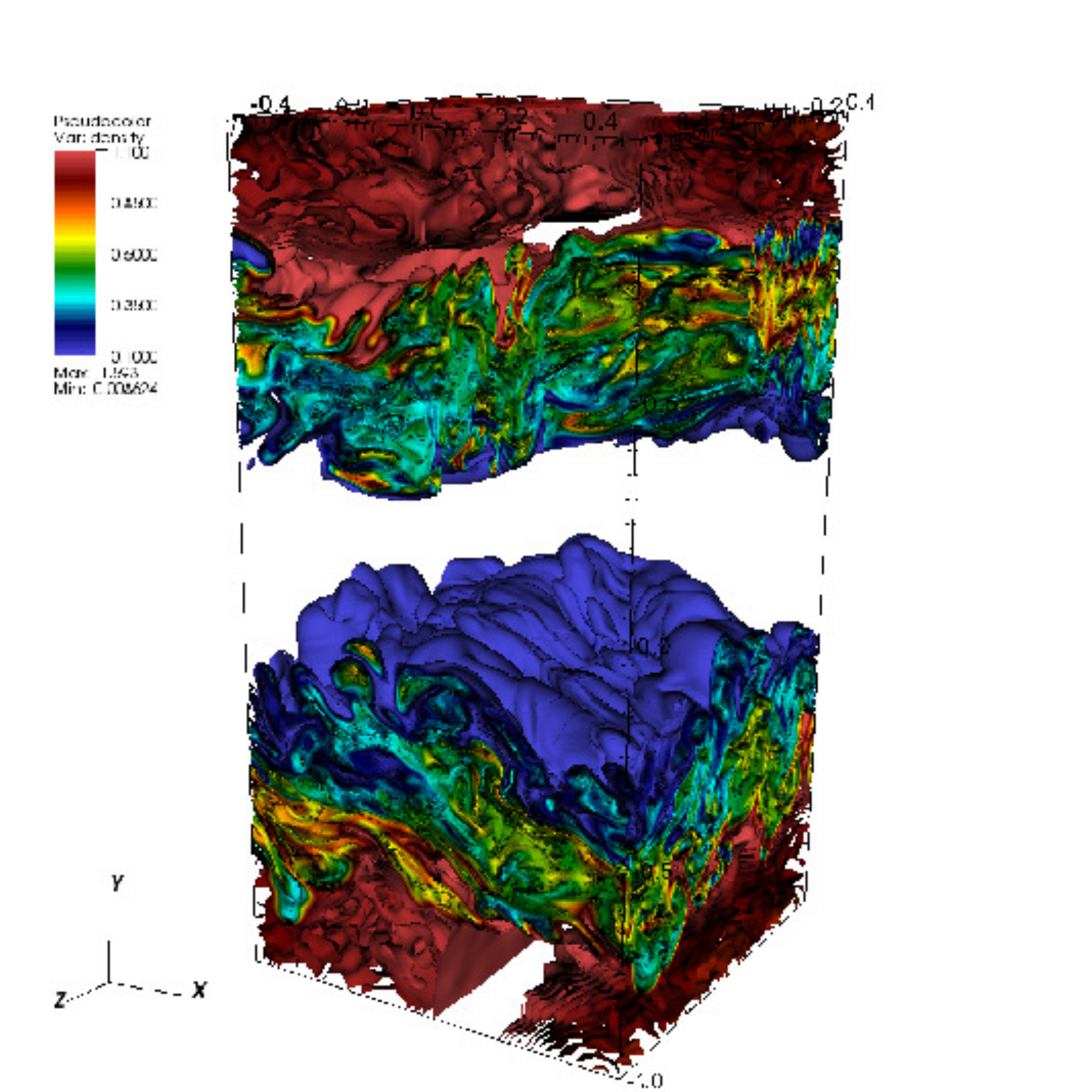}
\label{kh3d_rho_b}
}
\end{center}
\caption[]{(contin')}
\end{figure}

\begin{figure}
\ContinuedFloat
\begin{center}
\subfloat[][$t=30$, HLLE approximate Riemann solver]{
\includegraphics[width=\textwidth, viewport=0 0 500 600,clip]
{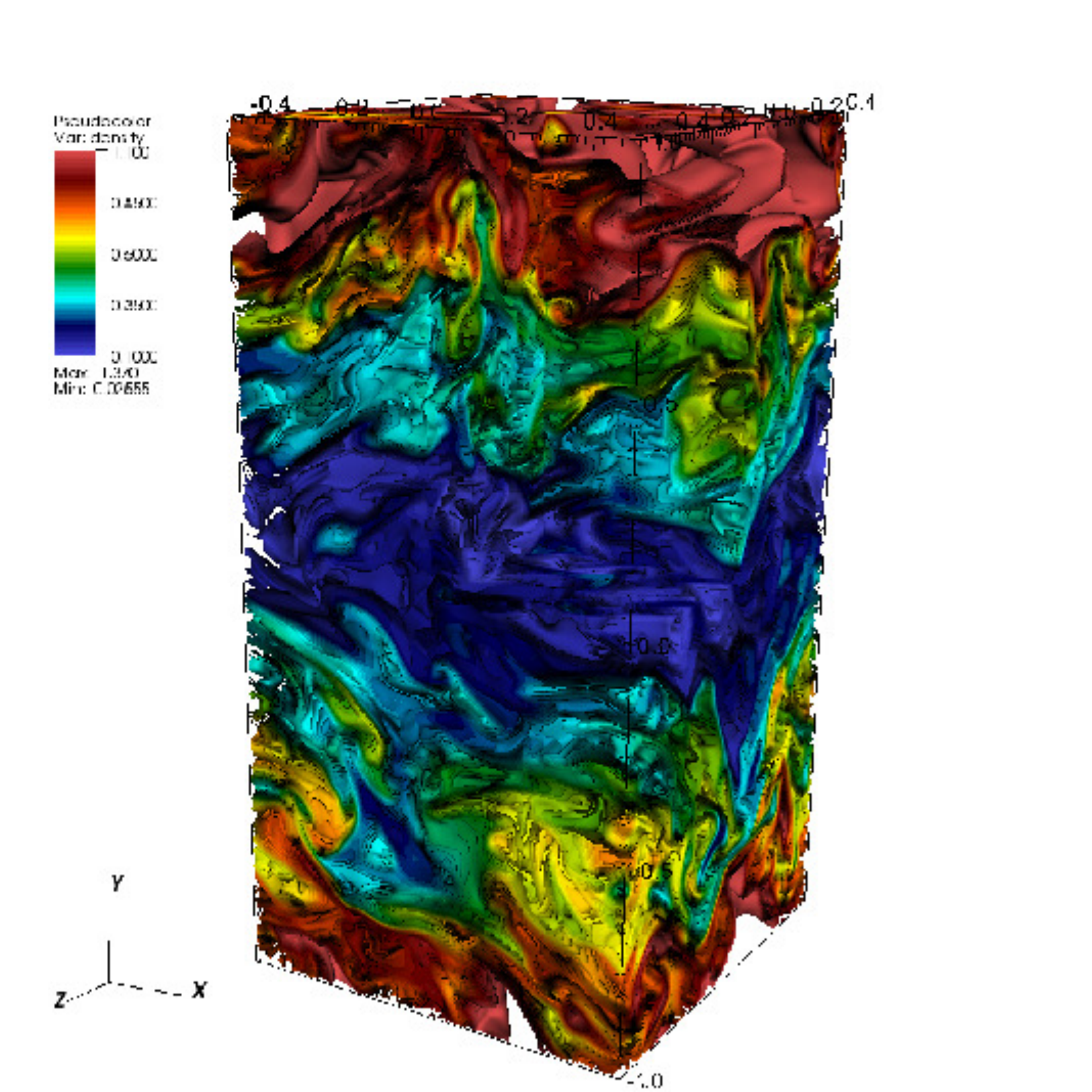}
\label{kh3d_rho_c} 
}
\end{center}
\caption[]{(contin')}
\end{figure}

\begin{figure}
\ContinuedFloat
\begin{center}
\subfloat[][$t=30$, HLLD approximate Riemann solver]{
\includegraphics[width=\textwidth, viewport=0 0 500 600,clip]
{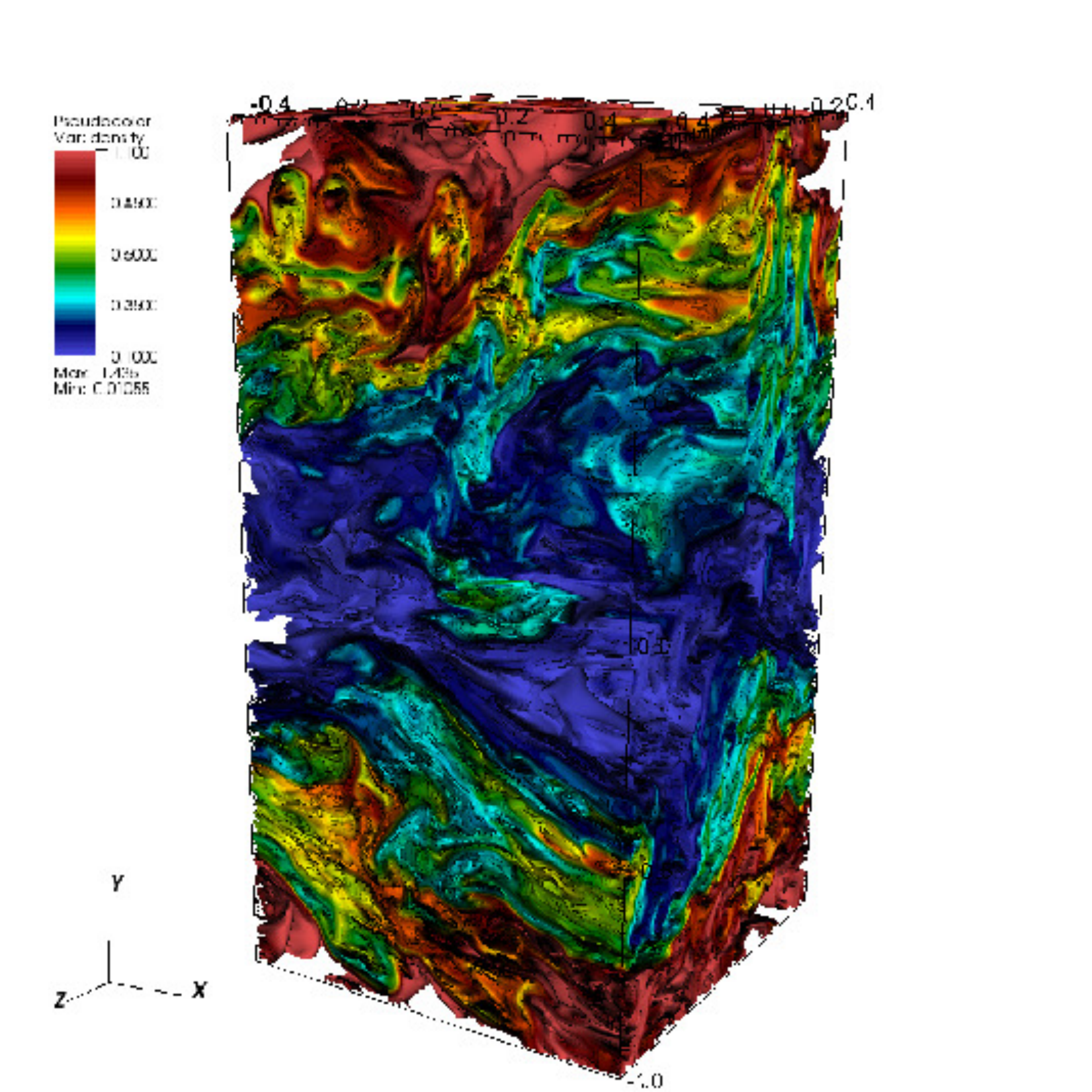}
\label{kh3d_rho_d}
}
\end{center}
\caption[]{(contin')}
\end{figure}

\begin{figure}
\begin{center}
\subfloat[][$t=10$, HLLE approximate Riemann solver]{
\includegraphics[width=\textwidth, viewport=0 0 500 600,clip]
{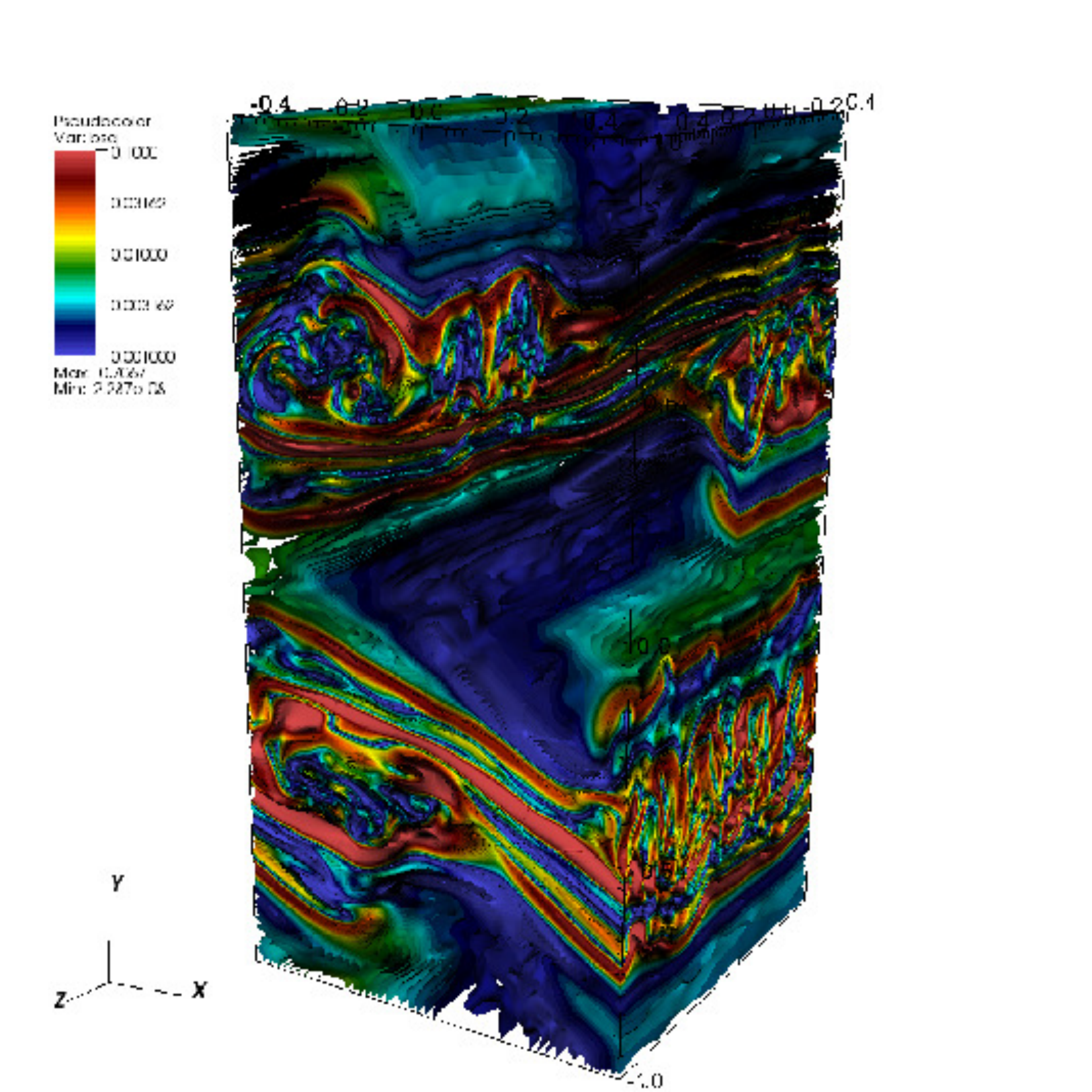}
\label{kh3d_bsq_a} 
}
\end{center}
\caption[]{Volumetric rendering of the magnetic field strength distribution for the three-dimensional Kelvin-Helmholtz instability test. The time and Riemann solver used is given on each panel.}
\label{kh3d_bsq} 
\end{figure}

\begin{figure}
\ContinuedFloat
\begin{center}
\subfloat[][$t=10$, HLLD approximate Riemann solver]{
\includegraphics[width=\textwidth, viewport=0 0 500 600,clip]
{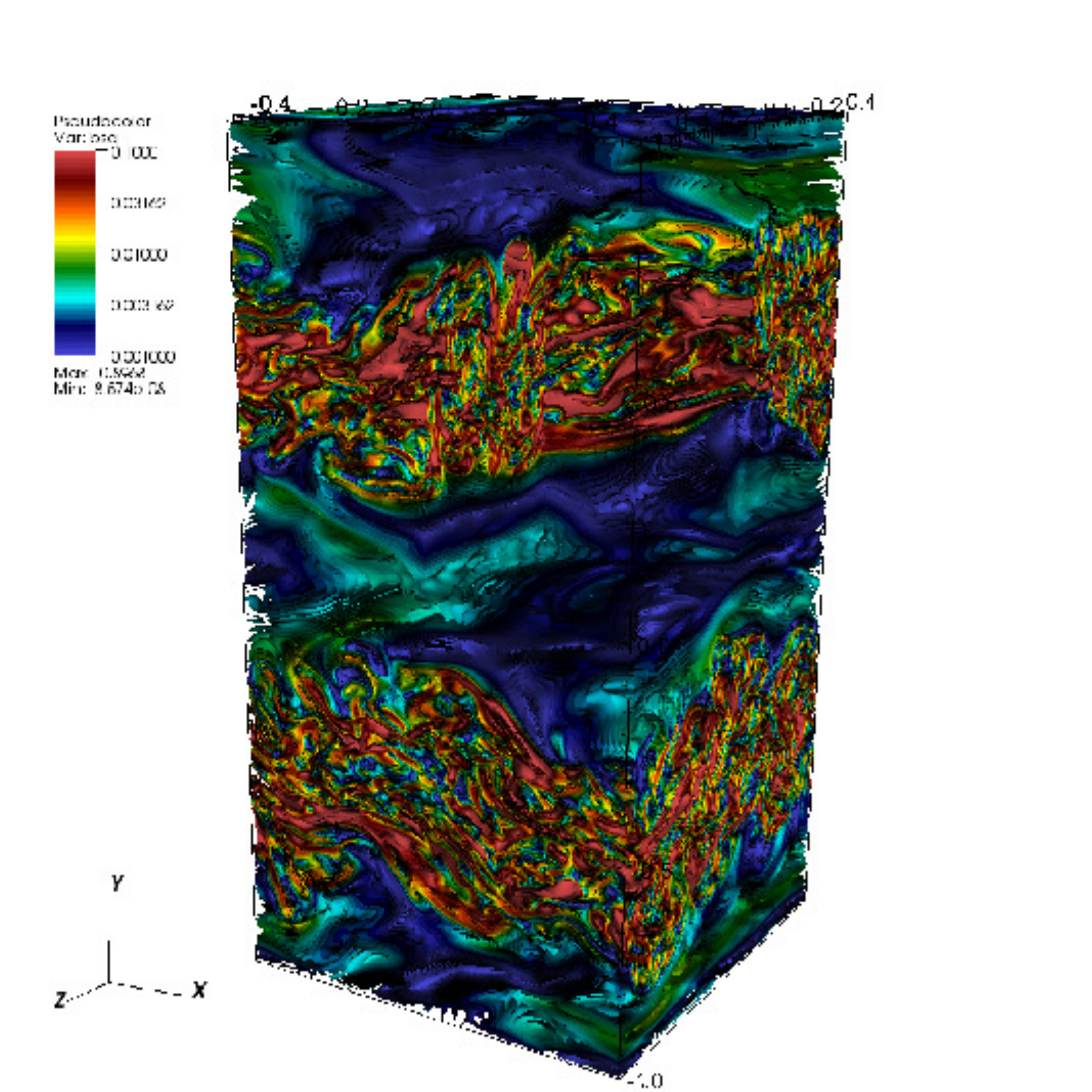}
\label{kh3d_bsq_b}
}
\end{center}
\caption[]{(contin')}
\end{figure}

\begin{figure}
\ContinuedFloat
\begin{center}
\subfloat[][$t=30$, HLLE approximate Riemann solver]{
\includegraphics[width=\textwidth, viewport=0 0 500 600,clip]
{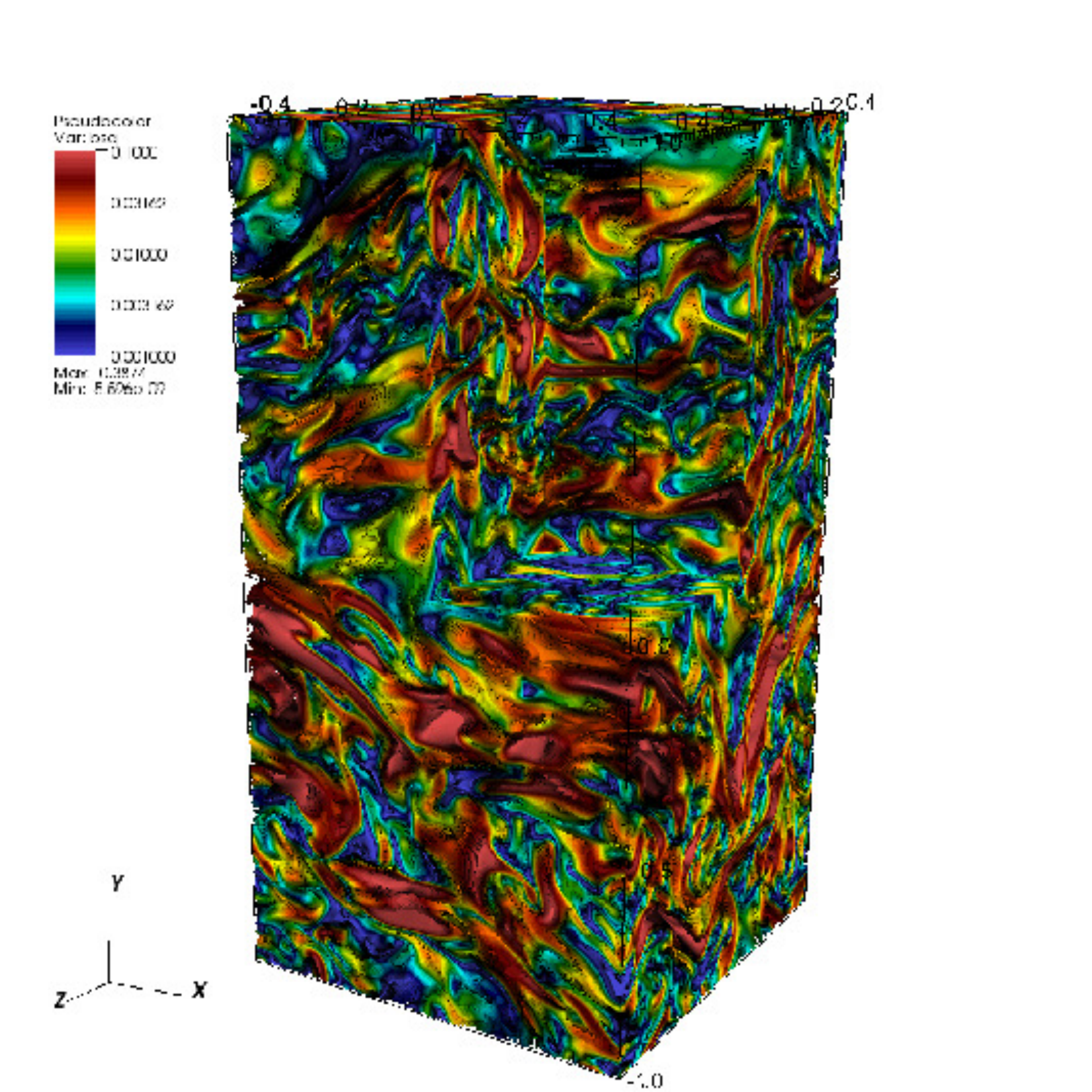}
\label{kh3d_bsq_c}
}
\end{center}
\caption[]{(contin')}
\end{figure}

\begin{figure}
\ContinuedFloat
\begin{center}
\subfloat[][$t=30$, HLLD approximate Riemann solver]{
\includegraphics[width=\textwidth, viewport=0 0 500 600,clip]
{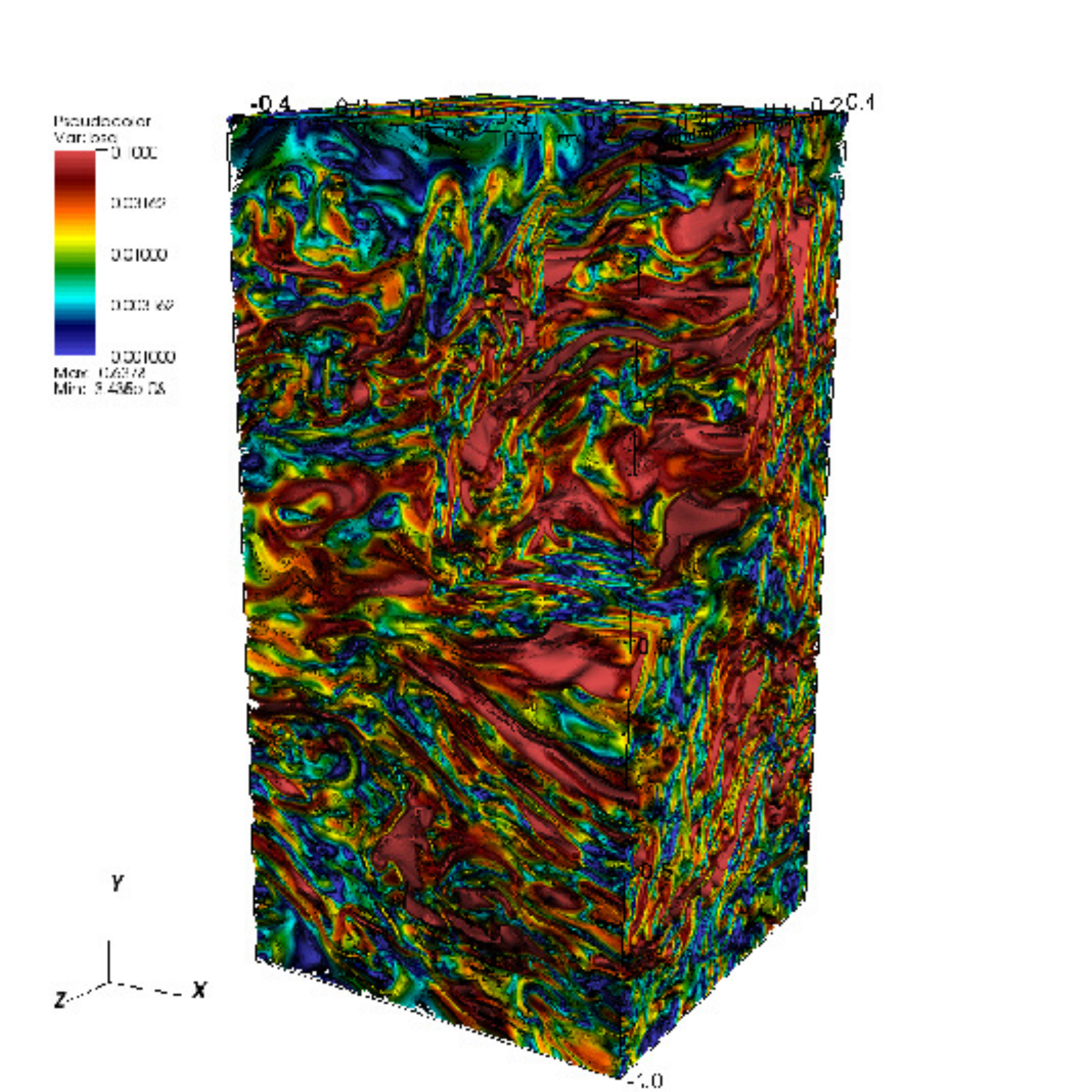}
\label{kh3d_bsq_d}
}
\end{center}
\caption[]{(contin')}
\end{figure}

\clearpage

\begin{figure}
\leavevmode
\begin{center}
\subfloat[][Solution computed using HLLC approximate Riemann solver]{
\includegraphics[width=0.85\textwidth, viewport=0 0 400 600,clip]
{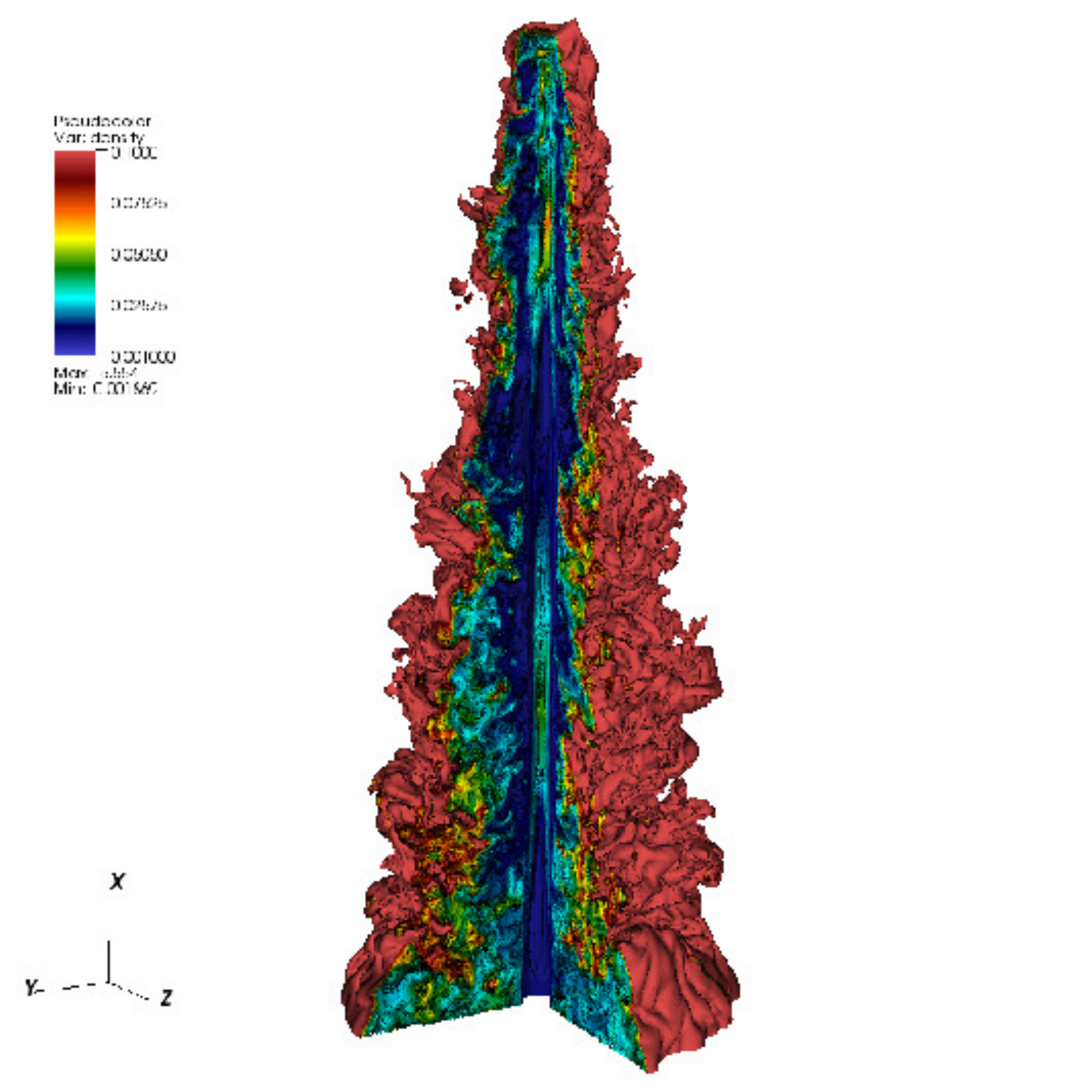}
}
\end{center}
\caption[]{Density distribution for $\Gamma=7$ jet propagating into a uniformly magnetized medium with $\beta=P_g/P_m=10.0$ at $t=100$.}
\label{SRjet_rho_3d} 
\end{figure}

\begin{figure}
\ContinuedFloat
\leavevmode
\begin{center}
\subfloat[][Solution computed using HLLD approximate Riemann solver]{
\includegraphics[width=0.85\textwidth, viewport=0 0 400 600,clip]
{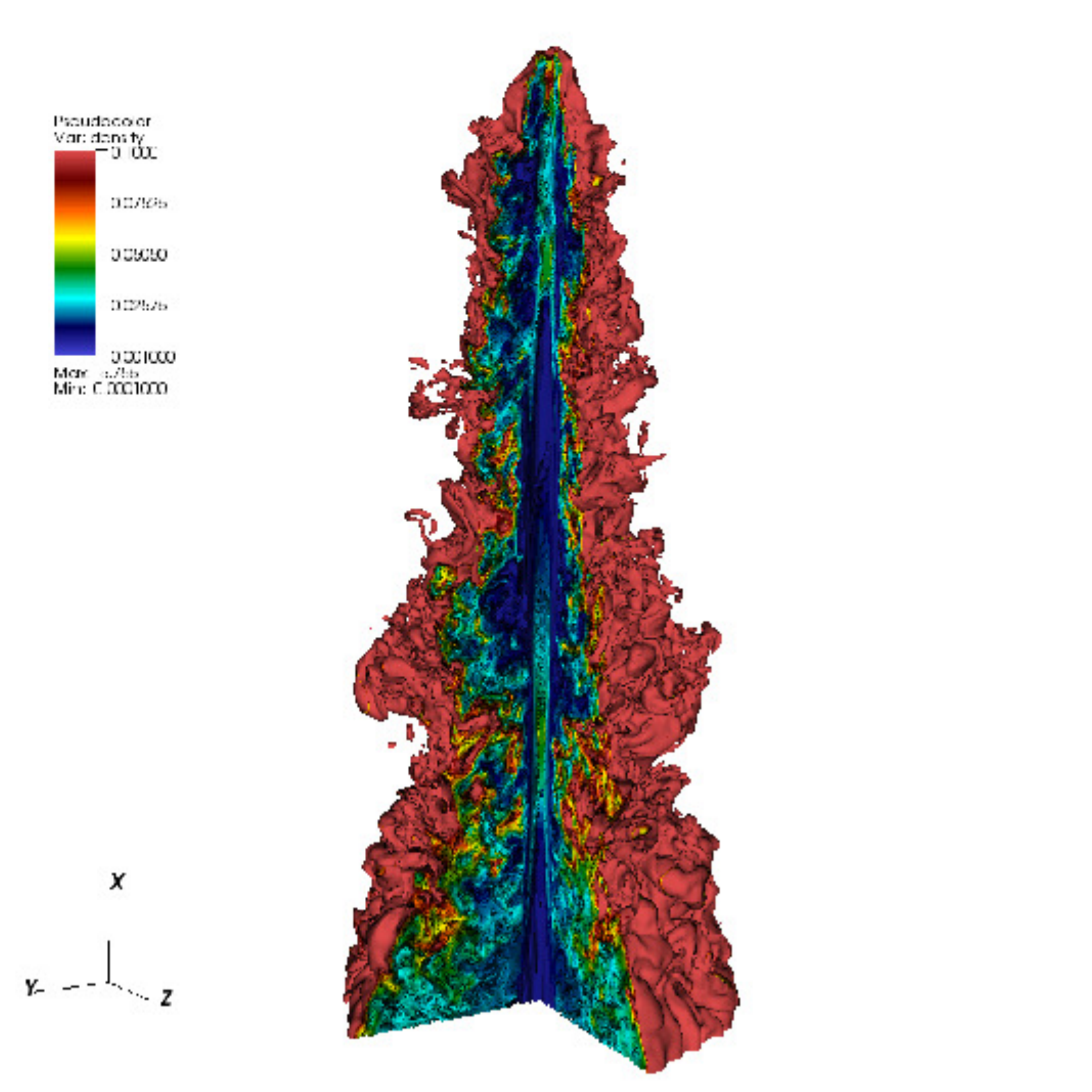}
}
\end{center}
\caption[]{(contin')}
\end{figure}

\begin{figure}
\leavevmode
\begin{center}
\subfloat[][Solution computed using HLLC approximate Riemann solver]{
\includegraphics[width=0.85\textwidth, viewport=0 0 400 600,clip]
{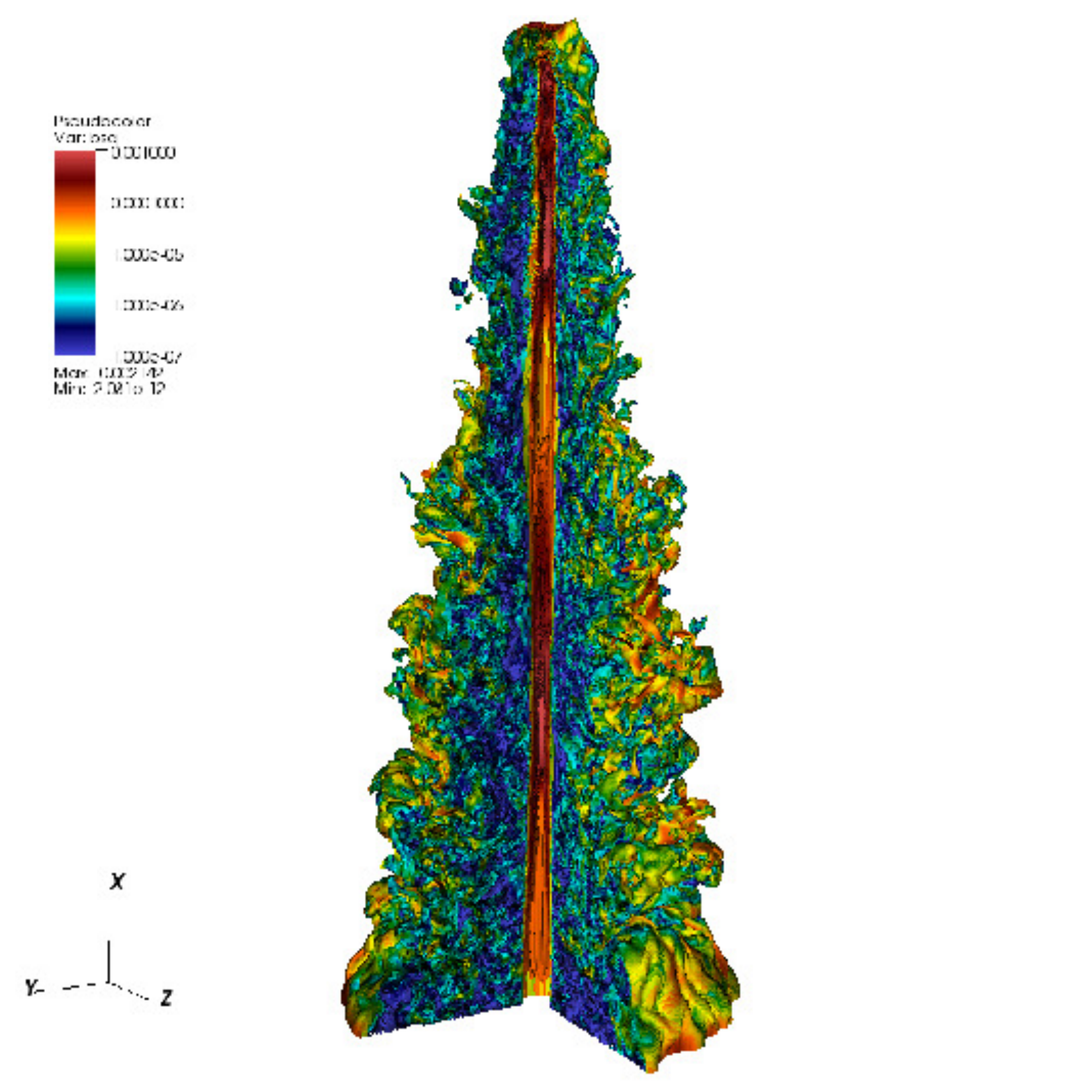}
}
\end{center}
\caption[]{Magnetic field strength distribution for $\Gamma=7$ jet propagating into a uniformly magnetized medium with $\beta=P_g/P_m=10.0$ at $t=100$.}
\label{SRjet_bsq_3d} 
\end{figure}

\begin{figure}
\ContinuedFloat
\leavevmode
\begin{center}
\subfloat[][Solution computed using HLLD approximate Riemann solver]{
\includegraphics[width=0.85\textwidth, viewport=0 0 400 600,clip]
{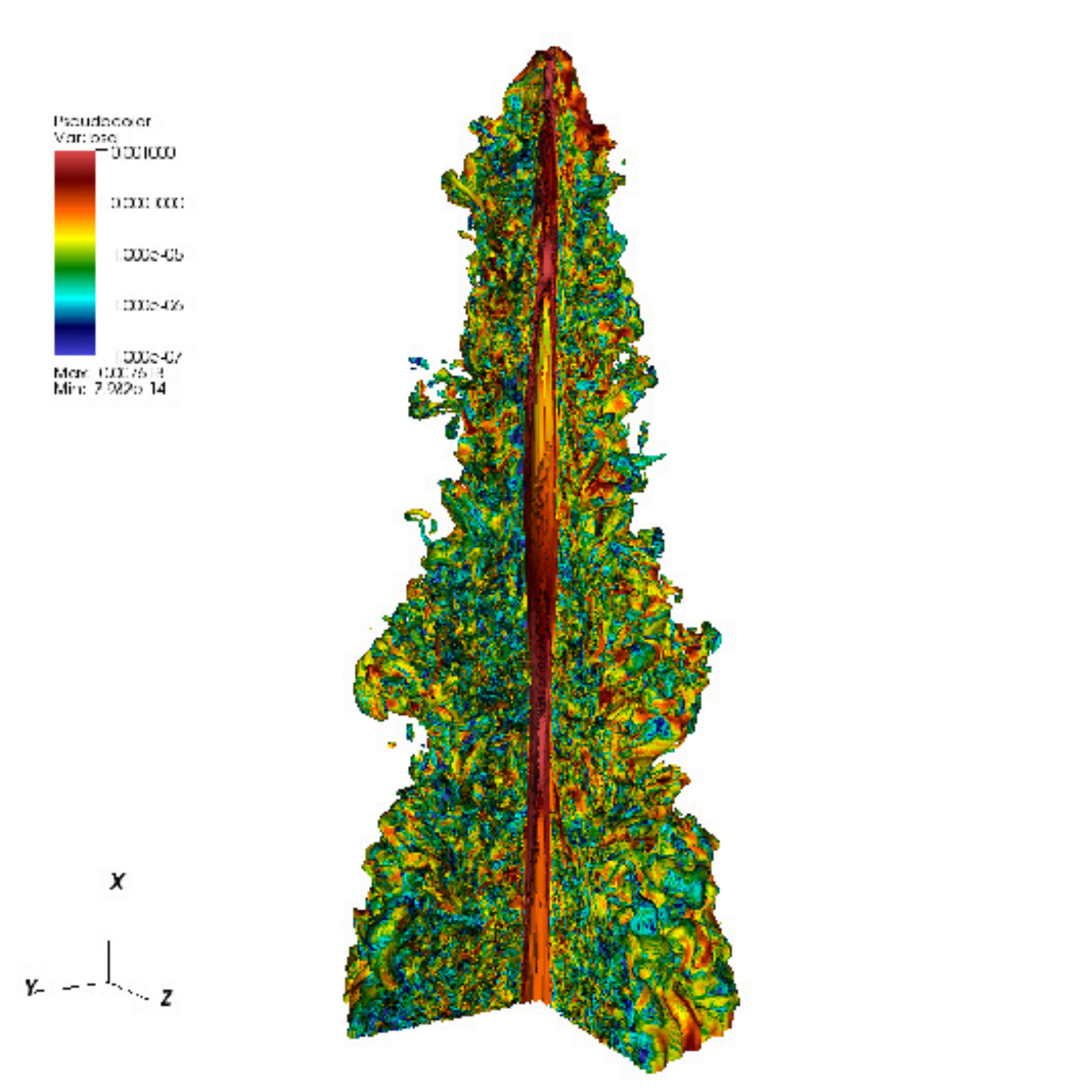}
}
\end{center}
\caption[]{(contin')}
\end{figure}

\end{document}